\documentclass[10pt]{iopart}
\usepackage{iopams}  
\usepackage{graphicx}
\usepackage[caption=false,font=footnotesize,position=top]{subfig}
\usepackage{float}
\usepackage{mathrsfs}
\usepackage[makeroom]{cancel}
\usepackage{changepage}
\usepackage{color}
\usepackage{cancel}
\usepackage[colorlinks=true,linkcolor=blue,citecolor=blue,urlcolor=blue]{hyperref}
\usepackage[euler]{textgreek}
\usepackage{tikz}
\usepackage{xcolor}

\newcommand*\circled[1]{\tikz[baseline=(char.base)]{
            \node[shape=circle,draw,inner sep=2pt] (char) {#1};}}
\newcommand\beq{\begin{equation}}
\newcommand\eeq{\end{equation}}
\newcommand{\bea}{\begin{eqnarray}}
\newcommand{\eea}{\end{eqnarray}}
\newcommand{\vpa}{v_{\parallel}}
\newcommand{\vpe}{v_{\perp}}
\newcommand{\vthi}{v_{\mathrm{th}i}}
\newcommand{\vthe}{v_{\mathrm{th}e}}

\newcommand{\eqref}[1]{Eq.~(\ref{#1})}
\newcommand{\uperp}{u_\perp}

\newcommand{\kpar}{k_{\parallel}}
\newcommand{\kperp}{k_{\perp}}

\newcommand{\apar}{A_{\parallel}}
\newcommand{\bpar}{B_{\parallel}}
\newcommand{\text}[1]{{\rm #1}}
\newcommand{\unit}[1]{\hat{\mathbf{#1}}}
\newcommand{\mbf}[1]{\mathbf{#1}}
\newcommand{\alfnc}{Alfv\'enic\text{ }}
\newcommand{\alf}{Alfv\'en\text{ }}
\newcommand{\delB}{\delta B_{\parallel}}
\newcommand{\deln}{\delta n}
\newcommand{\delT}{\delta T}

\newcommand{\dfrac}[2]{\displaystyle{\frac{#1}{#2}}}
\newcommand{\zphys}{z_{\text{phys}}}
\newcommand{\kparphys}{k_{\parallel,\text{phys}}}
\newcommand{\kzphys}{k_{z,\text{phys}}}
\newcommand{\omegaphys}{\omega_{\text{phys}}}

\newcommand{\sd}{\mathrm{d}}
\newcommand{\avg}[2]{\left<#1\right>_{#2}}

\renewcommand{\figurename}{Fig.}

\begin{document}

\title[Electromagnetic ITG high-beta runaway]{High-beta runaway transitions in a fluid model of electromagnetic ion-temperature-gradient turbulence}
\author[Yujia zhang et al.] 
{Y~Zhang,$^{1,2,3}$
M~Barnes,$^{1,4}$
A~A~Schekochihin,$^{1,5}$
P~G~Ivanov,$^{1,6,7}$
T~Adkins,$^{8}$      
}

\address{$^1$Rudolf Peierls Centre for Theoretical Physics, University of Oxford, Oxford OX1 3PU, UK}
\address{$^2$St Edmund Hall College, Oxford OX1 4AR, UK}
\address{$^3$Tokamak Energy Ltd, 173 Brook Dr, Milton, Abingdon OX14 4SD, UK}
\address{$^4$University College, Oxford OX1 4BH, UK}
\address{$^5$Merton College, Oxford OX1 4JD, UK}
\address{$^6$United Kingdom Atomic Energy Authority, Abingdon OX14 3DB, UK}
\address{$^7$Swiss Plasma Center, École polytechnique fédérale de Lausanne, Rte Cantonale, 1015 Lausanne, Switzerland}
\address{$^8$Princeton Plasma Physics Laboratory, 100 Stellarator Rd, Princeton, NJ 08540, USA}

\ead{yujia.zhang@seh.ox.ac.uk}

\begin{abstract}
Gyrokinetic simulations of tokamak turbulence indicate that fluctuation levels increase abruptly and dramatically when the plasma beta exceeds a certain critical value.  This increase in fluctuation levels
coincides with a transition from a state dominated by zonal flow to one in which turbulent eddies form radially-elongated `streamers'. Here we derive from gyrokinetics a minimal fluid model for electromagnetic ion-temperature-gradient (ITG) turbulence that captures the key features of this transition.  Due to the relative simplicity of the model, we are able to conduct a detailed numerical study of the interplay between the turbulence and the zonal flow across a broad range of values of the plasma beta and the ITG. We find that the transition occurs when the Reynolds stress, which tends to strengthen zonal flows, is overwhelmed by the Maxwell and diamagnetic stresses, which tend to weaken them. Power-law scalings of the stress ratios with plasma beta and ITG are obtained, indicating a possible means by which the location of the transition could be predicted with minimal computational cost.
\end{abstract}

\submitto{\PPCF\rm \!\!; this draft is of \today}

%
%
%
\maketitle
%
\ioptwocol

\section{Introduction}
\label{sec:intro}
In most magnetic-confinement-fusion plasmas, energy confinement is limited by turbulent mixing.  The applied magnetic fields used to confine such plasmas
are strong -- on the order of a few Tesla -- and the plasma pressure is sufficiently weak that the turbulent fluctuations are often well-approximated as electrostatic in nature.
While not all aspects of these electrostatic turbulence fluctuations are fully understood, many of their basic properties have been identified. In particular, 
the turbulence is usually driven by one of a number of instabilities with a characteristic wavelength comparable to the ion gyro-radius and feeds off a combination of gradients 
in the plasma density, flow and temperature.
The archetypal example is the curvature-mediated ion-temperature-gradient, or ITG, instability~\cite{hortonPoF_1981, Waltz_1988, romanelli_1989, Kotschenreuther_1995, evensenNF_1998, Roger_2005, Zocco_2018}, which has been observed experimentally to dominate
turbulent transport in a range of scenarios~\cite{Citrin_2017,van_Wyk_2017}. It has been extensively characterised by solving the gyrokinetic-Poisson system of equations, most commonly via numerical
simulations~\cite{Dimits_2007, Saarelma_2012, Navarro_2020, Mazzi_2022}, but also via analytical calculations in a handful of simplified limits~\cite{Anderson_2002,ivanov_2020, Ivanov_2022}.  

However, a growing number of current experiments, as well as many proposed fusion-reactor plasmas, reach large enough plasma pressures for the confining magnetic fields to be perturbed significantly. In such cases, the magnetic fluctuations can qualitatively change the turbulence and associated transport levels. The quantity that parameterises the relative importance of magnetic-field fluctuations is the plasma beta, $\beta$, defined as the ratio of the plasma pressure to the magnetic pressure.  Previous studies have shown that the inclusion of finite-$\beta$ effects can lead to stabilisation of
the linear ITG instability~\cite{Kim_1993,AJarmen,Kim_2019} and to nonlinear suppression of turbulence~\cite{SnyderGroF, YChen,Transport,citrin_non, Citrin_2014_EMstab, whelan_2018} even
for $\beta$ values on the order of a percent.  This 
makes physical sense: the bending of field lines by turbulent fluctuations produces a restoring force that resists the motions induced by the fluctuations.
Consequently, it has been proposed that going to higher beta via increasing both thermal plasma pressure and by introducing high-energy, or fast, particles into the plasma could improve confinement~\cite{Transport, Citrin_2014_EMstab, Sama_2024}.

\begin{figure*}[!t]
\centering
\begin{minipage}{0.45\textwidth}
    \centering
    \includegraphics[width=8cm]{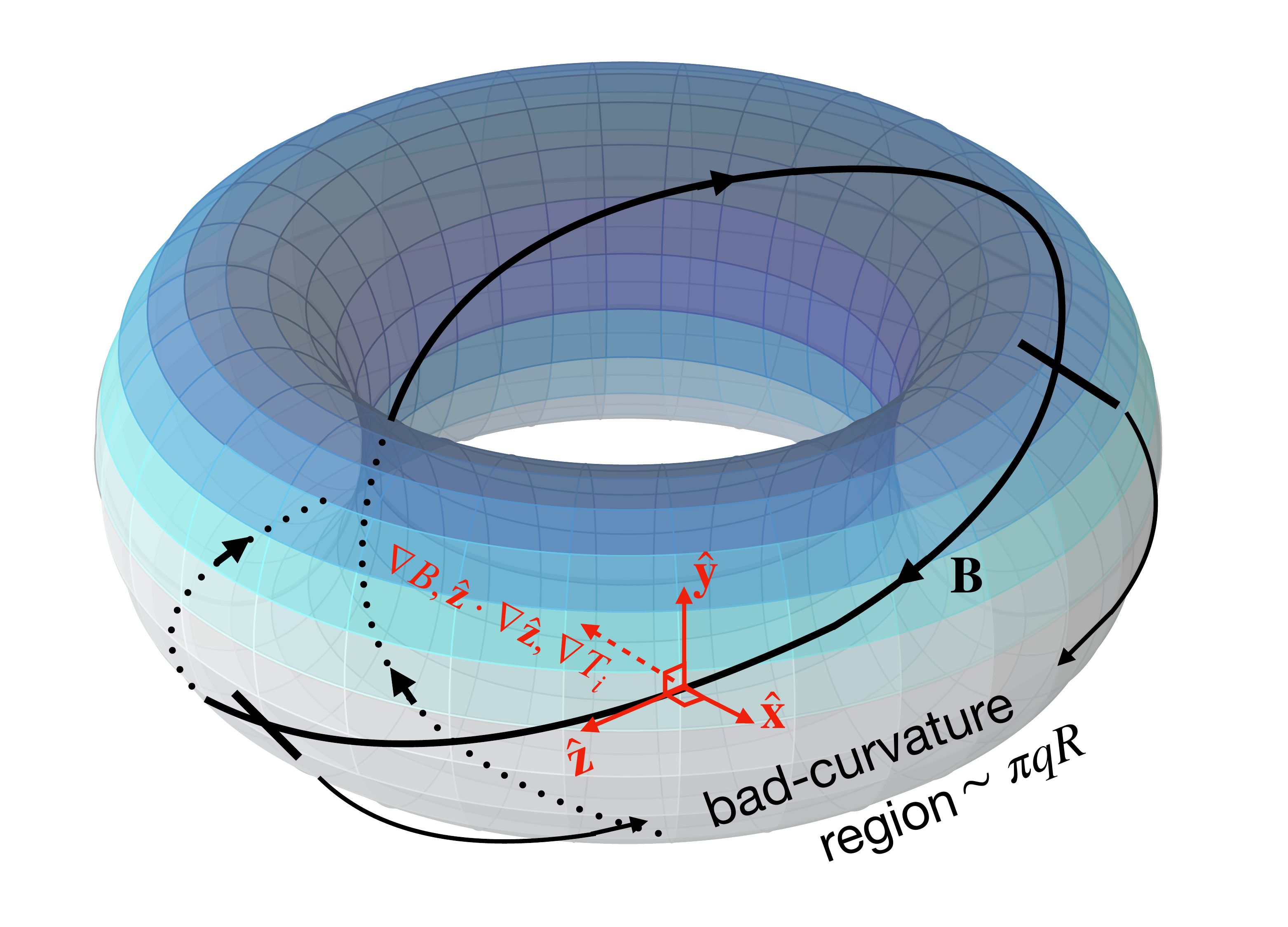}
\end{minipage}%
\hspace{0.0\textwidth} 
\begin{minipage}{0.45\textwidth}
    \centering
    \includegraphics[width=10cm]{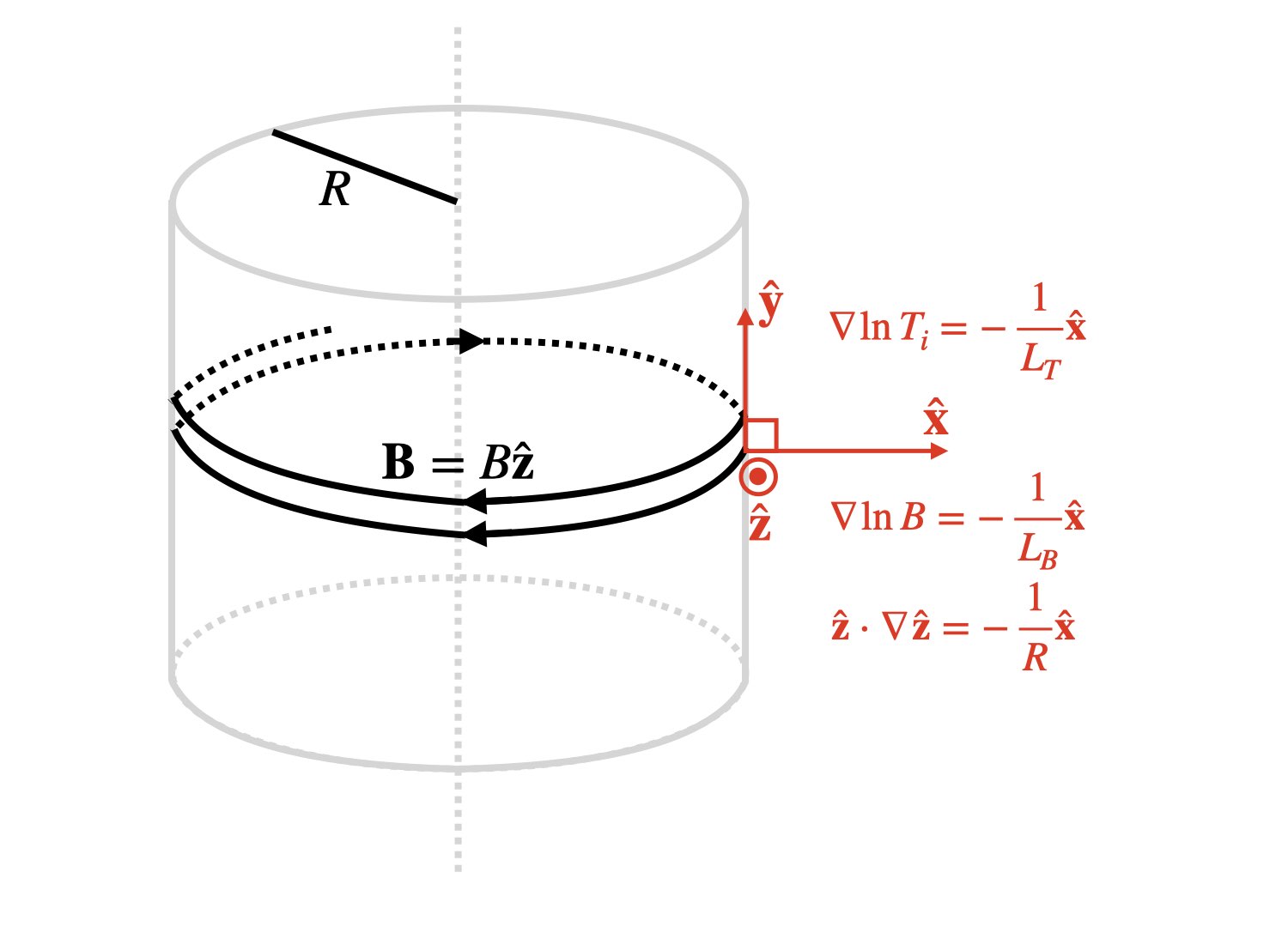}
\end{minipage}
\caption{\textbf{Left panel}: An illustration of the out-board side of a tokamak, sometimes called the bad-curvature region. The length of that region is roughly $\pi qR$ where $R$ is the major radius. Here, $\unit{z}$ is the direction of the equilibrium magnetic field, so the equilibrium field is $\mathbf{B} = B \unit{z}$ where $B$ is the field strength, $\unit{x}$ is the radial direction, and $\unit{y}$ is the poloidal direction. In a general tokamak geometry $\unit{x}$ and $\unit{y}$ are not orthonormal~\cite{Beer_1995_paper, highcock2012}. \textbf{Right panel}: $Z$-pinch equilibrium. In the limit of large-aspect ratio, large $q$ and small magnetic shear $\hat{s}$, the out-board side of the flux-surface resembles a $Z$-pinch geometry where $\unit{x}$, $\unit{y}$, and $\unit{z}$ form an orthonormal basis; the major radius $R$ becomes the radius of curvature of the field line.}
\label{fig:zpinch}
\end{figure*}

There is, however, an emerging body of evidence from gyrofluid~\cite{Scott_2006} and gyrokinetic~\cite{waltz_2010, Pueschel_2013_PhysRev,hysteresis, Giacomin} simulations that at even higher $\beta$ values, there is an abrupt transition from a zonally-dominated state with a modest heat flux to a state with radially-elongated streamers and a very large heat flux. The heat fluxes found numerically are orders of magnitude larger than the gyro-Bohm estimate in some cases~\cite{Giacomin}, and in others the turbulence does not appear to saturate at all~\cite{waltz_2010, Pueschel_2013_PhysRev,hysteresis}. This is found across a wide range of plasma parameters and simulation codes. Explanations for these high (or diverging) fluxes are varied: \cite{waltz_2010} claimed that this was due to tertiary electromagnetic instabilities that feed on the zonal structure; \cite{Mulholland_2025} showed that subdominant linear electromagnetic instabilities like kinetic ballooning modes (KBMs) \cite{Tang_1980,cheng_1982, Aleynikova} alone might be sufficient to drive the high flux levels via their interaction with tearing-parity modes; other studies showed that a marked increase in field-line wandering was correlated with the transition \cite{Pueschel_2013_PhysRev, nonzonal}, and hypothesised that electrons streaming along the stochastic field could `short out' zonal structures~\cite{Pueschel_2013_PhysRev}; more recently, a direct study of zonal-flow dynamics for electromagnetic turbulence highlighted the importance of the competition between Reynolds and Maxwell stresses for the transition~\cite{hysteresis}. 

The transition described above is strikingly similar to one that is observed in fluid simulations of electrostatic ITG turbulence~\cite{ivanov_2020, Ivanov_2022} when the temperature gradient is increased beyond a critical value. The latter transition, believed to be the minimal model of the so called Dimits transition~\cite{Dimits} in gyrokinetic plasmas, correlates well with a change in sign of the effective turbulent viscosity for the zonal flow (in~\cite{ivanov_2020, Ivanov_2022}, the negative and positive viscosities are identified as due to the Reynolds and diamagnetic stresses, respectively). A negative viscosity at small temperature gradients supports zonal flows whereas positive viscosities beyond some critical temperature gradient opposes them.  As weakening of zonal flows has also been observed in gyrokinetic simulations of the non-zonal transitions \cite{Pueschel_2013_PhysRev, nonzonal}, this motivates us to explore whether the same logic applies there.

In the spirit of~\cite{ivanov_2020}, we formulate here a minimal local model for electromagnetic ITG turbulence in a plasma with curved equilibrium magnetic field and find numerically that transitions from low- to high-transport states occur at certain critical values of the plasma beta and ion temperature gradient, with the latter turning out to be significantly lower than in the electrostatic limit of $\beta = 0$.  These values do indeed turn out to correspond to locations in parameter space where the turbulent viscosity, which now has an additional contribution due to the Maxwell stress, changes sign, and it is shown that the viscosity associated with the Maxwell stress tends to be positive, i.e., it reduces zonal flow shear. 

The rest of this paper is organized as follows.  We present the fluid model equations, along with the relevant ordering assumptions in Section~\ref{sec:fluid_model}, and we analyse the linear physics contained in the model in Section~\ref{sec:linear_physics}. In Section~\ref{sec:nonlin_sim}, we present results from nonlinear simulations that exhibit transitions to high-flux states beyond critical values of $\beta$ and temperature gradient. We also show that the sign of the turbulent viscosity determines the transition boundary. We conclude in Section~\ref{sec:discussion} with a discussion of the key results.

\section{A fluid model of electromagnetic ITG turbulence}
\label{sec:fluid_model}
Our aim is to construct an asymptotically correct, minimal model for electromagnetic ITG turbulence that exhibits the key features of the transitions from low to high transport observed in nonlinear gyrokinetic simulations. 

\subsection{Magnetic geometry}
\label{sec:mag_geo}
We consider a plasma consisting of electrons of mass $m_e$ and charge $-e$ and a single ion species of mass $m_i$ and charge $Z e$, immersed in a static magnetic field of constant curvature.  Such a configuration, often referred to as a $Z$-pinch or `curvy slab', is illustrated in the right panel of Fig.~\ref{fig:zpinch}.  The equilibrium magnetic field $\mathbf{B}=B(x)\unit{z}$ varies in magnitude along its curvature vector $(-\unit{x}/R)$ with the characteristic length scale \mbox{$L_B \equiv (-\sd \ln B/\sd x)^{-1}$}.  The equilibrium ion temperature $T_i$ is also allowed to vary in $x$ with characteristic length scale \mbox{$L_T \equiv (-\sd\ln T_i/\sd x)^{-1}$}, while the equilibrium ion density $n_i$, electron density $n_e$, and electron temperature $T_e$ are assumed constant. Similar electromagnetic turbulence models have been studied previously for electron-scale turbulence~\cite{Adkins_2022}. This geometry can be viewed as a local slice taken from the out-board side of a large-aspect-ratio tokamak with nearly toroidal magnetic field (corresponding to a large safety factor, $q$), as illustrated in Fig.~\ref{fig:zpinch}. The out-board side is where the ITG instability is most vigorous due to the alignment between the magnetic-field gradient (as well as curvature) and the ion-temperature gradient. 

There exists a constraint relating $L_B$, $R$ and $L_T$ that arises from the equilibrium force balance:
\begin{equation}
\label{equilibrium_force_balance}
\nabla p = \frac{\mathbf{j} \times \mathbf{B}}{c}, 
\end{equation}
where $p = n_iT_i + n_eT_e$ is the total equilibrium pressure and $\mathbf{j} = (c/4\pi)\nabla \times \mathbf{B}$ is the equilibrium current. Substituting these expressions into (\ref{equilibrium_force_balance}), and assuming zero density and electron-temperature gradients, we obtain
\beq
n_i \nabla T_i  =   \frac{1}{4\pi}\left[ \left(\mathbf{B}\cdot\nabla\right)\mathbf{B} - \frac{1}{2}\nabla B^2\right].
\eeq
Applying this constraint to our $Z$-pinch magnetic field yields
\beq
\label{eq:equilibrium constraint}
\frac{\beta_i}{2}\frac{1}{L_T} + \frac{1}{L_B}  =  \frac{1}{R}.
\eeq

\subsection{Orderings}
\label{Orderings}
As shown in \ref{derivation}, the electrons can be modelled as an isothermal fluid by exploiting the smallness of the electron-ion mass ratio~\cite{Schekochihin_2009}. A fluid model for the ions is obtained from the gyrokinetic-Maxwell system of equations~\cite{Frieman_1982, Abel_2013} by taking the ion-ion collision frequency $\nu_{ii}$ to be large and by restricting attention to wavelengths much larger than the ion thermal gyroradius $\rho_i = \vthi/\Omega_i$, where $\vthi=\sqrt{2T_i/m_i}$, $\Omega_i = ZeB/m_ic$, $m_i$ is the mass of ions, and $c$ is the speed of light. The fastest-growing ITG modes reside at scales comparable to the ion sound radius $\rho_s = c_s/\Omega_i$, with the ion sound speed $c_s = \sqrt{ZT_e/m_i}$. We thus require the ion-electron temperature ratio to be small so that these scales are included in the model. The specific orderings that we adopt are
\beq
\label{eq:para_ordering}
\sqrt{\frac{m_e}{m_i}} \ll \frac{L_T}{L_B} \sim \tau \sim \kperp^2\rho^2_i\ll \beta_e  \ll 1 \sim  k_{\perp}\rho_s,
\eeq
where $\tau = T_i / Z T_e$, $k_{\perp}=|k_x \unit{x} + k_y\unit{y}|$, with $k_x$ and $k_y$
the wave numbers in the $x$ and $y$ directions, respectively, and $\beta_{\alpha} = 8\pi n_{\alpha} T_{\alpha} / B^2$ is the ratio of thermal pressure for species $\alpha$ to magnetic pressure.

The frequencies present in the system are ordered to ensure that both the ITG and \alfnc fluctuations are included and to allow (in principle) for turbulence that is in critical balance, viz.,
\bea
\label{eq:frequency ordering}
\omega_{*T}\sim \omega \sim k_{\parallel} v_A \sim k_{\perp}|\mbf{v}_E| \sim \nu_{ii} \tau,
\eea
where 
\bea
\label{eq:omega_star}
\omega_{*T} \equiv k_y\rho_s\frac{c_s}{L_T}\tau \sim \frac{c_s}{L_B}
\eea
is the ion diamagnetic frequency, $\omega$ is a characteristic frequency of the turbulence, $k_{\parallel}$ is a characteristic wave number along the total magnetic field, 
\bea
v_A = \frac{\vthi}{\sqrt{\beta_i}}
\eea
is the \alf speed and
\beq
\mbf{v}_E = \frac{c}{B}\unit{z}\times\nabla\phi
\eeq
is the $\mbf{E}\times\mbf{B}$ velocity, with $\phi$ the electrostatic potential. 

Note that the orderings (\ref{eq:frequency ordering}) and the expression~(\ref{eq:omega_star}) constrain $k_{\parallel}L_B\sim \sqrt{\beta_e} \ll 1$ and $k_{\parallel} c_s \ll \omega$, with the latter restriction removing sound waves from the model. The observant reader might note that the ordering $k_{\parallel}L_B \sim \sqrt{\beta_e} \ll 1$ is not consistent with a purely azimuthal equilibrium magnetic field of radius $R=L_B$, as this implies that parallel wavelengths are longer than the circumference of the field. This inconsistency is overcome by allowing for an infinitesimal $y$-component of the equilibrium magnetic field such that it no longer connects exactly to itself after one turn. This picture is consistent with our attempt to construct a local approximation of the out-board side of a tokamak with a large $q$. It also implies that the ion drift frequency, 
\beq
\omega_{di} \equiv k_y\rho_s \frac{c_s}{L_B}\tau \sim \frac{c_s}{L_B}\tau,
\eeq
is much smaller than the ion diamagnetic frequency~$\omega_{*T}$. 

All that is left is to order the amplitudes of the fluctuations in the electromagnetic fields and in the ion fluid moments. To recover the Boltzmann response of electrons in the electrostatic limit we take $e\phi/T_e \sim \deln_e/n_e$, with $\deln_e$ the electron density fluctuation. The ion temperature fluctuation is ordered so that $\omega_{di} \delT_i/T_i\sim \omega \deln_e/n_e$, which ensures that the physics of curvature-ITG instability is captured.~We further demand that the bending of the magnetic field due to fluctuations $\delta \mbf{B}_{\perp}$ of the field vector perpendicular to $\unit{z}$ is dynamically important so that $(\delta \mbf{B}_{\perp}/B) \cdot
\nabla \sim \unit{z}\cdot\nabla$ when operating on fluctuations of any quantity. In combination with $\kperp |\mbf{v}_E| \sim \kpar v_A$, this is equivalent to allowing \alf waves to exist in our system: $e\phi/T_e \sim \apar/\rho_s B\sqrt{\beta_e}$, where $\apar$ is the fluctuation in the $z$-component of the magnetic vector potential, $\delta \mbf{B}_{\perp} = - \unit{z}\times\nabla_{\perp}\apar$. Finally, we note that, at low beta, the removal of sound waves from the system coincides with the removal of all compressive fluctuations, including of the slow-wave kind, and thus fluctuations in the parallel component of the magnetic field $\delB$ can be neglected. Piecing all of these constraints together results in the ordering
\beq
\label{eq:fields_ordering}
\frac{e\phi}{T_e}\sim \frac{\deln_e}{n_e}\sim \tau\frac{\delT_i}{T_i}\sim \frac{\apar}{\rho_sB\sqrt{\beta_e}}.
\eeq

\begin{table*}[!t]
\caption{Normalisations of fields, coordinates and parameters} 
\centering
\setlength{\tabcolsep}{6pt} 
\renewcommand{\arraystretch}{2.0} 
\begin{tabular}{|c|c|c|c|c|c|c|c|c|c|c|}
\hline\hline
variable & $\varphi$ & $A$ & $T$ & $\hat{n}$ & $\hat{t}$ & $\hat{x}$ & $\hat{y}$ & $\hat{z}$ & $\kappa_T$ & $\hat{\chi}$ \\
\hline
definition & $\dfrac{e\phi}{T_e}\dfrac{L_B}{2\rho_s}$ & $\dfrac{A_{\parallel}}{\rho_s B}\frac{L_B}{\sqrt{2\beta_e}\rho_s}$ & $\tau\dfrac{\delta T_i}{T_i} \dfrac{L_B}{2\rho_s}$ 
& $\dfrac{\delta n_e}{n_e}\dfrac{L_B}{2\rho_s}$ & $\dfrac{2c_s}{L_B}t$ & $\dfrac{x}{\rho_s}$ & $\dfrac{y}{\rho_s}$ 
& $\sqrt{2\beta_e}\dfrac{z}{L_B}$ & $\dfrac{\tau}{2} \dfrac{L_B}{L_T}$ & $\dfrac{\chi}{2\rho_s^2}\dfrac{L_B}{c_s}$ \\ [2.5ex]
\hline
\end{tabular}
\label{tab:norms}
\end{table*}

\subsection{System of equations}
\label{sec:system_of_equations}
Application of the orderings (\ref{eq:para_ordering}), (\ref{eq:frequency ordering}) and (\ref{eq:fields_ordering}) to the gyrokinetic-Maxwell's system of equations leads to a closed fluid model for electromagnetic ITG turbulence. The details of its derivation are provided in \mbox{\ref{derivation}}; here we merely state the final fluid equations: the parallel electron force balance (the ``ideal Ohm's law"),
\beq
\label{eq:apareq}
\frac{e}{T_e c} \frac{\partial A_{\parallel}}{\partial t} = \nabla_{\parallel} \left(\frac{\delta n_e}{n_e} - \frac{e\phi}{T_e}\right),
\eeq
the electron continuity equation incorporating the divergence of the parallel electron flow (current) and the effect of the curvature drift,
\bea
\label{eq:densityeq}
 \frac{\sd}{\sd t}\frac{\delta n_e}{n_e}  + \nabla_{\parallel}\frac{c}{4\pi e n_e}\nabla^2_{\perp}A_{\parallel} & + \frac{2\rho_s c_s}{L_B}\frac{\partial}{\partial y}\left(\frac{\delta n_e}{n_e} - \frac{e\phi}{T_e}\right)\nonumber \\ & = 0,
\eea
the equation for the $z$-component of the $\mbf{E}\times\mbf{B}$ vorticity, $\unit{z}\cdot(\nabla \times\mbf{v}_E) = (c/B)\nabla^2_{\perp}\phi$, obtained by subtraction of (\ref{eq:densityeq}) from the ion density equation,
\bea
\label{eq:phieq}
 - \frac{\sd}{\sd t}\rho^2_s \nabla^2_{\perp}\frac{e\phi}{T_e}- \nabla_{\parallel}\frac{c}{4\pi e n_e}\nabla^2_{\perp}A_{\parallel} \nonumber \\ + \tau\frac{\rho_s c_s}{L_{T}}\frac{\partial}{\partial y}\rho^2_s\nabla^2_{\perp}\frac{e\phi}{T_e}  - \frac{2\rho_s c_s}{L_B}\frac{\partial}{\partial y}\left(\frac{\delta n_e}{n_e} + \tau\frac{\delta T_i}{T_i}\right) \nonumber \\ +  c_s\rho^3_s \nabla_{\perp}\cdot\left\lbrace \nabla_{\perp}\frac{e\phi}{T_e}, \tau\frac{\delta T_i}{T_i}\right\rbrace \nonumber \\ = - \chi\rho^2_s\nabla^4_{\perp}\left(a\frac{e\phi}{T_e} - b\tau\frac{\delta T_i}{T_i}\right),
\eea
and the ion temperature equation including the effect of the advection of the equilibrium ion temperature gradient by the fluctuating $\mbf{E}\times\mbf{B}$ flow -- the ITG drive,
\beq
\label{eq:Teq}
\frac{\sd}{\sd t}\frac{\delta T_i}{T_i} + \frac{\rho_s c_s}{L_{T}}\frac{\partial}{\partial y}\frac{e\phi}{T_e} = \chi \nabla^2_{\perp}\frac{\delta T_i}{T_i}.
\eeq
In the above equations, for any fields $f$ and $g$, 
\beq
\frac{\sd f}{\sd t} \equiv \left(\frac{\partial}{\partial t} + \mbf{v}_E\cdot\nabla\right)f = \frac{\partial f}{\partial t} + \frac{c}{B}\left\lbrace \phi, f \right\rbrace
\eeq
is the convective time derivative,
\beq
\nabla_{\parallel}f \equiv \left(\frac{\partial}{\partial z} + \frac{\delta \mbf{B}_{\perp}}{B}\cdot\nabla\right)f = \frac{\partial f}{\partial z} - \frac{1}{B}\left\lbrace \apar, f \right\rbrace
\eeq
is the spatial derivative along the total magnetic field,
\beq
\left\lbrace f, g\right\rbrace \equiv \frac{\partial f}{\partial x}\frac{\partial g}{\partial y} - \frac{\partial f}{\partial y}\frac{\partial g}{\partial x}
\eeq
is the Poisson bracket, $\nabla_{\perp} \equiv \unit{x}\partial/\partial x + \unit{y}\partial/\partial y$ is the gradient operator in the plane perpendicular to $\mbf{B} = B \unit{z}$, $\nabla^2_{\perp} = \partial^2/\partial x^2 + \partial^2/\partial y^2$ is the perpendicular Laplacian, $\nabla^4_{\perp} = (\nabla^2_{\perp})^2$, and 
\bea
\nabla_{\perp}\cdot\left\lbrace \nabla_{\perp}f, g\right\rbrace \nonumber \\ = \left\lbrace\frac{\partial f}{\partial x},\frac{\partial g}{\partial x} \right\rbrace + \left\lbrace\frac{\partial f}{\partial y},\frac{\partial g}{\partial y} \right\rbrace + \left\lbrace\nabla^2_{\perp} f, g \right\rbrace.
\eea
The collisional diffusion coefficient appearing in the right-hand sides of (\ref{eq:phieq}) and (\ref{eq:Teq}) was obtained from the linearised Landau  operator for ion-ion collisions~\cite{Newton_2010, ivanov_2020}, where
\beq
\label{eq:definition_of_chi}
\chi = \frac{16}{9}\sqrt{\frac{2}{\pi}}\tau\rho^2_s\nu_{ii},
\eeq
$a = 9/40$ and $b = 67/160$.

Equations~(\ref{eq:apareq})-(\ref{eq:Teq}) form a closed system describing the evolution of the four fields $\deln_e$, $\delT_i$, $\phi$, and~$\apar$. These equations simplify to those of reduced MHD in the limit of a homogeneous plasma in a slab at long perpendicular wavelengths, $\kperp \rho_s \ll 1$, and also to the electrostatic ITG equations used by \cite{ivanov_2020} in the limit of short parallel wavelengths, which turns out to be equivalent to taking $\beta_e \rightarrow 0$ (see Section~\ref{sec:kpar_electrostatic}). Thus, they provide the desired bridge between electrostatic and electromagnetic (``finite-beta") turbulence.


Before proceeding, we simplify the form of our equations by introducing a convenient set of normalisations, given in Table~\ref{tab:norms}. Upon application of these normalisations, (\ref{eq:apareq})-(\ref{eq:Teq}) reduce to 
\beq
\label{eq:A_red}
\frac{\partial A}{\partial t} = \nabla_{\parallel}\left(n - \varphi\right),
\eeq
\beq
\label{eq:n_red}
\frac{\sd n}{\sd t} + \nabla_{\parallel}\nabla^2_{\perp}A + \frac{\partial\left(n - \varphi\right)}{\partial y} = 0,
\eeq
\bea
\label{eq:phi_red}
-\frac{\sd}{\sd t}\nabla^2_{\perp}\varphi - \nabla_{\parallel}\nabla^2_{\perp}A + \kappa_T\frac{\partial}{\partial y}\nabla^2_{\perp}\varphi - \frac{\partial \left(n + T\right)}{\partial y} \nonumber\\ + \nabla_{\perp}\cdot\left\lbrace \nabla_{\perp}\varphi, T \right\rbrace = - \chi \nabla^4_{\perp}\left(a\varphi - bT\right),
\eea
\beq
\label{eq:T_red}
\frac{\sd T}{\sd t} + \kappa_T\frac{\partial \varphi}{\partial y} = \chi \nabla^2_{\perp} T,
\eeq
where
\beq
\frac{\sd f}{\sd t} = \frac{\partial f}{\partial t} + \left\lbrace \varphi, f \right\rbrace\;\;\mathrm{and}\;\;\nabla_{\parallel} f = \frac{\partial f}{\partial z} - \left\lbrace A, f \right\rbrace
\eeq
are now defined with respect to the normalised coordinates. For notational simplicity, we have dropped the hats on the normalised quantities introduced in Table~\ref{tab:norms}.~In the following sections, we use the subscript `phys' on unnormalised versions of these quantities to avoid confusion.

\subsection{Parameters of the model}
There are effectively five free parameters in the system (\ref{eq:n_red})-(\ref{eq:T_red}): the normalised ion temperature gradient $\kappa_T$ and
collisionality $\chi$, and the box lengths in the $x$, $y$, and $z$ directions, denoted $L_x$, $L_y$ and $L_z$, respectively.  The box lengths $L_x$ and $L_y$ do not matter provided they are large enough that, for a given $\kappa_T$, there are well-defined turbulence outer scales in $x$ and $y$ that are smaller than the corresponding box scales (the outer scales depend on $\kappa_T$, so the limits $\kappa_T\rightarrow \infty$ and $L_x$,$L_y\rightarrow \infty$ do not commute, but we stay at $\kappa_T\sim 1$). We thus remove these two free parameters from our study.
The box scale in $z$ can be viewed as a proxy for $\beta_e$.~This is evident from the definition of the normalised $z$
as $\sqrt{2\beta_e}\zphys /L_B$ and is a consequence of ordering the fluctuation frequency $\omegaphys$ to be comparable to both the diamagnetic frequency $\omega_{*T}$ and
the Alfvén frequency $\kparphys v_A$: it is only possible to satisfy both orderings simultaneously if $\kparphys$ decreases with decreasing $\beta_e$. It is important to draw a distinction between $L_z$, which is the normalised simulation-box size in $z$, and any physical scale $L_{z,\text{phys}}$ that may exist in the system under consideration. In the case of a tokamak, $L_{z,\text{phys}}$ could be the length of the field line within the bad-curvature region as shown in Fig.~\ref{fig:zpinch}, $L_{z,\text{phys}} \sim \pi qR$. Thus, for a system where $L_{z,\text{phys}}$ is fixed, the simulation-box size $L_z$ should be regarded as a direct proxy for $\beta_e$.

\subsection{Role of $k_{\parallel}$ and recovery of the electrostatic limit}
\label{sec:kpar_electrostatic}
As our aim is to capture the finite-$\beta_e$ modification of the electrostatic ITG turbulence, let us demonstrate that our model recovers correctly the electrostatic limit,
corresponding to $\beta_e \rightarrow 0$. As explained above, in our model, this is formally equivalent to taking $\kpar \rightarrow \infty$. Physically, this means that \alf waves are assumed fast compared to the drift modes, and thus any magnetic perturbation caused by the latter will be quickly radiated away by the former (a useful analogy is pressure perturbations in ``incompressible" hydrodynamics being quickly radiated away by fast sound waves). 

The induction equation~(\ref{eq:A_red}) reduces in this limit to $n' = \varphi'$, where the prime denotes the difference
between the un-primed quantity and its flux-surface average: e.g., $\varphi' \equiv \varphi - \overline{\varphi}$, where
\beq
\overline{\varphi} \equiv \left\langle \varphi\right\rangle_{y,z}= \frac{1}{L_y L_z} \int \varphi  \ \sd y \sd z
\eeq
is the flux-surface average of $\varphi$, also referred to as the zonal component of $\varphi$. The angle brackets denote an average over the subscripted variables: e.g.,
\beq
\left\langle{f}\right\rangle_x = \frac{1}{L_x} \int_0^{L_x} f \sd x.
\eeq 
In a true $Z$-pinch, field lines close on themselves, so that a field-line average is not equivalent to a flux-surface average, but we have assumed the presence of an infinitesimal pitch angle for the magnetic field, to recover the equivalence of the two averages.

In the electrostatic limit, there is no generation of zonal density $\overline{n}$ by our equations, as seen by taking the flux-surface average of (\ref{eq:n_red}):
\beq
\frac{\partial \overline{n}}{\partial t} = \frac{\partial}{\partial x}\overline{\left(n'\frac{\partial \varphi'}{\partial y} - \frac{\partial^2 A'}{\partial x^2}\frac{\partial A'}{\partial y}\right)} = 0,
\eeq
where the nonlinear particle fluxes vanish as $A^{\prime} \rightarrow 0$ and $n'\rightarrow \varphi'$. We thus recover the familiar modified Boltzmann response for electrons~\cite{Hammett_1993, Roger_2000}: 
\beq
\label{eq:boltzmann}
n=n'=\varphi'.  
\eeq
Using (\ref{eq:boltzmann}) and adding (\ref{eq:phi_red}) to (\ref{eq:n_red}) results in a two-field set of equations for $\varphi$ and $T$:
\bea
\frac{\sd}{\sd t}\left(\varphi^{\prime} - \nabla^2_{\perp}\varphi\right) + \kappa_T \frac{\partial}{\partial y}\nabla_{\perp}^2 \varphi - \frac{\partial \left(\varphi+T\right)}{\partial y} 
\nonumber \\ \nabla_{\perp}\cdot\left\lbrace \nabla_{\perp}\varphi, T \right\rbrace  = -\chi \nabla_{\perp}^4\left(a\varphi-bT\right),
\eea
\beq
\frac{\sd T}{\sd t} + \kappa_T\frac{\partial \varphi}{\partial y} = \chi \nabla^2_{\perp}T.
\eeq
These are the equations derived by \cite{ivanov_2020} to study two-dimensional, electrostatic ITG turbulence. The three-dimensional physics of electrostatic ITG modes described in~\cite{Ivanov_2022} are absent, as these require
$\kparphys L_B \sim 1$, whereas in our model, $\kparphys L_B \sim \sqrt{\beta_e} \ll 1$. Thus, physically, the limit of two-dimensianl, electrostatic ITG turbulence corresponds to parallel wave numbers satisfying $\sqrt{\beta_e} \ll \kparphys L_B \ll 1$. In \cite{ivanov_2020}, this turbulence, and, in particular, the transition from low to high transport in it, was studied as a function of $\kappa_T$ and $\chi$. The addition of the parameter $L_z$ in our model, as explained above, gives us the ability to study this transition as a function of $\beta_e$.

\section{Linear physics}
\label{sec:linear_physics}
Let us first consider infinitesimal perturbations of the four fields $n$, $T$, $\varphi$, and $A$ of the form, e.g., $n\left(\mbf{r}, t\right) = \tilde{n}_{\mbf{k}} \exp{(-i\omega t + i\mbf{k}\cdot\mbf{r})}$ and neglect all nonlinear terms. Dropping tildes and subscripts on Fourier coefficients for notational simplicity, the linearised versions of (\ref{eq:A_red})-(\ref{eq:T_red}) are
\beq
\label{eq:A_lin}
-i\omega A = ik_z\left(n - \varphi\right),
\eeq
\beq
\label{eq:n_lin}
-i\omega n - ik_zk^2_{\perp}A + ik_y\left(n - \varphi\right)  = 0,
\eeq
\bea
\label{eq:phi_lin}
-i\omega k^2_{\perp}\varphi + ik_zk^2_{\perp}A - i\kappa_T k_yk^2_{\perp}\varphi - ik_y\left(n + T\right) \nonumber\\ = - \chi k^4_{\perp}\left(a\varphi - bT\right),
\eea
\beq
\label{eq:T_lin}
-i\omega T + i\kappa_T k_y\varphi = - \chi k^2_{\perp} T.
\eeq
In Sections~\ref{sec:lin_alfven}-\ref{sec:lin_dd}, we consider various limits in which a relatively simple, closed-form analytical
solution is obtainable.  In particular, we demonstrate that our equations contain the physics needed to describe the curvature ITG instability in the electrostatic limit,
its stabilisation with increasing $\beta_e$, and the excitation of unstable MHD interchange (IC) modes and double-diffusive modes at high~$\beta_e$. Numerical solutions of the full set of linearised equations are presented in Section~\ref{sec:lin_num}.
\subsection{Straight-field (high-frequency) limit: \alfnc modes}
\label{sec:lin_alfven}
Consider first the limit of straight field, usually studied in the astrophysical and space-physical contexts \cite{Schekochihin_2009}. This is formally recovered as $L_B \rightarrow \infty$, or, in terms of normalised quantities, $\omega \sim k_z \gg \kperp \sim 1$, $\kappa_T \sim \chi \sim 1$, and $A \sim \varphi \sim n \sim T$. With these orderings, (\ref{eq:A_lin})-(\ref{eq:T_lin}) reduce to 
\beq
\label{eq:A_lin_1}
-i\omega A = ik_z\left(n - \varphi\right),
\eeq
\beq
-i\omega n - ik_zk^2_{\perp}A = 0,
\eeq
\beq
\label{eq:phi_lin_1}
-i\omega k^2_{\perp}\varphi + ik_zk^2_{\perp}A = 0.
\eeq
\beq
-i\omega T = 0.
\eeq
The non-zero solutions for $\omega$ are
\beq
\omega = \pm k_z\sqrt{1 + k^2_{\perp}},
\eeq
with corresponding eigenvectors
\beq
A = \pm \sqrt{1+k_{\perp}^2}\varphi,\;\;n = - \kperp^2 \varphi,\;\; T = 0.
\eeq
In the MHD limit $k_{\perp}\rightarrow 0$, we have $\omega \rightarrow \pm k_z$, and $A \rightarrow \pm \varphi$, $n = 0$, recovering the Alfvén waves: 
\beq
\omega_{\text{phys}} = \pm k_{z,\text{phys}}v_A. 
\eeq
In the opposite limit of $k_{\perp} \gg 1$, we have $\omega \approx \pm k_zk_{\perp}$, or
\beq
\omega_{\text{phys}} = \pm k_{\perp,\text{phys}}\rho_s k_{z,\text{phys}} v_A, 
\eeq
which is the dispersion relation for kinetic Alfv\'en waves in the cold ion, low-$\beta_e$ limit~\cite{Schekochihin_2009}. Nonlinearly, in this limit, if energy were injected at large scales ($\kperp \ll 1$), our equations would support a standard MHD Alfv\'enic cascade towards $\kperp \sim 1$ and a kinetic Alfv\'enic wave cascade at $\kperp \gtrsim 1$ \cite{Schekochihin_2009, Schekochihin_2019} -- a classic outcome for a low-beta solar wind. This is not a relevant limit for a fusion device.

\subsection{Electrostatic (low-frequency) limit: curvature-ITG instability}
\label{sec:lin_ITG}
If we allow magnetic field to be curved, but still take $k_z\gg \kperp \sim 1$, this time as $\beta_e \rightarrow 0$, we order $\omega \sim 1$. As anticipated by the discussion at the end of Section~\ref{sec:kpar_electrostatic}, (\ref{eq:A_lin}) then gives us the Boltzmann electron response, $n=\varphi$. We may now eliminate~$A$ (which is now a spectator quantity) by summing (\ref{eq:n_lin}) and (\ref{eq:phi_lin}), and, together with (\ref{eq:T_lin}), we obtain an electrostatic system:
\bea
\label{eq:n_lin_2}
-i\omega (1 + k^2_{\perp})\varphi  - i\kappa_T k_yk^2_{\perp}\varphi - ik_y\left(\varphi + T\right) \nonumber\\  = -\chi k^4_{\perp}\left(a\varphi - bT \right),
\eea
\beq
\label{eq:T_lin_2}
-i\omega T + i\kappa_T k_y\varphi = -\chi k^2_{\perp}T.
\eeq
In the large-gradient, long-wavelength limit, $\kappa_T k_y \gg 1 \gg \kappa_Tk_yk^2_{\perp}$, the `finite-Larmor-radius' (FLR) terms (those containing powers of $\kperp^2$) are negligible and $T \gg \varphi$, so (\ref{eq:n_lin_2}) and (\ref{eq:T_lin_2}) reduce to 
\begin{equation}
\label{eq:n_lin_3}
-i\omega \varphi - ik_y T = 0,
\end{equation}
\begin{equation}
\label{eq:T_lin_3}
-i\omega T + i\kappa_T k_y\varphi = 0.
\end{equation}
The resulting dispersion relation is
\begin{equation}
\label{eq:itg_mode_gamma}
\omega = \pm ik_y\sqrt{\kappa_T}.
\end{equation}
The solution with a positive imaginary part is the archetypal curvature-ITG-instability growth rate,~cf.~\cite{Beer_1995, ivanov_2020},
\beq
\mathrm{Im}(\omega_{\text{phys}}) = \frac{k_{y,\text{phys}} \rho_i c_s}{ \sqrt{L_B L_T}}.
\eeq

The nonlinear version of the limit represented by (\ref{eq:n_lin_2}) and (\ref{eq:T_lin_2}) (with FLR terms restored to capture the short-wavelength stabilisation of the ITG modes, as well as correct zonal-flow dynamics) was studied in~\cite{ivanov_2020}, who proposed this as the minimal model for the Dimits transition from low to high turbulent transport. By design, this electrostatic limit is contained in our model.

\subsection{Finite-$\beta_e$ stabilisation of long-wavelength ITG modes}
\label{sec:lin_finite_beta}
We can now introduce finite-$\beta_e$ effects by relaxing the assumption of large $k_z$. Let us again take the long-wavelength limit, $k_y \ll 1$, and let $k_x = 0$ (linearly, these are always the fastest-growing modes), but leave $\kappa_T \sim 1$. The collisional terms involving $\chi$ will, for the time being, be neglected in this limit. From (\ref{eq:n_lin})-(\ref{eq:T_lin}), we find the quartic dispersion relation
\bea
\label{eq:kz1_disp}
\omega^4 + \left(\kappa_T - 1\right)k_y\omega^3 - \left(1-\kappa_T+k^2_z\right)\omega^2 \nonumber \\ - \left(\kappa_T+k^2_z\right)k_y\omega - \kappa_T k^2_zk^2_y = 0.
\eea 
Motivated by (\ref{eq:itg_mode_gamma}), we seek solutions to (\ref{eq:kz1_disp}) with $\omega \sim k_y$, so the first two terms can be dropped. The remaining dispersion relation is quadratic in $\omega$ and can be easily solved. In the limit of $k_z \rightarrow \infty$ and $\kappa_T \gg 1$, we recover the ITG mode (\ref{eq:itg_mode_gamma}), but for $\kappa_T \sim 1$ and $k_z \sim 1$, we find the condition of stability to be 
\begin{equation}
\label{eq:stab_itg}
\left[\left(4\kappa_T -1 \right)k^2_z + \kappa_T \right](\kappa_T - k^2_z) \geq 0.
\end{equation}
In the electrostatic limit, $k_z \rightarrow \infty$, this inequality is satisfied when $\kappa_T \leq 1/4$ -- this critical temperature gradient also appeared in the electrostatic model of~\cite{ivanov_2020}.  As we are interested in the finite-$\beta_e$ stabilisation of the electrostatic ITG modes, we restrict our attention to $\kappa_T > 1/4$.  With this proviso, we find from (\ref{eq:stab_itg}) that the curvature-ITG modes are stabilised for
\begin{equation}
\label{eq:stab_itg_es_kz_kappaT}
k^2_z \leq \kappa_T.
\end{equation}
This corresponds to a critical ion plasma beta 
\beq
\beta_i = \kzphys^2 L_B L_T. 
\eeq
For values of $\beta_i$ in excess of this threshold, the curvature-ITG instability is quenched. 

In a conventional tokamak plasma, the parallel wave number is often set by the system size: $\kzphys \sim 1/qL_B$, where $\pi q L_B$ is the connection length. This implies that the long-wavelength ITG modes in a tokamak should be stabilised if 
\beq
\label{eq:beta_condition}
\beta_i \gtrsim \frac{L_T}{q^2L_B}.
\eeq
Note that $\beta_i$ needed for stabilisation is smaller for larger temperature gradients, in agreement with previous observations~\cite{AJarmen,Zocco_2015} from both numerical and analytical calculations. Since the stabilization of ITG modes is often followed by the onset of electromagnetic instabilities, condition (\ref{eq:beta_condition}) also marks the threshold for these electromagnetic modes, as noted by simple arguments from \cite{Jenko_2001}.

\subsection{High-$\beta_e$ (2D) limit: interchange modes}
\label{sec:lin_mhd}

Finally, let us consider the high-$\beta_e$ limit, for which $k_z \rightarrow 0$, so the dynamics are effectively two-dimensional. Then (\ref{eq:A_lin}) gives $A \rightarrow 0$ and
the remaining equations (\ref{eq:n_lin}), (\ref{eq:phi_lin}) and (\ref{eq:T_lin}) simplify to
\beq
\label{eq:n_lin_4}
-i\omega n + ik_y\left(n - \varphi\right) = 0,
\eeq
\bea
\label{eq:phi_lin_4}
-i\omega k^2_{\perp}\varphi - i\kappa_T k_yk^2_{\perp}\varphi - ik_y\left(n + T\right) \nonumber \\ = -\chi k^4_{\perp}\left(a\varphi - bT \right),
\eea
\beq
\label{eq:T_lin_4}
-i\omega T + i\kappa_T k_y\varphi = - \chi k^2_{\perp}T.
\eeq
Consider $k_{\perp}\ll 1$ and (in this long-wavelength limit) neglect dissipation terms. This leaves us with
\begin{equation}
\label{eq:n_lin_5}
-i\omega n + ik_y (n-\varphi) = 0,
\end{equation}
\begin{equation}
\label{eq:phi_lin_5}
-i\omega \kperp^2\varphi - ik_y(n + T) = 0,
\end{equation}
\begin{equation}
\label{eq:T_lin_5}
-i\omega T + i\kappa_T k_y\varphi = 0.
\end{equation}
This gives us the dispersion relation
\begin{equation}
\kperp^2\omega^3 - k_y\kperp^2\omega^2 - (1 - \kappa_T)k^2_y\omega - \kappa_Tk^3_y = 0.
\end{equation}
Anticipating the appearance of MHD modes at \mbox{$\kperp\ll 1$}, assume $\omega \sim 1$. The dispersion relation can then be simplified to
\begin{equation}
\label{eq:MHD_omega}
\omega^2 = (1 - \kappa_T)\frac{k^2_y}{\kperp^2}.
\end{equation}
When $\kappa_T > 1$, or $L_B/L_T > 2/\tau$, there is an unstable mode with a physical growth rate in the limit of large~$\kappa_T$ and $k_x = 0$ given by
\beq
\mathrm{Im}(\omega_{\text{phys}}) = \frac{\vthi}{\sqrt{L_B L_T}}.
\eeq
This mode exists at an arbitrarily large perpendicular scale, is driven by a pressure gradient in the presence of magnetic curvature and is constant on a flux surface -- it is the familiar MHD (in fact, at $k_z = 0$, effectively electrostatic) interchange (IC) mode \cite{Mercier_1964, Freidberg_2014}. In tokamaks, at macroscopic scales, such modes are stabilised by the magnetic shear, and generally for any viable equilibrium, MHD modes must be stabilised -- this leads to the so-called beta limit \cite{Mercier_1964, Freidberg_2014, Troyon_1984}. Later, we will explore how close to (or far from) the MHD stability boundary the nonlinear high-beta runaway occurs. 

\subsection{High-$\beta_e$ (2D) limit: double-diffusive instabilities}
\label{sec:lin_dd}
The calculation in Section~\ref{sec:lin_finite_beta} suggests that when $\kappa_T < 1$, the system is stable. This is in fact not the case: while it is indeed ideally stable (to the IC mode), another (set of) unstable mode(s) kicks in, associated with the presence of diffusion and a destabilisation mechanism due to different fields diffusing at different rates -- in geophysical fluid dynamics, such unstable modes are called double-diffusive (DD) instabilities \cite{Stern_1960, Nield_1967, RADKO_2003}. The easiest way to explain them is to go back to (\ref{eq:n_lin_4})-(\ref{eq:T_lin_4}), assuming now $\kappa_T < 1$ and, again, $\kperp \ll 1$, and non-rigorously, to simplify calculations, set $a=b=0$, but keep $\chi \neq 0$; we will also set $k_x = 0$ as these will, still, as always, be the most unstable modes. The resulting simplified equations are again (\ref{eq:n_lin_5})-(\ref{eq:T_lin_5}) but (\ref{eq:T_lin_5}) now has the diffusion term, so 
\begin{equation}
\label{eq:dd_T_phi}
T = \frac{i\kappa_T k_y\varphi}{i\omega - \chi k^2_y} \approx \frac{\kappa_Tk_y\varphi}{\omega}\left(1 - i\frac{\chi k^2_y}{\omega}\right).    
\end{equation}
We shall expect the (small) dissipative correction that appears here to provide a growth rate to go with what to lowest order will be a real frequency. Indeed it does: amending the calculation that led to (\ref{eq:MHD_omega}), we find
\begin{equation}
\label{eq:dd_omega_a=0}
\omega \approx \pm \omega_r + i \frac{\chi \kappa_T k^2_y}{2\omega^2_r},
\end{equation}
where $\omega_r$ is given by (\ref{eq:MHD_omega}) with $\kappa_T < 1$. 

The physical picture of this instability is as follows. Imagine a blob of plasma oscillating with the drift wave at frequency $\omega_r$ and thus carried back and forth between hotter and colder regions (its displacement being in the $x$-direction). Due to the presence of heat diffusivity, some of the heat carried by the hot blob from a hotter region is diffused locally in the colder region before the blob has the time to swap back to its hotter origin. This gives rise to a net radial transport of heat.

A more rigorous calculation of these instabilities is given in \ref{more linear physics}, where we also show that, similarly to the way in which collisional effects destabilise the plasma on the stable side of the ideal-IC stability boundary, they also destabilise it on the stable side of the ITG stability boundary, giving rise to two further instabilities, which we call the $\chi$ITG$+$ and $\chi$ITG$-$ modes. We do not dwell on these here because, while interesting, they are, in fact, not of great consequence as far as the main subject of this paper is concerned.

\subsection{Numerical calculation of linear stability}
\label{sec:lin_num}
In this section, we present results obtained by applying an eigenvalue solver to the linear system of equations~(\ref{eq:A_lin})-(\ref{eq:T_lin}). Our aim is to recover the analytically tractable limits obtained in Sections~\ref{sec:lin_alfven}-\ref{sec:lin_dd} and to connect them via regions of the parameter space -- in $k_y$, $k_z$ (or $\beta_e$), and $\kappa_T$ -- where a simple analytical solution is not possible. 

\begin{figure}[htb]%
    \centering
    \subfloat[\centering ]{{\includegraphics[width=7cm]{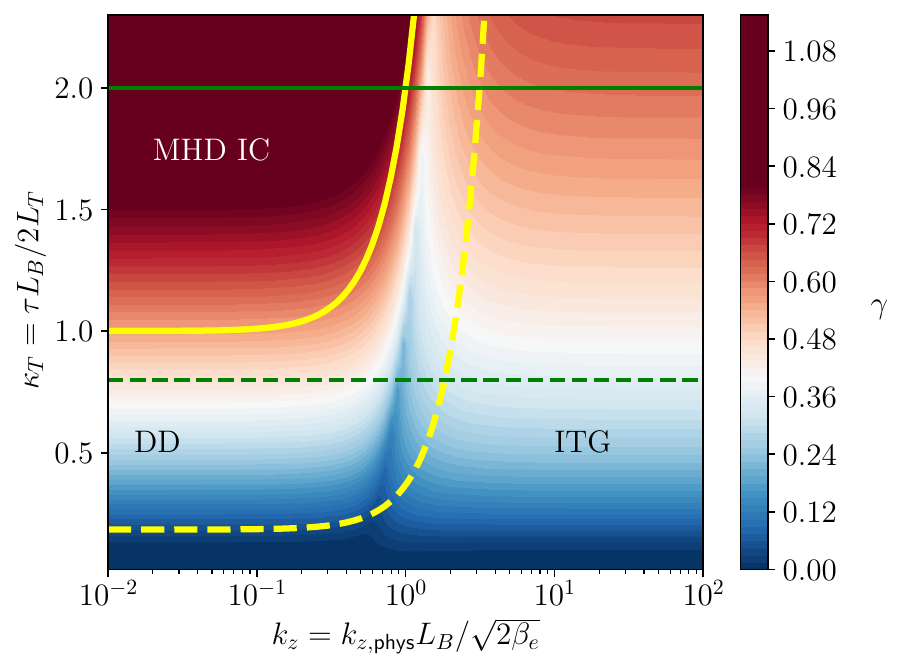}}  \label{fig:Growth_rate_spec_kTkz}}%
    \vfill
    \subfloat[\centering  ]{{\includegraphics[width=7cm]{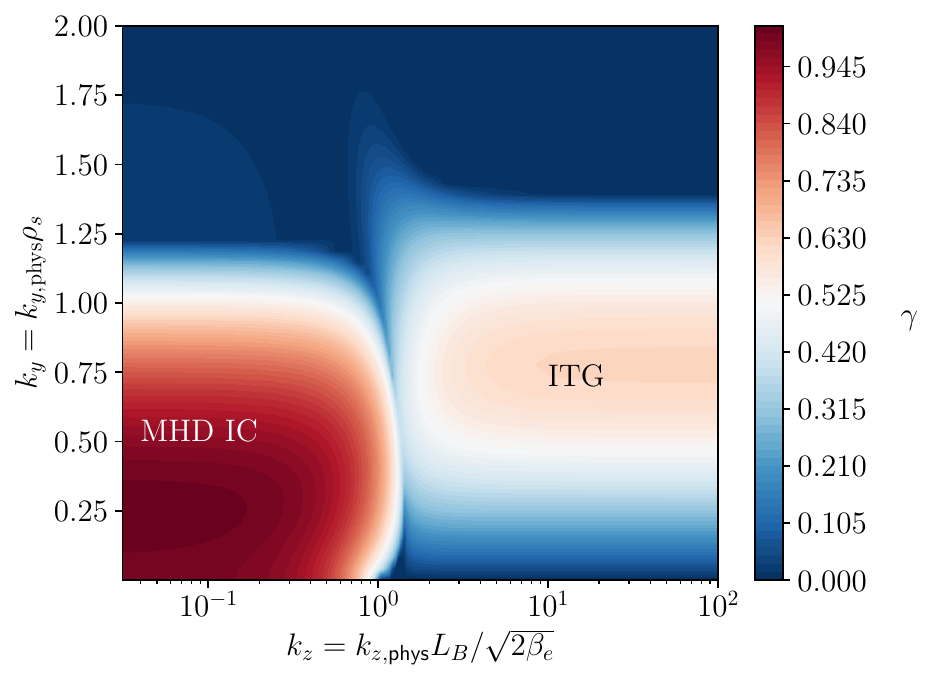}}      \label{fig:Growth_rate_spec_kykz_kT=2.0}}%
    \vfill
    \subfloat[\centering  ]{{\includegraphics[width=7cm]{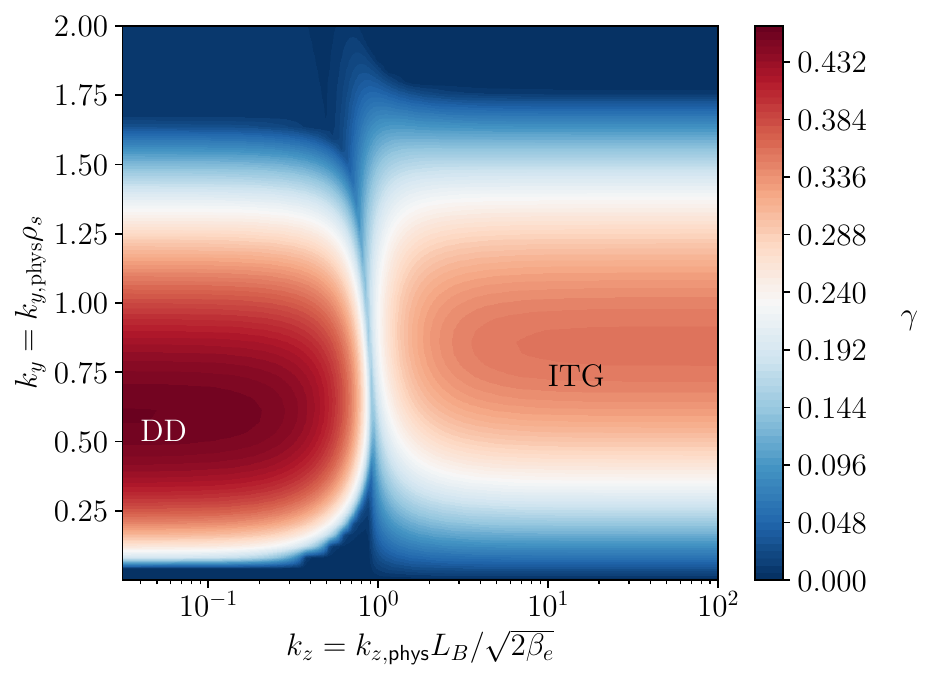}}      \label{fig:Growth_rate_spec_kykz_kT=0.8}}%
    \caption{Linear growth rates for modes with $k_x = 0$ and $\chi = 0.2$. (a) Growth rate versus ($k_z$, $\kappa_T$) for the $k_y$ corresponding to the fastest-growing mode. The solid yellow curve is the onset boundary given by (\ref{eq:ic_stability_condition}) for the MHD IC mode and the dashed yellow curve is the onset boundary given by (\ref{eq:dd_stability_boundary}) for the DD mode. The boundaries are independent of $\chi$, and they are calculated analytically for the case with $k_y \ll 1$ (see \mbox{\ref{more linear physics}}). (b)~Growth rate versus ($k_z$, $k_y$), for $\kappa_T = 2.0$ (green solid line in panel a). The MHD IC mode is unstable (the growth rate at $k_y = 0$ is finite). (c)~Growth rate versus ($k_z$, $k_y$), for $\kappa_T = 0.8$ (green dashed line in panel a). Now, the DD modes are unstable at low $k_z$.}
    \label{fig:plot_gspec}
\end{figure}

Growth rates computed numerically are shown in Fig.~\ref{fig:plot_gspec}. The results from scans in $k_z$ and $\kappa_T$ for the fastest-growing modes at $\chi = 0.2$ show the key features obtained analytically in the long-wavelength limit: the electrostatic curvature-ITG mode is unstable when $k_z \gg 1$, with a growth rate that increases roughly as $\sqrt{\kappa_T}$; this ITG mode is partially stabilised at $k_z$ values of order unity; and IC and DD modes are driven unstable for small enough $k_z$. As anticipated in Section~\ref{sec:lin_dd}, the DD modes appear below the critical $\kappa_T$ for the onset of unstable IC modes, and they extend from $k_z=0$ to $k_z \sim 1$ at $k_T\sim 1$, corresponding to the area between the solid yellow and dashed yellow curves in the figure. 

The solid green line in Fig.~\ref{fig:plot_gspec}(a) denotes a cut at $\kappa_T = 2.0$, with the corresponding growth-rate spectrum in $(k_z, k_y)$ presented in Fig.~\ref{fig:plot_gspec}(b). Similarly, the dashed green line represents a cut at $\kappa_T = 0.8$, and the associated growth-rate spectrum is shown in Fig.~\ref{fig:plot_gspec}(c). A common feature of the linear stability at both temperature gradients is the emergence of ITG instability at high $k_z$, followed by its stabilisation around $k_z \sim 1$. The primary difference between the two cases is in the low-$k_z$ regime: at $\kappa_T = 0.8$, the DD instability is dominant, whereas at $\kappa_T = 2.0$, the IC instability dominates, as indicated by the finite growth rate in the limit $k_y \rightarrow 0$.

Fig.~\ref{fig:plot_gspec} only shows the dominant instabilities, but multiple instabilities can co-exist in certain parameter regimes. For $k_z \gg 1$, only the ITG mode is unstable. When $k_z$ is decreased beyond the dashed yellow line in Fig.~\ref{fig:plot_gspec}(a), additional modes are destabilised. Both the DD mode and another collisional mode (the $\chi$ITG$-$ mode, discussed in \ref{sec:ITG instability and its collisional extension}) are sub-dominant to the ITG mode near the boundary. As $k_z$ is decreased further, the DD mode becomes the dominant instability. When the solid yellow line in Fig.~\ref{fig:plot_gspec}(a) is crossed, the IC mode becomes the dominant instability and yet another collisional mode (the $\chi$ITG$+$ mode, discussed in \ref{sec:ITG instability and its collisional extension}) is sub-dominant. 

In Section~\ref{sec:lin_finite_beta}, we found a complete stabilisation of ITG at long parallel wavelengths. The conclusion there was that any instability whose growth rate satisfied $\gamma \sim k_y$ was stabilised above a critical $\beta_e$. In that analytical calculation, the collisional terms were ignored. It turns out that for finite $\chi$, the growth rate at low $k_y$ can be restored and scales as $k^2_y$, similarly to the case of the DD mode. This can be seen both numerically and analytically (\ref{sec:ITG instability and its collisional extension}) --- these are the $\chi$ITG$\pm$ modes. So, at finite $\chi$, $\beta_e$ above the critical value worked out in Section~\ref{sec:lin_finite_beta} does not completely stabilise ITG but only adds a double diffusive feature to the modes at low $k_y$, hence changing the growth rate's scaling at low $k_y$. A detailed discussion of all these instabilities can be found in \ref{more linear physics} (the summary plot is Fig.~\ref{fig:stability_boundary}).

As we have discussed, between the solid and dashed yellow curves in Fig.~\ref{fig:plot_gspec}(a), the DD and $\chi$ITG$-$ modes overlap with the ITG mode around $k_z \sim 1$, with a growth rate that is typically sub-dominant to that of the ITG mode near the dashed yellow line. While this complicates the interpretation of the nonlinear simulation data presented in Section~\ref{sec:nonlin_sim}, we will show that the appearance of these sub-dominant modes does not correlate well with the observed transitions from low- to high-heat-flux states.

\section{Nonlinear dynamics}
\label{sec:nonlin_sim}
We are interested primarily in determining how the amplitudes of the turbulent fluctuations in our model depend on $\beta_e$ and $\kappa_T$.  We will show in Section~\ref{sec:fluxes} that our simulations exhibit abrupt transitions from low-amplitude, zonally dominated saturated states to high-amplitude, `streamer'-dominated
turbulence once a certain critical boundary in the $(\beta_e,\kappa_T)$ plane is crossed. These transitions are reminiscent of the `non-zonal transitions' observed in gyrokinetic simulations of electromagnetic turbulence in tokamaks~\cite{waltz_2010,hysteresis,Properties} and of the `Dimits transition' in the fluid model of electrostatic turbulence driven by the curvature-ITG instability in a $Z$-pinch~\cite{ivanov_2020}.  Drawing inspiration
from these studies, we consider in Section~\ref{sec:stresses} the balance of the turbulent stresses in the evolution of the zonal flow and find that, as in \cite{ivanov_2020}, the transition
is correlated with a change in the sign of an effective turbulent viscosity, but this time it is aided and abetted by the Maxwell stress.

\subsection{Numerical setup}
\label{sec:numerics}
The fluid equations~(\ref{eq:A_red})-(\ref{eq:T_red}) are solved in a triply periodic box on a single GPU by an electromagnetic version
of the code described in~\cite{ivanov_2020}.  The Crank-Nicholson scheme is used for the time evolution of the linear terms, and the three-step Adams-Bashforth scheme is used
for the time evolution of the nonlinear terms. A pseudo-spectral method with de-aliasing according to the `2/3 rule' is used to evaluate nonlinearities.

As is clear from the analysis in Section~\ref{sec:linear_physics}, there is no cut-off of linear growth rates at high $k_z$ in our model, because there are no parallel heat fluxes. To avoid an accumulation of spectral energy at $k_{z,\mathrm{max}}$, the maximum $k_z$ in our simulation, we include a hyper-dissipation term of the form $0.1(k_z/k_{z,\mathrm{max}})^4$ in all equations. This form of the hyper-dissipation ensures that, as we scan in $L_z$, the dissipation rate at the Nyquist scale remains fixed as long as the resolution is kept fixed. It is this scan in $L_z$ that provides the variation in $\beta_e$: increasing $L_z$ corresponds to increasing~$\beta_e$, as explained in Section~\ref{sec:kpar_electrostatic}. In principle, besides $L_z$ and $\kappa_T$, the collisionality $\chi$ is also a parameter that controls the Dimits transition \cite{ivanov_2020}, but we shall not study this here, and fix $\chi=0.2$ in all our nonlinear runs.

Unless stated otherwise, the perpendicular box sizes used for nonlinear simulations are $L_x=100$ and $L_y=150$ in units of $\rho_s$, with $N_{k_x}=171$ $k_x$-values, $N_{k_y}=255$ $k_y$-values
and $N_{k_z}=11$ $k_z$-values. Each simulation is initialised with low-amplitude noise and run to a normalised time $t=10000$, with statistical quantities
averaged over the time interval $t\in [5000,10000]$. We found this
duration of simulation to be sufficient in all cases to determine whether the system would saturate in a zonally dominated state.

Finally, to emulate a tokamak plasma where finite magnetic shear and the presence of good- and bad-curvature regions along the magnetic field force instabilities to have $k_z \neq 0$, we zero out the modes with $k_z = 0$ and $k_y \neq 0$ but not those with $k_z = 0$ and $k_y = 0$, $k_x \neq 0$ (the zonal modes).

\subsection{Heat fluxes and the Dimits transition}
\label{sec:fluxes}

We characterise the saturated state of our simulations primarily via the time- and volume-averaged radial ion heat flux
\beq
\label{eq:Q_def}
Q \equiv \left\langle{-\frac{\partial \varphi}{\partial y} T}\right\rangle_{x,y,z} =\tau \frac{Q_{\text{phys}}}{6n_iT_i c_s} \left(\frac{L_B}{\rho_s}\right)^2,
\eeq
where $Q_{\mathrm{phys}}$ is the dimensional ion heat flux.
\begin{figure}[!t]%
    \centering
    \includegraphics[width=7cm]{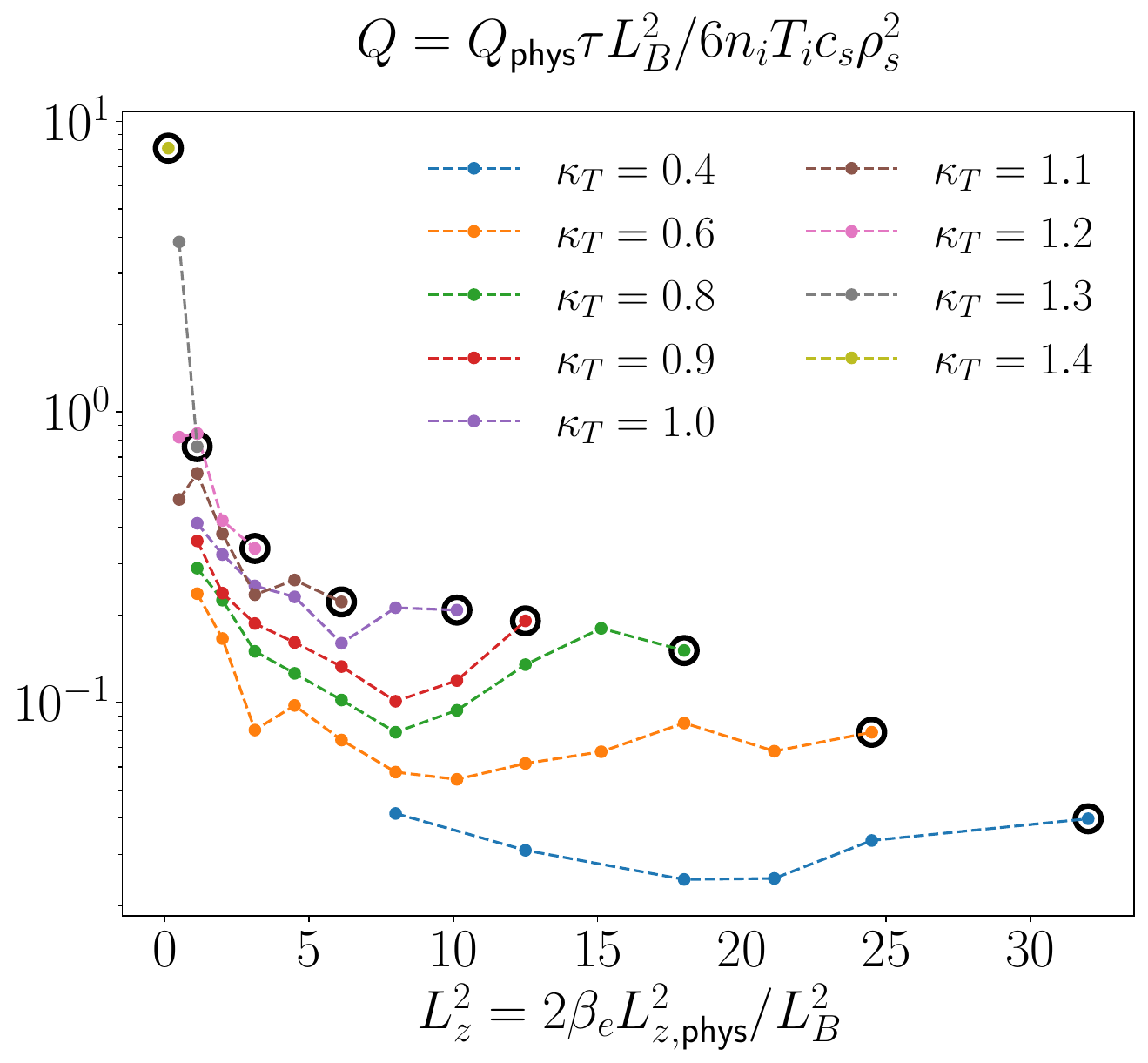}         
    \caption{Normalized heat flux $Q$, defined in (\ref{eq:Q_def}), versus $L^2_z$ (equivalently, $\beta_e$) for a sequence of values of $\kappa_T$. Black circled points denote the boundary beyond which the fluxes obtained are approximately two orders of magnitude higher than those within the boundary.}
    \label{fig:hf_beta_scan}
\end{figure}
The variation of $Q$ versus $L^2_z$ (equivalently, $\beta_e$) for a number of values of $\kappa_T$ is shown in Fig.~\ref{fig:hf_beta_scan} and corresponding examples of the heat-flux time traces are provided in~\ref{time traces}.  The heat flux initially decreases with increasing $L_z$ for small values of $L_z$ but starts increasing again when $L_z$ becomes sufficiently large. For each value of $\kappa_T$, there is a critical value of $L_z$ above which the heat flux abruptly increases by several orders of magnitude.  The boundary in the $(L_z,\kappa_T)$ plane where 
this transition occurs is indicated by black circles in Fig.~\ref{fig:hf_beta_scan}. Simulations to the left of the black circles (for example, $\kappa_T = 0.8,\,L^2_z = 7.8$) saturate with modest heat fluxes in zonal-flow-dominated states, as can be seen in the representative snapshots of the $y$-component of the $\mbf{E}\times\mbf{B}$ velocity (top right panel of Fig.~\ref{fig:plot_vy}) and of the non-zonal component of the ion-temperature fluctuations (top left panel of Fig.~\ref{fig:plot_vy}). The corresponding time traces of the fields are shown in Fig.~\ref{fig:plot_timetrace_0.8_4.0}(a), from which it is evident that the zonal component of $\varphi$ dominates over the non-zonal modes. In contrast, simulations to the right of the black circles (for example, $\kappa_T = 0.8,\,L^2_z = 19.4$) have no clear zonal structure (bottom right panel of Fig.~\ref{fig:plot_vy}), and the ion-temperature fluctuations take the form of box-scale `streamers' oriented along the $x$-direction (bottom left panel of Fig.~\ref{fig:plot_vy}). The absence of strong zonal flows in these states is also evident in Fig.~\ref{fig:plot_timetrace_0.8_6.25}(a), where the zonal mode remains comparatively weak. As a result, the system exhibits much higher heat fluxes and enters a regime that is qualitatively distinct from the zonal-flow-dominated states.

The crucial conclusion from these numerical experiments is that the critical $L_z$ and, therefore, the critical $\beta_e$, is smaller for larger temperature gradients. For $\kappa_T \approx 1.4$, it is zero -- this coincides with the `Dimits transition' identified
in the 2D, electrostatic model of ITG turbulence studied in~\cite{ivanov_2020}. To rephrase this in terms that best describe the nature of the phenomenon of `high-beta runaway', as $\beta_e$ is increased, the critical temperature gradient at which the Dimits transition occurs becomes lower.

\begin{figure*}[htp]%
    \centering
    \captionsetup[subfloat]{labelformat=empty}
    \subfloat[\centering]{{\includegraphics[width=8cm]{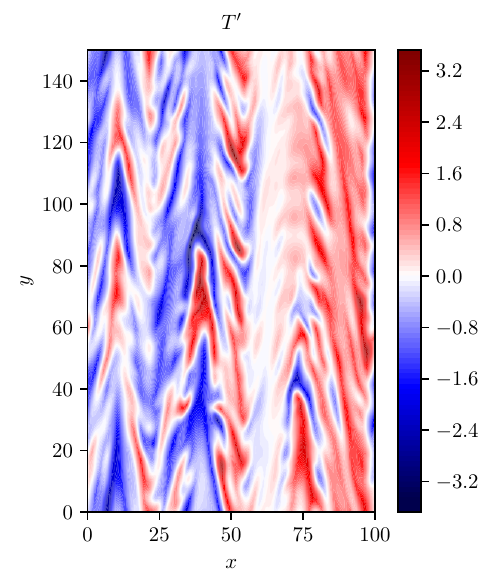}} }%
    \quad
    \subfloat[\centering ]{{\includegraphics[width=8cm]{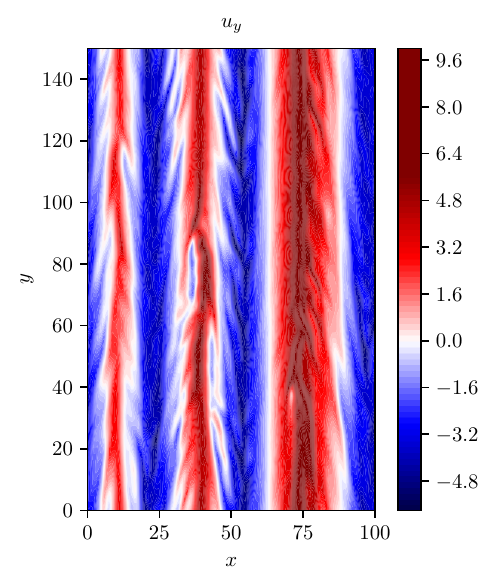}}  \label{fig:2D_plot_xy_vy_0.8_4.0}}%
    \vspace{-20pt}
    \quad
    \subfloat[\centering]{{\includegraphics[width=8cm]{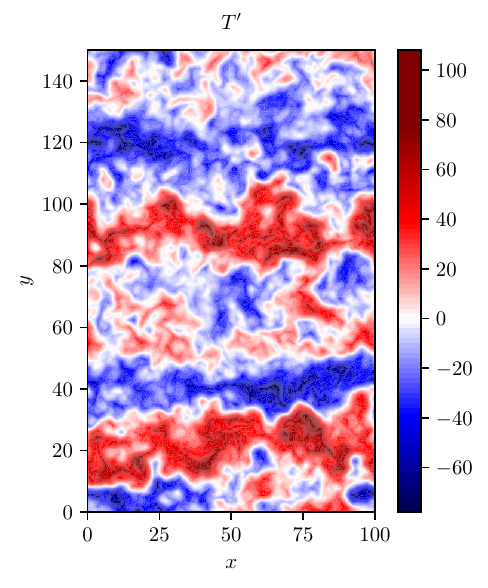}} }%
    \quad
    \subfloat[\centering ]{{\includegraphics[width=8cm]{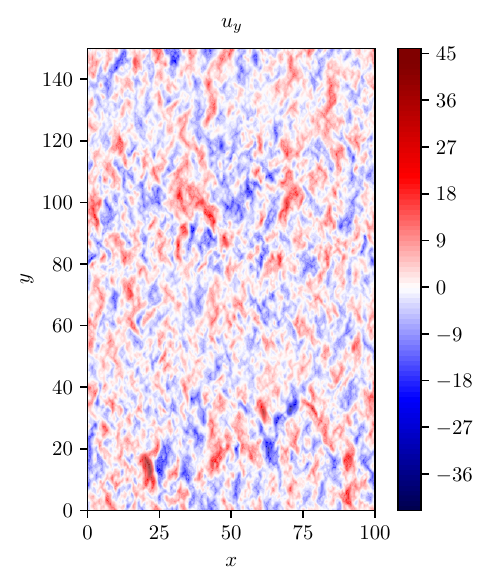}}  \label{fig:2D_plot_xy_vy_0.8_6.25}}%
    \caption{Snapshots of fields at the $z=0$ plane for a pair of runs with $\kappa_T = 0.8$ (the green line in Fig.~\ref{fig:hf_beta_scan}) and two values of $L_z$: $L^2_z = 7.8$ at $t = 9000$ (top) and $L^2_z = 19.4$ at $t = 450$ (bottom). The left panels show the non-zonal part of $T$, which indicates the level of turbulent fluctuation, and the right ones show the $y$-component of the $\mbf{E}\times \mbf{B}$ flow $u_y$, with a strong zonal flow manifest at the lower value of $L^2_z$.}
    \label{fig:plot_vy}
\end{figure*}

\subsection{A game of stresses}
\label{sec:stresses}
Given the striking correlation between the presence (or otherwise) of a strong zonal-flow and the way in which electromagnetic turbulence saturates, let us analyse the zonal flow's evolution. The evolution equation for the zonal vorticity is obtained by applying the flux-surface average to the vorticity equation~(\ref{eq:phi_red}):
\bea
\label{eq:zonal_vorticity}
 - \partial_t \partial^2_x\overline{\varphi} - \overline{\left\lbrace \varphi,\nabla^2_{\perp} \varphi\right\rbrace}& + \overline{\left\lbrace A,\nabla^2_{\perp} A\right\rbrace} + \overline{\partial_x\left\lbrace \partial_x\varphi,T \right\rbrace}\nonumber \\& = - \chi \partial^4_x\left(a\overline{\varphi} - b\overline{T}\right),
\eea
where $\overline{\varphi} \equiv \left\langle{\varphi}\right\rangle_{y,z}$ is the flux-surface average of $\varphi$.  
Integrating (\ref{eq:zonal_vorticity}) twice in $x$ results in an evolution equation for the zonal potential $\overline{\varphi}$:\footnote{The constants of integration are of the form $f(t)x + g(t)$, where $f$ vanishes due to the periodicity of the zonal mode and $g$ is safely set to zero because the physical effect only comes from the flow $\partial_x \overline{\phi}$.}
\beq
\label{zonal_t}
\partial_t\overline{\varphi} + \Pi_{\text{turb}} + \Pi_{\chi} =  0,
\eeq
where 
\beq
\Pi_{\chi} \equiv \chi \partial^2_x\left(a\overline{\varphi} - b\overline{T}\right)
\eeq
is the collisional stress, and $\Pi_{\text{turb}} = \Pi_{\varphi} + \Pi_A + \Pi_T$ is the turbulent stress, composed of the Reynolds stress
\beq
\label{eq:rey}
\Pi_{\varphi} \equiv -\overline{\left(\partial_x \varphi\right)( \partial_y \varphi)}, 
\eeq
the Maxwell stress
\beq
\label{eq:max}
\Pi_A \equiv \overline{\left(\partial_x A\right)( \partial_y A)},
\eeq
and the diamagnetic stress
\beq
\Pi_T \equiv -\overline{\left(\partial_x \varphi \right)(\partial_y T)}.
\eeq
We can now derive an evolution equation for the zonal flow's mean energy by multiplying both sides of (\ref{zonal_t}) by the zonal shear $S\equiv \partial_x^2\varphi$ and integrating
over the box in $x$:
\beq
\label{zonal_energy}
\partial_t \left\langle{\frac{\overline{u}^2_y}{2}}\right\rangle_x = \left\langle{S\Pi_{\text{turb}}}\right\rangle_x + \left\langle{S\Pi_{\chi}}\right\rangle_{x},
\eeq
where $\overline{u}_y = \partial_x \overline{\varphi}$ is the zonal flow velocity.

If a steady state with zonal flows is to be achieved, the contribution to (\ref{zonal_energy}) from the turbulent stress $\Pi_{\text{turb}}$ must balance the contribution
from the collisional stress $\Pi_{\chi}$.  As the latter is negative-definite, $\left\langle{S\Pi_{\text{turb}}}\right\rangle_x > 0$ is a necessary condition for a saturated zonally dominated state. This means that, cumulatively, the three turbulent stresses must have the same sign as $S$ -- reinforcing it, rather than opposing, as conventional viscosity (exemplified by $\Pi_{\chi}$) would do. 

Previous studies of electrostatic turbulence in a fluid model~\cite{ivanov_2020,  Ivanov_2022} have shown that the Reynolds ($\Pi_{\varphi}$) and diamagnetic ($\Pi_T$) stresses tend to support and to oppose zonal-flow generation, respectively\footnote{The fact that Reynolds stress in 2D will generate zonal flows is, of course, a well-established tenet of plasma physics -- this is described variously as a secondary Kelvin-Helmholtz instability of the ITG modes \cite{Roger_2000}, negative viscosity \cite{Diamond_2005}, or inverse cascade \cite{Balk_1990, Diamond_2005}. The latter two notions in fact originate in the theory of 2D hydrodynamic turbulence \cite{Kraichnan_1967, Batchelor_1969, Kraichnan_1976}}. Indeed, Ref.~\cite{ivanov_2020} argues that in the electrostatic limit, the transition from a zonally dominated, `Dimits' state to an un-saturated, streamer state occurs when the diamagnetic stress wins the battle with the Reynolds stress, for a certain definition of `win' that we will come to shortly. Since $\Pi_T$ always grows more important as $\kappa_T$ increases, the zonal-flow generation is eventually quenched after $\kappa_T$ reaches a critical value. Both gyrokinetic and fluid studies of electromagnetic turbulence~\cite{hysteresis, Anderson_2011, Naulin_2005, Scott_2005} have shown that the Maxwell stress tends to oppose zonal-flow generation, consistent with the decrease in critical $\kappa_T$ with increasing $L_z$ demonstrated in Fig.~\ref{fig:hf_beta_scan}. This is unsurprising, as the bending of magnetic field lines that accompanies a sheared zonal flow generates a restoring force that opposes the zonal flow. For purely Alfvénic fluctuations, where $\varphi=\pm A$, the Reynolds and Maxwell stresses annihilate one another due to the sign difference between (\ref{eq:rey}) and (\ref{eq:max}), leaving the diamagnetic stress the sole survivor.

\begin{figure}[!t]%
    \centering
    \includegraphics[width=8cm]{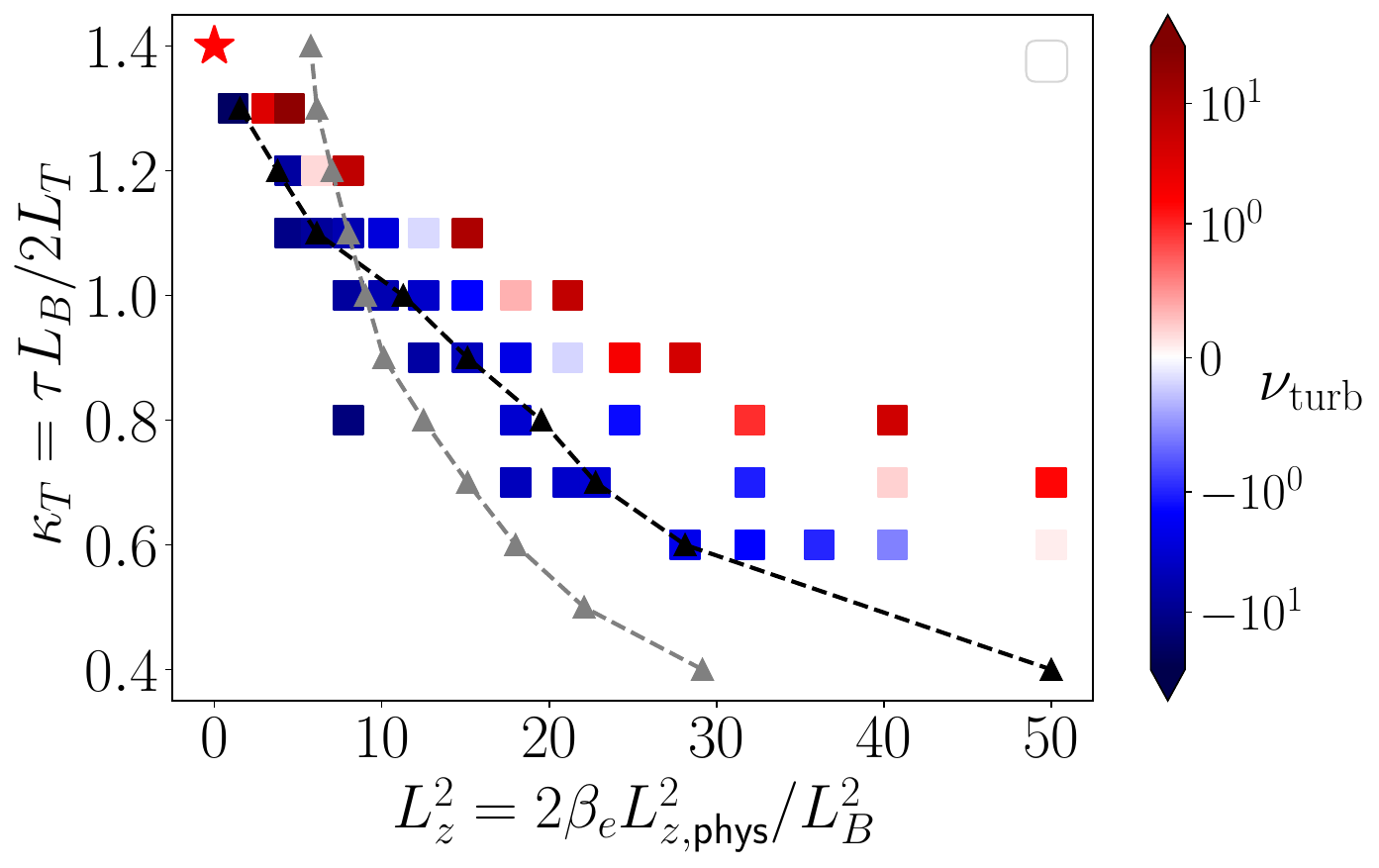}
    \caption{Turbulent viscosity $\nu_{\mathrm{turb}}$ versus ($\kappa_T$, $L^2_z$). The values of $\nu_{\mathrm{turb}}$ are calculated from simulations with fixed-in-time zonal flow. The black dashed line is the Dimits transition boundary calculated from simulations with self-consistently generated zonal flows, and the grey dashed line is the onset boundary of the DD and $\chi$ITG$-$ modes, which is also shown as the yellow dashed line in Fig.~\ref{fig:plot_gspec}(a). The red star is the Dimits transition threshold in the electrostatic limit~\cite{ivanov_2020}.}
    \label{fig:turbvis_beta_scan}
\end{figure}

When interpreting numerical results, it is convenient to characterise the zonal-flow dynamics in terms of turbulent viscosity $\nu_{\mathrm{turb}}$ \cite{ivanov_2020}. The standard definition of this quantity is in terms of an intuitively natural closure for the turbulent stress in (\ref{zonal_t}): $\Pi_{\mathrm{turb}} = - \nu_{\mathrm{turb}} S$. A positive $\nu_{\mathrm{turb}}$ corresponds to diffusive behaviour and the erosion of zonal flow, while negative $\nu_{\mathrm{turb}}$ corresponds to anti-diffusive behaviour and the generation of zonal flow. Of course, in general, $\nu_{\mathrm{turb}} = - \Pi_{\mathrm{turb}}(x,t)/\partial^2_x\overline{\varphi}(x,t)$ is a function both of time and space, but it is useful to construct a constant proxy quantity to characterise the propensity of turbulence to amplify (or otherwise) the zonal flows. This is easiest done on the basis of the zonal-flow energy equation (\ref{zonal_energy}), where the generation term is expressed as follows:\footnote{The term $b\chi \avg{S\partial^2_x\overline{T}}{x}$ has been neglected in (\ref{eq:zonal_energy2}). Numerically, it does indeed turn out always to be small.}
\begin{equation}
\label{eq:zonal_energy2}
\partial_t \avg{\frac{\overline{u}^2_y}{2}}{x} = - \nu_{\mathrm{turb}} \avg{S^2}{x} - a\chi \avg{S^2}{x},
\end{equation}
where we also assume that the zonal-flow evolution is much slower than the time variation of turbulent fluctuations and so define the turbulent viscosity as an average over many turnover times of the turbulence, during which the zonal flow remains approximately stationary:
\beq
\label{eq:nu_def}
\nu_{\mathrm{turb}} = - \left\langle \frac{\left\langle S \Pi_{\text{turb}} \right\rangle_x}{\left\langle S^2\right\rangle_x} \right\rangle_t.
\eeq
If this quantity is negative, that means that the stress-shear correlation is on average positive and so the zonal flows are reinforced by turbulence. 

The definition (\ref{eq:nu_def}) makes it possible to construct a series of numerical experiments in which $\nu_{\mathrm{turb}}$ can be measured directly as a function of $\kappa_T$ and $L^2_z$ (i.e., $\beta_e$). In these simulations, a fixed external zonal flow was applied. The shape of the imposed zonal flow is a triangle wave whose period is $L_x$; there is thus a constant, positive zonal shear across half of the box and an equal and opposite zonal shear across the other half. The aim is to determine how well the sign of the measured turbulent viscosity correlates with the ability of the turbulence to saturate.

A scatter plot of $\nu_{\mathrm{turb}}$ obtained from these simulations is shown in Fig. \ref{fig:turbvis_beta_scan}. For sufficiently small values of $\kappa_T$ and $L_z$, where the system saturates in the zonally dominated regime, $\nu_{\mathrm{turb}}$ is negative: as expected, the turbulent stresses reinforce the imposed zonal flow. In contrast, $\nu_{\mathrm{turb}}$ turns positive for sufficiently large $\kappa_T$ and $L_z$. This is unsurprising because the contribution from Maxwell stress increases with $L_z$ and the diamagnetic stress contribution increases with $\kappa_T$. The boundary in the $(\kappa_T, L_z^2)$ plane where the Dimits transition to high transport occurs is represented by black triangles in Fig. \ref{fig:turbvis_beta_scan}.  One can see that this transition boundary, traced out by the dashed black line in the figure, coincides approximately with the line where $\nu_{\mathrm{turb}}$ changes sign, albeit with a small offset.  In contrast, the onset of linear electromagnetic instabilities does not correlate well with the transition. In Fig. \ref{fig:turbvis_beta_scan}, this is illustrated by the grey triangles, also connected by a dashed line, indicating the $(\kappa_T, L_z^2)$ values where the DD and $\chi$ITG$-$ modes first go unstable.  The small offset of the transition boundary from the $\nu_{\mathrm{turb}}=0$ contour is likely due to the fact that $\nu_{\mathrm{turb}}$ was measured using the imposed zonal flows described above: as these imposed flows are not precisely the same as the flows that would be set up self-consistently by the turbulence, it is reasonable to expect $\nu_{\mathrm{turb}}$ to deviate from the true value unless $\nu_{\mathrm{turb}}$ is exactly independent of the zonal flow itself.

\begin{figure}[htb]%
    \centering
    \captionsetup[subfloat]{labelformat=empty}
    \subfloat[\centering ]{{\includegraphics[width=8cm]{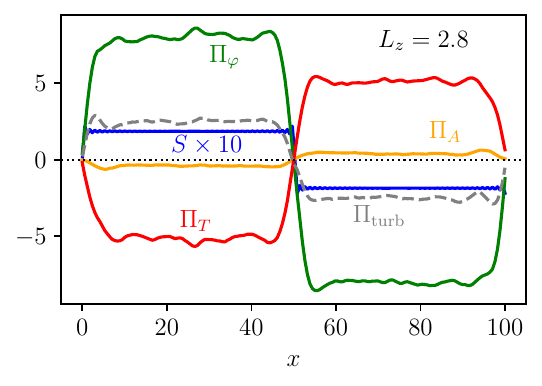}} }%
    \vfill
    \vspace{-30pt}
    \subfloat[\centering ]{{\includegraphics[width=8cm]{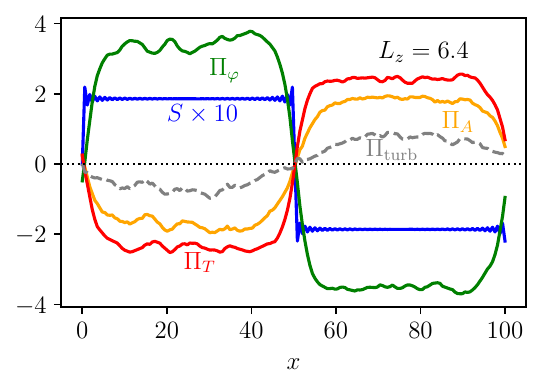} }}%
    \vspace{-10pt}
    \caption{Time-averaged profiles of the zonal shear $S$ (blue), the Reynolds stress $\Pi_{\varphi}$ (green), the Maxwell stress $\Pi_{A}$ (orange), the diamagnetic stress $\Pi_{T}$ (red) and the net turbulent stress $\Pi_{\mathrm{turb}}$ (grey) for $\kappa_T = 0.8$ and two different values of $L_z$: the $L_z=2.8$ case (top panel) corresponds to a zonal-flow-dominated state, whereas the $L_z=6.4$ case (bottom panel) lies beyond the runaway transition indicated in Fig.~\ref{fig:hf_beta_scan}.}
    \label{fig:plot_corre}
\end{figure}

\begin{figure}[htb]%
    \centering
    \includegraphics[width=8cm]{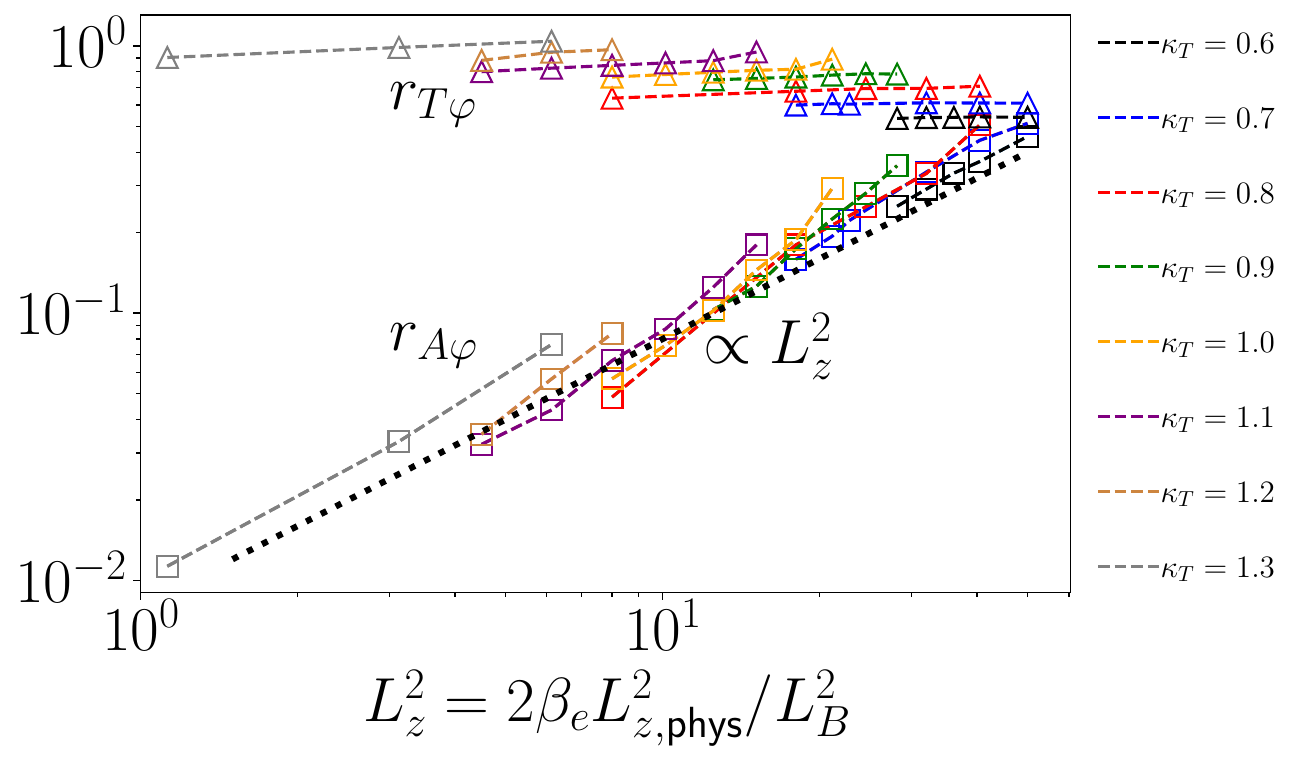}         
    \caption{The Maxwell-Reynolds stress ratio $r_{A\varphi}$ (squares) and the diamagnetic-Reynolds stress ratio $r_{T\varphi}$ (triangles). The different colours denote different values of $\kappa_T$. The data was obtained from simulations with frozen-in-time zonal flows. The scaling $r_{A\varphi} \propto L^2_z$ is given by the dotted line, for reference.}
    \label{fig:stress_ratio}
\end{figure}

As evident from (\ref{eq:nu_def}), the sign of $\nu_{\mathrm{turb}}$ is determined by the sign of $\Pi_{\text{turb}}$ relative to the zonal shear. To identify the role played by each component of the turbulent stress, in Fig.~\ref{fig:plot_corre} we plot indicative profiles of the time-averaged stresses and zonal shear from simulations with fixed-in-time zonal flows. As expected, the diamagnetic and Maxwell stresses both oppose the zonal flow, while the Reynolds stress supports it. As $L_z$ increases, the Maxwell stress increases in magnitude relative to the other stresses, flipping the sign of $\Pi_{\text{turb}}$ and leading to a state where zonal flows cannot be self-consistently reinforced. In Fig.~\ref{fig:stress_ratio}, we plot the ratios
\beq
r_{A\varphi} = \frac{\left\langle|\Pi_A| \right\rangle_x}{\left\langle|\Pi_{\varphi}|\right\rangle_x}, \;\; r_{T\varphi} = \frac{\left\langle |\Pi_T|\right\rangle_x}{\left\langle |\Pi_{\varphi}| \right\rangle_x}
\eeq
versus $L_z$ for a sequence of values of $\kappa_T$. One can see that the ratio between the Maxwell ($\Pi_A$) and Reynolds ($\Pi_{\varphi}$) stresses scales roughly as $L^2_z$ (or $\beta_e$) regardless of~$\kappa_T$. In contrast, the ratio between the diamagnetic stress ($\Pi_T$) and Reynolds stresses increases with $\kappa_T$, but remains relatively constant as $L_z$ is varied at a given~$\kappa_T$. Thus, the `high-beta runaway' is brought about entirely by the hostility of the Maxwell stress to zonal-flow generation.

A simple heuristic argument can be used to explain why the Maxwell stress tends to oppose zonal flows.  Consider a quasi-static zonal flow with a shearing rate $S$ that is constant in space.  The part of the turbulent viscosity due to the Maxwell stress is then
\bea
\nu_{t,A} = -\left<\frac{\left<S\Pi_{A}\right>_x}{\left<S^2\right>_x}\right>_t 
= -\frac{1}{S}\left<(\partial_x A) (\partial_y A)\right>_{x,y,z,t} \nonumber \\
= -\frac{1}{S}\sum_{\mathbf{k}}\left<k_x k_y \left|A_{\mathbf{k}}\right|^2\right>_{z,t}.
\eea
If we assume that $k_x$ is determined by the zonal shear, then $k_x \sim -S \tau k_y$, where $\tau$ is the characteristic turbulence decorrelation time.  Then the contribution 
of the Maxwell stress to the turbulent viscosity is
\beq
\nu_{t,A} \sim \sum_{\mathbf{k}}\left<\tau k_y^2 \left|A_{\mathbf{k}}\right|^2\right>_{z,t} > 0.
\eeq
All that is needed for this result is the fact that shear-induced $k_x \propto - S\tau k_y$ (by the same token, the Reynolds stress gives rise to negative turbulent viscosity, $\nu_{t,\mathrm{\varphi}}< 0$).

\section{Discussion}
\label{sec:discussion}
In this paper, we have proposed a minimal fluid model for electromagnetic, curvature-ITG-driven turbulence in a $Z$-pinch magnetic geometry.  Despite the simplicity
of our model, its numerical solution reproduces the key qualitative features observed in gyrokinetic simulations of electromagnetic turbulence in tokamaks.  In particular,
our simulations demonstrate that, for each value of the ion temperature gradient, there is a critical value of plasma beta beyond which the system transitions abruptly from a zonal-flow-dominated state with a low
heat flux to a streamer-dominated state with a much higher heat flux.
This transition, reminiscent of the finite-beta `non-zonal transitions', or `high-beta runaways', seen in gyrokinetic simulations \cite{waltz_2010, Transport, hysteresis}, occurs when the diamagnetic and Maxwell stresses -- which tend to oppose zonal-flow formation -- overcome the Reynolds stress, to which the zonal flows owe their existence.  Clear and simple scalings for the ratios of the diamagnetic and Maxwell stresses to the Reynolds stress are present
in the data, with the diamagnetic-Reynolds stress ratio independent of beta and the Maxwell-Reynolds stress ratio increasing linearly with $\beta_e$.  Consequently, the critical $\beta_e$ at which the transition occurs can be predicted with reasonable accuracy for each $\kappa_T$ using data from a single simulation at low $\beta_e$.

Nonlinear gyrokinetic simulations even at moderate values of $\beta_e$ are costly -- sometimes prohibitively so -- and thus the ability to predict the critical $\beta_e$ from a single, low-$\beta_e$ simulation would be of great utility.  An obvious next step is thus to extend the analysis carried out here to gyrokinetic simulations.  Moving to a gyrokinetic model and to tokamak
geometry will undoubtedly complicate matters: the inclusion of additional geometric parameters in tokamaks may introduce new dependences in the stress-ratio scalings, and the
form of the stresses themselves will require modification to account for kinetic and finite-Larmour-radius effects.  There is already evidence of these potential complications
in gyrokinetic simulations, where the inclusion of $\beta_e$ has been observed to leave the Dimits ITG threshold unaffected~\cite{Citrin_2014} or even increase it~\cite{Transport}.
Indeed, even the stress-balance picture for the Dimits transition in
electrostatic turbulence posited in~\cite{ivanov_2020} has yet to be verified in gyrokinetics -- while the stress-based interpretation may be relevant to the high-beta runaways, it may not always apply to the Dimits transition observed in gyrokinetic simulations in the same way as argued in \cite{ivanov_2020, Ivanov_2022}. Nonetheless, recent gyrokinetic simulation results indicating a
strong correlation between the high-beta runaway and stress balance for the zonal flow~\cite{hysteresis} provide a strong motivation to look for stress-ratio scalings similar to the ones obtained here.

\section*{Acknowledgments}
The work of Y.~Z. was supported in part by Tokamak Energy Ltd. His work, and the work of A.~A.~S. was also supported in part by the Simons Foundation via a Simons Investigator Award to A.~A.~S. The work of M.~B.,  A.~A.~S. and P.~G.~I. was supported in part by EPSRC (grant EPR034737/1). The work of T.~A. was supported in part by the Laboratory Directed Research and Development (LDRD) Program at the Princeton Plasma Physics Laboratory for the U.S. Department of Energy under Contract No. DE-AC02-09CH11466. The United States Government retains a non-exclusive, paid-up, irrevocable, world-wide license to publish or reproduce the published form of this manuscript, or allow others to do so, for United States Government purposes.

The authors would like to thank W.~Clarke, R.~Dutta, J.~Edmiston, M.~Hardman, Y.~Kawazura, D.~Kennedy, R.~Nies, M.~Romanelli, H.~Zhang, and the rest of the Oxford plasma-theory group for fruitful discussions and useful comments.

\section*{References}

\bibliography{References.bib}
\bibliographystyle{iopart-num}

\appendix

\section{Derivation of the fluid model equations of electromagnetic ITG turbulence}

\label{derivation}

\subsection{Gyrokinetics}

Our starting point is the gyrokinetic-Maxwell system of equations, obtained by imposing the gyrokinetic ordering on the Vlasov-Landau-Maxwell equations. The ordering is  
\bea
\label{eq:gk_ordering}
\frac{\omega}{\Omega_i}\sim \frac{\delta f_s}{f_s}\sim \frac{e\phi}{T_e}\sim \frac{|\delta \mathbf{B}|}{B}\sim\frac{k_{\parallel}}{k_{\perp}} \sim \frac{\rho_i}{L} \equiv \rho_* \ll 1,
\eea
where $\omega$ is the charateristic frequency of the plasma fluctuations, $\Omega_i$ is the ion cyclotron frequency, $f_s$ is the particle distribution function for species $s$, $\delta f_s$ is its fluctuating component, $e$ is the proton charge, $T_e \sim T_i$ is the electron (ion) temperature, $\phi$ is the electrostatic potential, $\delta \mathbf{B}$ is the magnetic-field perturbation, $B$ is the equilibrium magnetic field's strength, $\kpar$ and $\kperp$ are the characteristic wavenumbers along and across the equilibrium magnetic field, $\rho_i$ is the thermal ion Larmor radius, and $L \sim L_B \sim L_T \sim L_n$ is a typical scale length of the equilibrium, with the gradient length scale of the quantity $\alpha$ defined by $L_{\alpha} = |\nabla \ln{\alpha}|^{-1}$. The resulting gyrokinetic equation is~\cite{Abel_2013}:
\bea
\label{eq:gk_eqs}
\frac{\partial h_s}{\partial t} + \left(v_{\parallel}\hat{\mathbf{z}}\ + \left\langle \mathbf{V}_{\chi} \right\rangle_{\mathbf{R}_s} + \mathbf{V}_M\right) \cdot \frac{\partial h_s}{\partial \mathbf{R}_s} \nonumber \\ + \left\langle \mathbf{V}_{\chi}\right\rangle_{\mathbf{R}_s}\cdot\left[\frac{\partial \ln{n_{s}}}{\partial \mathbf{R}_s} + \left(\frac{\varepsilon}{T_s}-\frac{3}{2}\right)\frac{\partial \ln{T_s}}{\partial \mathbf{R}_s}\right] F_{s}\nonumber \\ = \frac{Z_s e}{T_s}\frac{\partial \left\langle \chi\right\rangle_{\mathbf{R}_s}}{\partial t} F_s + \left\langle C\left[h_s\right]\right\rangle_{\mathbf{R}_s},
\eea
which describes the time-evolution of the non-Boltzmann part of $\delta f_s = -Z_se\phi F_s/T_s + h_s$, with the following notation:
\begin{equation}
\label{eq:vchi}
\left\langle \mathbf{V}_{\chi} \right\rangle_{\mathbf{R}_s} = \frac{c}{B}\hat{\mathbf{z}}\times\frac{\partial \left\langle \chi\right\rangle_{\mathbf{R}_s}}{\partial \mathbf{R}_s},
\end{equation}
\begin{equation}
\label{eq:chi_definition}
\chi = \phi - \frac{\mathbf{v}\cdot \mathbf{A}}{c},
\end{equation}
\begin{equation}
\mathbf{V}_M = \frac{v^2_{\perp}}{2\Omega_s}\hat{\mathbf{z}}\times\nabla\ln{B} + \frac{v^2_{\parallel}}{\Omega_s}\hat{\mathbf{z}}\times \left(\hat{\mathbf{z}}\cdot\nabla \hat{\mathbf{z}}\right),
\end{equation}
\begin{equation}
F_s\left(\mathbf{R}_s,\varepsilon\right) = \frac{n_s}{\left(2\pi T_s/m_s\right)^{3/2}} \exp{\left(-\frac{\varepsilon}{T_s}\right)},
\end{equation}
where $\hat{\mathbf{z}}$ is the unit vector pointing in the direction of the equilibrium field, $\vpa$ and $\vpe$ are the velocity components parallel and perpendicular to $\unit{z}$, respectively, $Z_s$ is the atomic number of species $s$ ($Z_s = -1$ for the electrons), and $\Omega_s = Z_s eB/m_sc$. The gyrokinetic equation (\ref{eq:gk_eqs}) is written in phase-space coordinates that are the guiding-centre position $\mathbf{R}_s$ of the particles of species~$s$, the particle's kinetic energy $\varepsilon = m_s(\vpa^2 + \vpe^2)/2$, its magnetic moment $\mu = m_s\vpe^2/2B$, and the sign $\lambda = \pm 1$ of $\vpa$ (where the magnitude is set by $\varepsilon$ and $\mu$). The generalized potential $\chi$ (note that this symbol is also used to represent collisionally) is evaluated at the particle position $\mathbf{r}$ (before being gyro-averaged at fixed $\mathbf{R}_s$).

The gyrokinetic equation (\ref{eq:gk_eqs}) is solved together with the field equations for $\phi$ and $\mathbf{A}$. The quasi-neutrality constraint gives an equation for $\phi$:
\begin{equation}
\label{eq:75}
0 = \sum_s Z_s \left[-\frac{Z_s e \phi}{T_s}n_s + \int \sd^3 \mathbf{v} \left\langle h_s \right\rangle_{\mathbf{r}}\right].
\end{equation}
The parallel component $A_\parallel$ of the vector potential is calculated from parallel Amp\`{e}re's law:
\begin{equation}
\label{eq:parallel_ampere}
\nabla^2_{\perp}A_{\parallel} = -\frac{4\pi}{c}\sum_s Z_s e \int \sd^3 \mathbf{v}\, v_{\parallel} \left\langle h_s\right\rangle_{\mathbf{r}},
\end{equation}
while the perpendicular component $\mathbf{A}_{\perp}$ is related to the field-strength perturbation  $\delta B_{\parallel} = \delta \mathbf{B} \cdot \unit{z}$ by 
\begin{eqnarray}
\delta B_{\parallel} = \unit{z}\cdot \left(\nabla_{\perp}\times \mathbf{A}_{\perp}\right),
\end{eqnarray}
and the equation for $\delta B_{\parallel}$ is deduced from the perpendicular part of Amp\`{e}re's law:
\bea
\label{eq:perp_ampere}
&\nabla^2_{\perp} \delta B_{\parallel} =  \nonumber \\ & - \frac{4\pi}{c}\hat{\mathbf{z}}\cdot\left[\nabla_{\perp} \times \sum_s Z_s e \int \sd^3 \mathbf{v}\,\left\langle\mathbf{v}_{\perp}h_s\right\rangle_{\mathbf{r}}\right].
\eea
From these field equations, one obtains $\phi$, $A_{\parallel}$ and $\delta B_{\parallel}$, whence $\left\langle \mathbf{V}_{\chi} \right\rangle_{\mathbf{R}_s}$ in (\ref{eq:gk_eqs}) is calculated via (\ref{eq:vchi}):
\bea
\label{eq:chiR}
\left\langle \chi \right\rangle_{\mathbf{R}_s} = \sum_{\mathbf{k}} e^{i\mathbf{k}\cdot\mathbf{R}_s} \Bigg[J_0\left(k_{\perp}v_{\perp}/\Omega_s\right)\left(\phi_{\mathbf{k}} - \frac{v_{\parallel} A_{\parallel,\mathbf{k}}}{c}\right)\nonumber \\ + \frac{T_s}{Z_s e}\frac{2v^2_{\perp}}{v^2_{\text{th}s}}\frac{J_1\left(k_{\perp}v_{\perp}/\Omega_s \right)}{k_{\perp}v_{\perp}/\Omega_s}\frac{\delta B_{\parallel,\mathbf{k}}}{B}\Bigg],
\eea
where $J_0$ and $J_1$ are Bessel functions, which are the gyroaveraging operators in Fourier space. We stress that in this work, when written without arguments, $\phi$, $A_{\parallel}$ and $\delta B_{\parallel}$ are evaluated at the particle position $\mathbf{r}$, while $h_s$ is always evaluated at the guiding-centre position $\mathbf{R}_s$. In other cases, the function's arguments will be explicitly written to avoid confusion.

\subsection{Small-mass-ratio expansion}
\label{mass ratio expansion}
Here we use the smallness of the electron-ion mass ratio to derive a reduced model for electrons in which they are treated as an isothermal fluid. This is achieved by applying a subsidiary expansion in $\sqrt{m_e/m_i} \ll 1$ to the electron gyrokinetic equation~\cite{Schekochihin_2009}. We focus on the ion-scale fluctuations, for which $\kperp \rho_e \sim \sqrt{m_e/m_i} \ll 1$, but otherwise adopt the minimal ordering in which the plasma $\beta$, the ion-electron temperature ratio $T_i/T_e$, and the ion atomic number $Z$ are all of order unity.

The resulting equation for $h_e$, valid up to first order in $\sqrt{m_e/m_i}$, is 
\bea
\label{eq:GK he}
\underbrace{\frac{\partial h_e}{\partial t}}_{1} + \Bigg[\underbrace{v_{\parallel}\unit{z}}_{0} + \frac{c}{B}\unit{z}\times \frac{\partial}{\partial \mathbf{R}_e}\Bigg(\underbrace{\phi\left(\mathbf{R}_e\right)}_{1} - \underbrace{\frac{v_{\parallel} A_{\parallel}\left(\mathbf{R}_e\right)}{c}}_{0}\nonumber\\ - \underbrace{\frac{T_e v^2_{\perp} \delta B_{\parallel}\left(\mathbf{R}_e\right)}{e v^2_{\text{th}e}B}}_{1}\Bigg)\Bigg] \cdot\frac{\partial h_e}{\partial \mathbf{R}_e}  + \underbrace{\mathbf{V}_{M}\cdot \frac{\partial h_e}{\partial \mathbf{R}_e}}_{1} \nonumber\\ = - \frac{e}{T_e} \frac{\partial}{\partial t} \Bigg(\underbrace{\phi\left(\mathbf{R}_e\right)}_{1} - \underbrace{\frac{v_{\parallel}A_{\parallel}\left(\mathbf{R}_e\right)}{c}}_{0} - \underbrace{\frac{T_ev^2_{\perp}\delta B_{\parallel}\left(\mathbf{R}_e\right)}{e v^2_{\text{th}e}B}}_{1}\Bigg)F_e  \nonumber\\ + \underbrace{\left\langle C\left[h_e\right]\right\rangle_{\mathbf{R}_e}}_{0},
\eea
where the number below each term indicates its order in $\sqrt{m_e/m_i}$ relative to $(\vthe/L)h_e$. In ordering the terms in (\ref{eq:GK he}), we have assumed that the electron-electron collision frequency $\nu_{ee}$ is of order $\vthe/L$ and have used the fact that $\avg{\chi}{\mathbf{R}_e} \approx \chi(\mathbf{R}_e)$ to the order considered. 

We next expand $h_e = h^{(0)}_e + h^{(1)}_e + \cdots$, with $h^{(j+1)}_e/h^{(j)}_e \sim \sqrt{m_e/m_i}$. The lowest-order electron gyrokinetic equation is then
\bea
\label{eq:lowest order he}
v_{\parallel}\frac{\partial h^{(0)}_e(\mathbf{r})}{\partial z} - \frac{v_{\parallel}}{B}\left\lbrace \apar,  h^{(0)}_e\left(\mathbf{r}\right)\right\rbrace \nonumber \\=\frac{v_{\parallel}}{c}\frac{e}{T_e}\frac{\partial \apar}{\partial t}  F_e  + C\left[h^{(0)}_e(\mathbf{r})\right].
\eea
This equation constrains $h^{(0)}_e(\mathbf{r})$ to be a Maxwellian velocity distribution. To demonstrate this, we first multiply (\ref{eq:lowest order he}) by $h^{(0)}_e(\mathbf{r})/F_e$ and integrate over the ($\mathbf{r}, \mathbf{v}$) phase space\footnote{Note that the integral over the domain perpendicular to the mean magnetic field is defined so that its extent in both $x$ and $y$ coordinates, denoted $l_x$ and $l_y$, respectively, satisfies $\rho_i \ll l_x,\, l_y \ll L$.}.  On the left-hand side of (\ref{eq:lowest order he}), the first term vanishes upon integration over $z$, since it can be expressed as a total derivative with respect to $z$, following from the fact that $F_e$ is constant along the field line. The Poisson bracket in the second term also vanishes after integration by parts in ($x$,$y$). The resulting equation is 
\bea
\label{eq:two_integrals}
\frac{n_e e}{cT_e} \int d^3\mathbf{r}\, \frac{\partial A_{\parallel}}{\partial t}u^{(0)}_{\parallel e} \nonumber \\ + \int d^3\mathbf{v}d^3\mathbf{r}\, C\left[h^{(0)}_e(\mathbf{r})\right] \frac{h^{(0)}_e(\mathbf{r})}{F_e}  = 0,
\eea
where $u^{(0)}_{\parallel e} = (1/n_e)\int d^3\mathbf{v}\,v_{\parallel} h^{(0)}_e(\mathbf{r}) $ is the the lowest-order parallel flow of the electrons. The size of $u^{(0)}_{\parallel e}$ is constrained by parallel Amp\`{e}re's law (\ref{eq:parallel_ampere}):
\bea
\label{eq:parallel_ampere2}
\nabla^2_{\perp} A_{\parallel} = - \frac{4\pi e n_e}{c}\left(u_{\parallel i} - u_{\parallel e}\right),
\eea
where $u_{\parallel i} = (1/n_i)\int d^3\mathbf{v}\,v_{\parallel}\left\langle h_i\right\rangle_{\mathbf{r}}  \sim  \rho_* \vthi \ll u^{(0)}_{\parallel e} \sim \rho_*v_{\text{th}e}$, so $u_{\parallel i}$ can be dropped. So can $\nabla^2_{\perp} A_{\parallel}$ because
\bea
\frac{k^2_{\perp}A_{\parallel} c}{4\pi en_e u^{(0)}_{\parallel e}}& \sim k_{\perp}\rho_i \frac{\delta B_{\perp}}{B \rho^*} \frac{1}{\beta_i}\sqrt{\frac{\tau m_e}{m_i}} \sim \sqrt{\frac{m_e}{m_i}}.
\eea
Thus, as no term in (\ref{eq:parallel_ampere2}) can balance the $u_{\parallel e}$ term to lowest order, $u^{(0)}_{\parallel e} = 0$ and $u_{\parallel e} \approx u^{(1)}_{\parallel e} \sim \rho_*\vthi$. Using this result in (\ref{eq:two_integrals}) gives us
\begin{equation}
\int d^3\mathbf{v}d^3\mathbf{r}\, C\left[h^{(0)}_e(\mathbf{r})\right] \frac{h^{(0)}_e(\mathbf{r})}{F_e} = 0,
\end{equation}
whence, according to Boltzmann's $H$-theorem, $h^{(0)}_e\left(\mathbf{r}\right)$ is a perturbed Maxwellian (with no mean flow to lowest order):
\begin{equation}
\label{eq:h0}
h^{(0)}_e\left(\mathbf{r}\right) = \left[\frac{\delta n_e}{n_e} - \frac{e \phi}{T_e} + \left(\frac{\varepsilon}{T_e} - \frac{3}{2}\right)\frac{\delta T_e }{T_e}\right]F_e.
\end{equation}
Substituting (\ref{eq:h0}) back into (\ref{eq:lowest order he}), we obtain
\begin{equation}
\label{eq:84}
\frac{e}{T_e c} \frac{\partial A_{\parallel}}{\partial t} = \nabla_{\parallel} \left[\frac{\delta n_e}{n_e} - \frac{e\phi}{T_e} + \left(\frac{\varepsilon}{T_e} - \frac{3}{2}\right)\frac{\delta T_e}{T_e}\right].
\end{equation}
where $\nabla_{\parallel} (\cdots) = \partial (\cdots)/\partial z - (1/B)\left\lbrace \apar, (\cdots) \right\rbrace$ is the derivative taken along the exact field line. Since (\ref{eq:84}) holds true for all $\varepsilon$, we can split it into two separate equations:
\begin{equation}
\label{eq:A_e}
\frac{e}{T_e c} \frac{\partial A_{\parallel}}{\partial t} = \nabla_{\parallel} \left(\frac{\delta n_e}{n_e} - \frac{e\phi}{T_e}\right),
\end{equation}
and
\begin{equation}
\label{eq:Te}
\nabla_{\parallel} \delta T_e = 0.
\end{equation}
Since $\delta T_e = 0$ is a solution of (\ref{eq:Te}), and we have assumed that there are no mean electron temperature or density gradients, $\delta T_e$ will be set to zero from here onwards, i.e., the electrons are isothermal. The other equation that we have derived, (\ref{eq:A_e}), determines $\apar$ via what is readily recognised as the parallel force balance between the electric field and the electron pressure gradient.

Let us now take the density moment of (\ref{eq:GK he}), working to lowest order in $\sqrt{m_e/m_i}$ (i.e., ignoring electron FLR corrections):
\bea
\label{n_e}
\frac{\sd}{\sd t}\left(\frac{\delta n_e}{n_e} - \frac{\delta B_{\parallel}}{B}\right) + \nabla_{\parallel} u_{\parallel e} + \frac{cT_e}{B e} \left\lbrace \frac{\delta n_e}{n_e}, \frac{\delta B_{\parallel}}{B}\right\rbrace \nonumber\\ + \frac{\rho_ev_{\text{th}e}}{2}\left(\frac{1}{L_B} + \frac{1}{R}\right)\frac{\partial}{\partial y}\left(\frac{\delta n_e}{n_e} - \frac{e\phi}{T_e}\right) = 0,
\eea
where we have used the $Z$-pinch coordinates ($x$,$y$,$z$) defined in Fig.~\ref{fig:zpinch}. The relationship between $\delta n_e/n_e$ and $\phi$ is set by the quasi-neutrality condition (\ref{eq:75}), which, using (\ref{eq:h0}), can be written as
\begin{equation}
\label{quasi_e}
\frac{\delta n_e}{n_e} = - \frac{Ze\phi}{T_i} + \frac{1}{n_i}\int d^3\mathbf{v} \left\langle h_i\right\rangle_{\mathbf{r}},
\end{equation}
where $h_i$ still needs to be calculated from the ion gyrokinetic equation. Using (\ref{eq:h0}) also for the perpendicular Amp\`{e}re's law (\ref{eq:perp_ampere}), one finds
\bea
\label{perpendicular_ampere_e}
\frac{\delta B_{\parallel}}{B} + \frac{\beta_e}{2}\frac{\delta n_e}{n_e} - \frac{\beta_e}{2}\frac{e\phi}{T_e} \nonumber\\ = - \frac{\beta_i}{2}\sum_{\mathbf{k}}e^{i\mathbf{k}\cdot\mathbf{r}} \frac{1}{n_i} \int d^3\mathbf{v}\text{ }\frac{2v^2_{\perp}}{v^2_{\text{th}i}}\frac{J_1\left(k_{\perp}v_{\perp}/\Omega_i\right)}{k_{\perp}v_{\perp}/\Omega_i}h_{i\mathbf{k}}.
\eea

Equations (\ref{eq:parallel_ampere2}), (\ref{eq:A_e}) and (\ref{n_e})-(\ref{perpendicular_ampere_e}), together with the ion gyrokinetic equation, form a closed set of equations for the fields $u_{\parallel e}$, $A_{\parallel}$, $\delta n_e$, $\phi$, $\delta B_{\parallel}$, and $h_i$.

\subsection{Cold-ion fluid}
\label{cold ion expansion}
\subsubsection*{Orderings.}-- We shall now derive a set of closed equations in which the ions are treated as a cold fluid. This is achieved by performing a subsidiary expansion in $\tau = T_i/ZT_e\ll 1$ and enforcing a collisional closure on the moment hierarchy. We restrict our consideration to the long-wavelength limit $\kperp \rho_i \ll 1$, but retain finite sound radius: $\kperp \rho_s \sim 1 \gg \tau \sim \kperp^2\rho^2_i$, where $\rho_s = \rho_i /\sqrt{2\tau} \gg \rho_i$. This is necessary to capture the peak of the ITG mode's growth rate.

For now, we retain finite-$\beta_e$ effects by adopting the minimal ordering $\beta_e \sim 1$, which we will relax in \ref{low beta limit}. Because the ions are cold, $\beta_i = \tau \beta_e \ll 1$. For clarity, we gather the ordering assumptions here:
\begin{equation}
\label{eq:cold_ion_ordering}
\sqrt{\frac{m_e}{m_i}} \ll \beta_i \sim \kperp^2\rho^2_i \sim \tau \sim \frac{L_T}{L_B} \ll \beta_e \sim 1,
\end{equation}
so our system is also strongly driven ($L_T \ll L_B$). The various frequencies present in the system are therefore ordered as 
\bea
\label{eq:frequency ordering app}
\frac{1}{\tau}\omega_{di} \sim \omega_{de} \sim \omega_{*T}&\sim \omega \sim k_{\parallel}\frac{v_{\text{th}i}}{\sqrt{\tau}} \sim k_{\parallel} v_A \nonumber\\ &\sim k_{\parallel}c_s \sim k_{\perp}u_{\perp} \sim \nu_{ii} \tau,
\eea
and the fluctuations satisfy
\bea
\label{eq:cold_ion_ordering_full}
\tau\frac{Ze\phi}{T_i} \sim \tau\frac{\delta T_i}{T_i} \sim \frac{\delta n_e}{n_e} = \frac{\delta n_i}{n_i} &\sim \frac{k_{\perp}A_{\parallel}}{B}\nonumber\\& \sim \frac{\delta B_{\parallel}}{B} \sim \sqrt{\tau} \frac{u_{\parallel i}}{v_{\text{th}i}}.
\eea
Note that $\delta \bpar \sim \kperp \apar$, consistent with $\beta_e \sim 1$. The ordering (\ref{eq:cold_ion_ordering_full}) for the fluctuations will be verified as consistent a posteriori.

\subsubsection*{Expansion of the ion gyrokinetic equation.}-- We now expand the ion distribution function as $h_i = h^{(-1)}_i + h^{(0)}_i + h^{(1)}_i + \cdots$, with $h^{j+1}_i/h^{j}_i \sim \sqrt{\tau}$, and order all terms in the ion gyrokinetic equation:
\bea
\label{eq:ion_gk_eq}
\frac{\partial}{\partial t}\Bigg[\underbrace{h^{(-1)}_i}_{-1} + \underbrace{h^{(0)}_i}_{0} + \underbrace{h^{(1)}_i}_{1}\Bigg] + v_{\parallel}\hat{\mathbf{z}}\cdot\frac{\partial}{\partial \mathbf{R}_i}\Bigg[\underbrace{h^{(-1)}_i}_{0} + \underbrace{h^{(0)}_i}_{1}\Bigg] \nonumber\\ + \frac{c}{B}\left\lbrace \left\langle \phi\right\rangle_{\mathbf{R}_i}, \underbrace{h^{(-1)}_i}_{-1} + \underbrace{h^{(0)}_i}_{0} + \underbrace{h^{(1)}_i}_{1} \right\rbrace \nonumber\\ + \frac{c}{B}\left\lbrace \left\langle  \frac{-v_{\parallel}A_{\parallel}}{c}\right\rangle_{\mathbf{R}_i}, \underbrace{h^{(-1)}_i}_{0} + \underbrace{h^{(0)}_i}_{1}\right\rbrace \nonumber\\+ \underbrace{\frac{c}{B}\left\lbrace \left\langle \frac{-\mathbf{v_{\perp}}\cdot\mathbf{A}_{\perp}}{c}\right\rangle_{\mathbf{R}_i}, h^{(-1)}_i\right\rbrace}_{1} + \underbrace{\mathbf{V}_M\cdot\frac{\partial h^{(-1)}_i}{\partial \mathbf{R}_i}}_{1} \nonumber\\ + \left(\frac{v^{2}}{v^2_{\text{th}i}}-\frac{3}{2}\right)\frac{c}{B}\hat{\mathbf{z}}\times\frac{\partial}{\partial \mathbf{R}_i}\Bigg[\underbrace{\left\langle \phi\right\rangle_{\mathbf{R}_i}}_{-1} + \underbrace{\left\langle \frac{-v_{\parallel}A_{\parallel}}{c}\right\rangle_{\mathbf{R}_i}}_{0} \nonumber\\+ \underbrace{\left\langle \frac{-\mathbf{v_{\perp}}\cdot\mathbf{A}_{\perp}}{c}\right\rangle_{\mathbf{R}_i}}_{1}\Bigg]\cdot\frac{\partial \ln T_i}{\partial \mathbf{R}_i}F_i \nonumber\\ = \frac{Ze}{T_i}\frac{\partial}{\partial t}\Bigg[\underbrace{\left\langle \phi\right\rangle_{\mathbf{R}_i}}_{-1} + \underbrace{\left\langle \frac{-v_{\parallel}A_{\parallel}}{c}\right\rangle_{\mathbf{R}_i}}_{0} + \underbrace{\left\langle \frac{-\mathbf{v}_{\perp}\cdot\mathbf{A}_{\perp}}{c}\right\rangle_{\mathbf{R}_i}}_{1} \Bigg]F_i\nonumber \\ + \left\langle C\left[\underbrace{h^{(-1)}_i}_{-3} + \underbrace{h^{(0)}_i}_{-2} + \underbrace{h^{(1)}_i}_{-1}+ \underbrace{h^{(2)}_i}_{0} + \underbrace{h^{(3)}_i}_{1}\right]\right\rangle_{\mathbf{R}_i},
\eea
where the number below each term indicates its order in $\sqrt{\tau}$ relative to $\vthi h^{(-1)}_i/L_B$. Note that the electromagnetic-field terms contain contributions at multiple orders in $\sqrt{\tau}$ due to the presence of gyro-averages: e.g., $\avg{\phi}{\mathbf{R}_i} = \phi\left(\mathbf{R}_i\right) + (\rho^2_i/2)\partial^2\phi\left(\mathbf{R}_i\right)/\partial \mathbf{R}^2_i + \cdots = \phi\left(\mathbf{R}_i\right) (1 + O[\tau])$. These contributions will be disentangled where necessary in the calculations that follow.

Because the collision frequency is ordered to be much larger than the fluctuation frequency ($\nu_{ii} \sim \omega \tau^{-1}$), both $h^{(-1)}_i$ and $h^{(0)}_i$ are constrained by (\ref{eq:ion_gk_eq}) to be perturbed Maxwellians:
\begin{equation}
\label{eq:vanishing_collision_op}
C\left[h^{(-1)}_i\right] = C\left[h^{(0)}_i\right] = 0,
\end{equation}
where the gyro-averages can be dropped to this order due to the fact that $\kperp^2\rho^2_i \sim \tau \ll 1$. It is thus useful to decompose $h_i$ as 
\bea
\label{eq:h_iR}
h_i = \Bigg[\frac{Ze\phi\left(\mathbf{R}_i\right)}{T_i} + \frac{\delta n_i\left(\mathbf{R}_i\right)}{n_i} + \frac{2v_{\parallel}u_{\parallel i}\left(\mathbf{R}_i\right)}{v^2_{\text{th}i}}\nonumber\\ + \left(\frac{v^2}{v^2_{\text{th}i}} - \frac{3}{2}\right)\frac{\delta T_i\left(\mathbf{R}_i\right)}{T_i}\Bigg]F_i\left(\mathbf{R}_i\right) + \cdots
\eea
Using (\ref{eq:cold_ion_ordering_full}), we compare the sizes of each of the terms inside the square brackets in (\ref{eq:h_iR}) to obtain expressions for $h^{\left(-1\right)}_i$, $h^{\left(0\right)}_i$, and $h^{\left(1\right)}_i$:
\bea
\label{eq:hi-1}
h^{\left(-1\right)}_i & = \Bigg[\frac{Ze\phi\left(\mathbf{R}_i\right)}{T_i} \nonumber\\& + \left(\frac{v^2}{v^2_{\text{th}i}} - \frac{3}{2}\right)\frac{\delta T_i\left(\mathbf{R}_i\right)}{T_i}\Bigg]F_i\left(\mathbf{R}_i\right),
\eea
\bea
\label{eq:hi_0}
h^{\left(0\right)}_i =  \frac{2v_{\parallel}u_{\parallel i}\left(\mathbf{R}_i\right)}{v^2_{\text{th}i}} F_i\left(\mathbf{R}_i\right),
\eea
\bea
\label{eq:hi_1}
h^{\left(1\right)}_i = \frac{\delta n_i\left(\mathbf{R}_i\right)}{n_i} F_i\left(\mathbf{R}_i\right) + \tilde{h}^{\left(1\right)}_i.
\eea
Note that we have included an additional contribution $\tilde{h}^{\left(1\right)}_i$ to $h^{\left(1\right)}_i$ because the latter is not constrained by (\ref{eq:vanishing_collision_op}) to be a perturbed Maxwellian. 

\subsubsection*{Expansion of the field equations.}--
We now demonstrate that the fields obtained by substituting (\ref{eq:hi-1})-(\ref{eq:hi_1}) into Maxwell's equations are consistent with the ordering (\ref{eq:cold_ion_ordering_full}). We first consider the quasi-neutrality condition (\ref{quasi_e}):
\bea
\label{quasi_i}
\frac{\delta n_e}{n_e}  & = - \frac{Ze\phi}{T_i} + \frac{1}{n_i} \int \sd^3\mathbf{v}\,\Bigg[1 + \frac{1}{4}\frac{v^2_{\perp}\rho^2_i}{v^2_{\text{th}i}}\nabla^2_{\perp} \nonumber\\ & + O\left(\tau^2\right)\Bigg]h^{(-1)}_i\left(\mathbf{r}\right) + \frac{1}{n_i}\int \sd^3\mathbf{v}\, h^{(1)}_i\left(\mathbf{r}\right),
\eea
where we have expanded the gyro-average. Using (\ref{eq:hi-1}) and (\ref{eq:hi_1}) for $h^{(-1)}_i$ and $h^{(1)}_i$ in (\ref{quasi_i}) results in a constraint on $\tilde{h}^{\left(1\right)}_i$ 
\bea
\label{eq:htilde_constraint}
\frac{1}{n_i}\int \sd^3 \mathbf{v}\, \tilde{h}^{\left(1\right)}_i\left(\mathbf{r}\right) = - \frac{1}{4}\rho^2_i\nabla^2_{\perp}\left(\frac{Ze\phi}{T_i} + \frac{\delta T_i}{T_i}\right) .
\eea
The left-hand side of (\ref{eq:htilde_constraint}) is, according to (\ref{eq:hi_1}), of size $\delta n_i/ n_i$, and the right-hand side is of size $\tau (e\phi/T_i)$. This is consistent with the ordering (\ref{eq:cold_ion_ordering_full}).

Next, consider the parallel Amp\`{e}re's law (\ref{eq:parallel_ampere2}). Since $h^{(-1)}_i$ according to (\ref{eq:hi-1}) has no $\vpa$ moment (to any order) and the $\vpa$ moment of $h^{(0)}_i$ according to~(\ref{eq:hi_0}) is $u_{\parallel i}$, (\ref{eq:parallel_ampere2}) holds, and confirms the relative ordering of $u_{\parallel i}$ and $\apar$ in (\ref{eq:cold_ion_ordering_full}).

The relative size of $\delta \bpar$ is determined from the perpendicular Amp\`{e}re's law (\ref{perpendicular_ampere_e}):
\bea
\frac{\delta B_{\parallel}}{B} + \frac{\beta_e}{2}\frac{\delta n_e}{n_e} - \frac{\beta_e}{2}\frac{e\phi}{T_e} \nonumber\\ = - \frac{\beta_i}{2}\sum_{\mathbf{k}}e^{i\mathbf{k}\cdot\mathbf{r}} \frac{1}{n_i} \int \sd^3\mathbf{v}\,\frac{2v^2_{\perp}}{v^2_{\text{th}i}}\frac{J_1\left(k_{\perp}v_{\perp}/\Omega_i\right)}{k_{\perp}v_{\perp}/\Omega_i}h_{i\mathbf{k}}, \nonumber\\  \approx -\frac{\beta_i}{2}\int \sd^3\mathbf{v} \,\frac{v^2_{\perp}}{v^2_{\text{th}i}}h^{(-1)}_i\left(\mathbf{r}\right),
\eea
leading to 
\beq
\label{eq:pressure_balance}
\frac{\delta B_{\parallel}}{B}  = - \frac{\beta_e}{2}\left(\frac{\delta n_e}{n_e} + \tau\frac{\delta T_i}{T_i}\right),
\eeq
to lowest order in $\sqrt{\tau}$, which satisfies (\ref{eq:cold_ion_ordering_full}). This equation describes the balance between the magnetic and plasma pressures (the contribution from $\delta n_i$ is smaller than that from $\delta T_i$ by a factor of $\beta_i/\beta_e = \tau$.).

\subsubsection*{Moments of the ion gyrokinetic equation.}--
Let us now use (\ref{eq:ion_gk_eq}) to derive the evolution equations for the moments of $h_i$. Consider first the density moment,
\beq
\frac{1}{n_i}\int\sd^3\mathbf{v}\,\avg{(\mathrm{\ref{eq:ion_gk_eq})}}{\mathbf{r}}.
\eeq
All terms in (\ref{eq:ion_gk_eq}) that are odd in $\vpa$ will vanish upon velocity integration, so we need only consider the contributions even in $\vpa$:
\bea
\label{eq:ion_gk_eq_density}
\underbrace{\frac{\partial}{\partial t}\Bigg[h^{(-1)}_i + h^{(1)}_i\Bigg]}_{\circled{1}} + \underbrace{v_{\parallel}\hat{\mathbf{z}}\cdot\frac{\partial h^{(0)}_i}{\partial \mathbf{R}_i}}_{\circled{2}} \nonumber\\+ \underbrace{\frac{c}{B}\left\lbrace \left\langle \phi\right\rangle_{\mathbf{R}_i}, h^{(-1)}_i + h^{(1)}_i \right\rbrace}_{\circled{3}} + \underbrace{\frac{c}{B}\left\lbrace \left\langle  \frac{-v_{\parallel}A_{\parallel}}{c}\right\rangle_{\mathbf{R}_i}, h^{(0)}_i\right\rbrace}_{\circled{4}}\nonumber\\ + \underbrace{\frac{c}{B}\left\lbrace \left\langle \frac{-\mathbf{v_{\perp}}\cdot\mathbf{A}_{\perp}}{c}\right\rangle_{\mathbf{R}_i}, h^{(-1)}_i\right\rbrace}_{\circled{5}} + \underbrace{\mathbf{V}_M\cdot\frac{\partial h^{(-1)}_i}{\partial \mathbf{R}_i}}_{\circled{6}} \nonumber\\ + \underbrace{\left(\frac{v^{2}}{v^2_{\text{th}i}}-\frac{3}{2}\right)\frac{c}{B}\hat{\mathbf{z}}\times\frac{\partial}{\partial \mathbf{R}_i}\Bigg[\left\langle \phi\right\rangle_{\mathbf{R}_i} + \left\langle \frac{-\mathbf{v_{\perp}}\cdot\mathbf{A}_{\perp}}{c}\right\rangle_{\mathbf{R}_i}\Bigg]}_{\circled{7}} \nonumber\\ \cdot\frac{\partial \ln T_i}{\partial \mathbf{R}_i}F_i  = \underbrace{\frac{Ze}{T_i}\frac{\partial}{\partial t}\Bigg[\left\langle \phi\right\rangle_{\mathbf{R}_i}  +\left\langle \frac{-\mathbf{v}_{\perp}\cdot\mathbf{A}_{\perp}}{c}\right\rangle_{\mathbf{R}_i} \Bigg]F_i}_{\circled{8}} \nonumber\\+ \left\langle C\left[h^{(-1)}_i + h^{(1)}_i\right]\right\rangle_{\mathbf{R}_i} + (\mathrm{terms\;odd\;in\;}\vpa),
\eea
where we have kept the collisional term, which will contain non-zero FLR corrections to the order $\sqrt{\tau}\vthi h_i/L_B$. The terms are of mixed orders at this point, but cancellations will occur upon taking the density moment so that all remaining terms will end up of the same order. Let us calculate the velocity integral of each term, keeping in mind that we only need to retain contributions to lowest order, up to first order ($\sim \sqrt{\tau}\vthi h_i/L_B$).

\noindent\textbf{Term} $\circled{1}$ :
\bea
\label{eq:n_term1}
\frac{\partial}{\partial t} \frac{1}{n_i}\int \sd^3\mathbf{v} \,\avg{h^{(-1)}_i + h^{(1)}_i}{\mathrm{r}} \nonumber \\ = \frac{\partial}{\partial t}\left(\frac{Ze\phi}{T_i} + \frac{\delta n_e}{n_e}\right),
\eea
which follows from (\ref{eq:hi-1}), (\ref{eq:hi_1}) and (\ref{quasi_i}).

\noindent\textbf{Term} $\circled{2}$ : as this term is first order in (\ref{eq:ion_gk_eq}), we can safely ignore the ion FLR corrections: to lowest order,
\bea
\label{eq:n_term2}
\frac{\partial}{\partial z} \frac{1}{n_i}\int \sd^3 \mathbf{v}\, v_{\parallel} \left\langle h^{(0)}_i \right\rangle_{\mathbf{r}} & = \frac{\partial}{\partial z} \frac{1}{n_i}\int \sd^3 \mathbf{v}\, v_{\parallel} h^{(0)}_i\left(\mathbf{r}\right)  \nonumber\\ &= \frac{\partial u_{\parallel i}}{\partial z}.
\eea

\noindent\textbf{Term} $\circled{3}$ :
before calculating the velocity integral, let us write the Poisson bracket in Fourier space:
\bea
\label{eq:poisson_in_fourier}
\left\lbrace \left\langle \phi\right\rangle_{\mathbf{R}_i}, h_i \right\rbrace \nonumber \approx \sum_{\mathbf{k},\mathbf{k}^{\prime}} e^{i\mathbf{k}\cdot\mathbf{r} + i\mathbf{k^{\prime}}\cdot\mathbf{r}} \left[-\hat{\mathbf{z}}\cdot\left(\mathbf{k}\times\mathbf{k^{\prime}}\right)\right]\Bigg[1 - \mathbf{k}\mathbf{k^{\prime}}:\boldsymbol{\rho}\boldsymbol{\rho} \nonumber\\ - \frac{\left(\mathbf{k}\cdot\boldsymbol{\rho}\right)^2}{2} - \frac{\left(\mathbf{k^{\prime}}\cdot\boldsymbol{\rho}\right)^2}{2} -  \frac{1}{4}\frac{v^2_{\perp}}{v^2_{\text{th}i}} k^2_{\perp}\rho^2_i\Bigg] \phi_{\mathbf{k}}h_{i,\mathbf{k^{\prime}}},
\eea
where we have expanded $e^{i\mathbf{k}\cdot\boldsymbol{\rho}}$ and $J_0(\kperp \rho_i\vpe/\vthi)$ in $\kperp \rho_i \ll 1$ with $\boldsymbol{\rho} = \mathbf{r} - \mathbf{R}_i$. Since 
\bea
\label{eq:gyro_property}
\left\langle \boldsymbol{\rho}\right\rangle_{\mathbf{r}} = 0,\text{ }\text{ }\text{ }  \left\langle \boldsymbol{\rho}\boldsymbol{\rho}\right\rangle_{\mathbf{r}} = \frac{1}{2}\frac{v^2_{\perp}}{v^2_{\text{th}i}}\rho^2_i\textbf{I},
\eea
where $\textbf{I}$ is the identity matrix $\hat{\mathbf{x}}\hat{\mathbf{x}} + \hat{\mathbf{y}}\hat{\mathbf{y}}$, the gyro-average of (\ref{eq:poisson_in_fourier}) at fixed $\mathbf{r}$ is
\bea
\label{eq:113}
\left\langle \left\lbrace \left\langle \phi\right\rangle_{\mathbf{R}_i}, h_i \right\rbrace \right\rangle_{\mathbf{r}} \approx \sum_{\mathbf{k},\mathbf{k}^{\prime}} e^{i\mathbf{k}\cdot\mathbf{r} + i\mathbf{k^{\prime}}\cdot\mathbf{r}} \left[-\hat{\mathbf{z}}\cdot\left(\mathbf{k}\times\mathbf{k^{\prime}}\right)\right]\times \nonumber\\ \Bigg[1 - \mathbf{k}\mathbf{k^{\prime}}:\left(\frac{1}{2}\frac{v^2_{\perp}}{v^2_{\text{th}i}}\rho^2_i\textbf{I}\right) - \mathbf{k}\mathbf{k}:\left(\frac{1}{4}\frac{v^2_{\perp}}{v^2_{\text{th}i}}\rho^2_i\textbf{I}\right) \nonumber\\- \mathbf{k^{\prime}}\mathbf{k^{\prime}}:\left(\frac{1}{4}\frac{v^2_{\perp}}{v^2_{\text{th}i}}\rho^2_i\textbf{I}\right) - \frac{1}{4}\frac{v^2_{\perp}}{v^2_{\text{th}i}}k^2_{\perp}\rho^2_i\Bigg]\phi_{\mathbf{k}}h_{i,\mathbf{k^{\prime}}}\nonumber \\ \approx \left\lbrace \phi, h_i\left(\mathbf{r}\right) \right\rbrace + \frac{1}{2} \frac{v^2_{\perp}}{v^2_{\text{th}i}}\rho^2_i \nabla_{\perp} \cdot\left\lbrace \nabla_{\perp} \phi, h_i\left(\mathbf{r}\right) \right\rbrace \nonumber \\+ \frac{1}{4}\frac{v^2_{\perp}}{v^2_{\text{th}i}}\rho^2_i\left\lbrace \phi, \nabla^2_{\perp} h_i\left(\mathbf{r}\right) \right\rbrace \nonumber\\ \approx \left\lbrace \phi, h^{(-1)}_i\left(\mathbf{r}\right) + h^{(1)}_i\left(\mathbf{r}\right) \right\rbrace \nonumber\\+ \frac{1}{2} \frac{v^2_{\perp}}{v^2_{\text{th}i}}\rho^2_i \nabla_{\perp} \cdot\left\lbrace \nabla_{\perp} \phi, h^{(-1)}_i\left(\mathbf{r}\right) \right\rbrace \nonumber\\ + \frac{1}{4}\frac{v^2_{\perp}}{v^2_{\text{th}i}}\rho^2_i\left\lbrace \phi, \nabla^2_{\perp} h^{(-1)}_i\left(\mathbf{r}\right) \right\rbrace.
\eea
Upon velocity integration, the first and last term of (\ref{eq:113}) can be combined to give $\left\lbrace \phi, \delta n_e/n_e\right\rbrace$ according to~(\ref{quasi_i}). The density moment of the $\phi$-nonlinearity is thus
\bea
\label{eq:n_term3}
\frac{1}{n_i}\int \sd^3\mathbf{v}\,\left\langle \frac{c}{B}\left\lbrace \left\langle \phi\right\rangle_{\mathbf{R}_i}, h^{(-1)}_i + h^{(1)}_i \right\rbrace \right\rangle_{\mathbf{r}}\nonumber \\ \approx \frac{c}{B}\left\lbrace \phi, \frac{\delta n_e}{n_e} \right\rbrace + \frac{c}{B}\frac{\rho^2_i}{2}\nabla_{\perp}\cdot\left\lbrace \nabla_{\perp}\phi, \frac{\delta T_i}{T_i}\right\rbrace \nonumber\\+ \frac{c}{B}\frac{\rho^2_i}{2}\left\lbrace \nabla^2_{\perp} \phi, \frac{Ze\phi}{T_i} \right\rbrace.
\eea

\noindent\textbf{Term} $\circled{4}$ : to lowest order,
\bea
\label{eq:n_term4}
\frac{1}{n_i}\int \sd^3\mathbf{v}\,\left\langle \frac{c}{B}\left\lbrace \left\langle  \frac{-v_{\parallel}A_{\parallel}}{c}\right\rangle_{\mathbf{R}_i}, h^{(0)}_i\right\rbrace \right\rangle_{\mathbf{r}}\nonumber\\ \approx - \frac{1}{B}\left\lbrace A_{\parallel}, \frac{1}{n_i}\int \sd^3\mathbf{v}\, v_{\parallel} h^{(0)}_i\left(\mathbf{r}\right) \right\rbrace \nonumber\\ = - \frac{1}{B} \left\lbrace A_{\parallel}, u_{\parallel i}\right\rbrace,
\eea
where we have used $\avg{\apar}{\mathbf{R}_i} = \apar$ for $\kperp \rho_i \ll 1$.

\noindent\textbf{Term} $\circled{5}$ : to lowest order,
\bea
\label{eq:n_term5}
&\frac{1}{n_i}\int \sd^3\mathbf{v}\,\left\langle \frac{c}{B}\left\lbrace \left\langle  \frac{-\mathbf{v}_{\perp}\cdot\mathbf{A}_{\perp}}{c}\right\rangle_{\mathbf{R}_i}, h^{(-1)}_i\right\rbrace \right\rangle_{\mathbf{r}} \nonumber\\ & \approx \frac{c}{B}\frac{T_i}{Ze}\left\lbrace \frac{\delta B_{\parallel}}{B}, \frac{1}{n_i}\int \sd^3\mathbf{v}\,\frac{v^2_{\perp}}{v^2_{\text{th}i}}h^{(-1)}_i\left(\mathbf{r}\right) \right\rbrace \\ & =  \frac{c}{B}\frac{T_i}{Ze}\left\lbrace \frac{\delta B_{\parallel}}{B}, \frac{Ze\phi}{T_i}+\frac{\delta T_i}{T_i}\right\rbrace,
\eea
where we have used (\ref{eq:hi-1}) and 
\begin{equation}
\label{eq:lowest_order_vaperp}
\avg{\mathbf{\vpe} \cdot \mathbf{A}_{\perp}}{\mathbf{R}_i} \approx \frac{\delta \bpar}{B}\frac{\vpe^2}{\vthi^2}\frac{T_i}{Ze}
\end{equation}
for $\kperp \rho_i \ll 1$.

\noindent\textbf{Term} $\circled{6}$ : to lowest order,
\bea
\label{eq:n_term6}
\frac{1}{n_i}\int \sd^3\mathbf{v}\,\left\langle \mathbf{V}_M\cdot\frac{\partial h^{(-1)}_i}{\partial \mathbf{R}_i}  \right\rangle_{\mathbf{r}} \nonumber\\ \approx \frac{1}{n_i}\int \sd^3\mathbf{v}\,\Bigg[ \frac{v^2_{\perp}}{2\Omega_i}\hat{\mathbf{z}}\times \nabla \ln B \nonumber \\ + \frac{v^2_{\parallel}}{\Omega_i}\hat{\mathbf{z}}\times\left(\hat{\mathbf{z}}\cdot\nabla \hat{\mathbf{z}}\right)\Bigg] \cdot \nabla h^{(-1)}_i\left(\mathbf{r}\right), \nonumber\\  = - \frac{v^2_{\text{th}i}}{2\Omega_i} \left(\frac{1}{R} + \frac{1}{L_B}\right)\frac{\partial}{\partial y}\left(\frac{Ze\phi}{T_i} + \frac{\delta T_i}{T_i}\right),
\eea
where we have used (\ref{eq:hi-1}), $\hat{\mathbf{x}}\times \hat{\mathbf{y}} = \hat{\mathbf{z}}$ and the expression for the magnetic curvature $\hat{\mathbf{z}}\cdot\nabla \hat{\mathbf{z}} = -\hat{\mathbf{x}}/R$.

\noindent\textbf{Term} $\circled{7}$ :
the lowest-order expressions for $\avg{\phi}{\mathbf{R}_i}$ and $\avg{\mathbf{\vpe} \cdot \mathbf{A}_{\perp}}{\mathbf{R}_i}$ are $\phi$ and (\ref{eq:lowest_order_vaperp}), respectively; however, we have to retain the next-order correction for $\avg{\phi}{\mathbf{R}_i}$, because its lowest-order contribution vanishes upon velocity integration. To obtain this correction, we perform the double gyro-average:
\bea
\label{eq:double_gyro}
\avg{\frac{\partial \avg{\phi}{\mathbf{R}_i}}{\partial \mathbf{R}_i}}{\mathbf{r}} = \left\langle \sum_{\mathbf{k}} i\mathbf{k} e^{i\mathbf{k}\cdot\mathbf{R}_i} J_0\left(k_{\perp}\rho_i\frac{v_{\perp}}{v_{\text{th}i}}\right) \phi_{\mathbf{k}}   \right\rangle_{\mathbf{r}} \nonumber \\ \approx  \sum_{\mathbf{k}} i\mathbf{k} \left\langle e^{i\mathbf{k}\cdot\mathbf{r}}\left(1 - i\mathbf{k}\cdot\boldsymbol{\rho} - \frac{1}{2}\mathbf{k}\mathbf{k}:\boldsymbol{\rho}\boldsymbol{\rho} \right) \right\rangle_{\mathbf{r}} \nonumber\\  \cdot \left( 1- \frac{\kperp^2\rho^2_i}{4}\frac{\vpe^2}{\vthi^2} \right) \phi_{\mathbf{k}}\nonumber\\ \approx  \nabla\phi + \frac{\rho^2_i}{2}\frac{v^2_{\perp}}{v^2_{\text{th}i}} \nabla \nabla^2_{\perp}\phi,
\eea
where we have expanded $e^{i\mathbf{k}\cdot\boldsymbol{\rho}}$ and $J_0$ in $\kperp \rho_i \ll 1$ and used (\ref{eq:gyro_property}). The density moment of \circled{7} is thus
\bea
\label{eq:n_term7}
\frac{1}{n_i}\int \sd^3\mathbf{v}\, \left\langle\circled{7} \right\rangle_{\mathbf{r}} \nonumber\\\approx \frac{cT_i}{BZe}\frac{1}{L_T}\frac{\partial}{\partial y}\left( \frac{\rho^2_i}{2}\nabla^2_{\perp}\frac{Ze\phi}{T_i} + \frac{\delta B_{\parallel}}{B} \right).
\eea

\noindent\textbf{Term} $\circled{8}$ :
the treatment of this term is analogous to that of term \circled{7}, viz.,
\bea
\label{eq:n_term8}
\frac{1}{n_i}\int \sd^3\mathbf{v}\, \left\langle\circled{8} \right\rangle_{\mathbf{r}} \nonumber \\  \approx \frac{1}{n_i}\int \sd^3\mathbf{v}\, \frac{\partial}{\partial t} \left[\frac{Ze\phi}{T_i} + \frac{\rho^2_i}{2}\frac{v^2_{\perp}}{v^2_{\text{th}i}}\nabla^2_{\perp}\frac{Ze\phi}{T_i} + \frac{v^2_{\perp}}{v^2_{\text{th}i}}\frac{\delta B_{\parallel}}{B} \right]F_i \nonumber\\  =  \frac{\partial}{\partial t}\left(\frac{Ze\phi}{T_i} +\frac{\rho^2_i}{2}\nabla^2_{\perp} \frac{Ze\phi}{T_i} + \frac{\delta B_{\parallel}}{B}\right),
\eea
where the double gyro-average has been evaluated as in (\ref{eq:double_gyro}). 

Collecting all eight of the terms (\ref{eq:n_term1}), (\ref{eq:n_term2}), (\ref{eq:n_term3}), (\ref{eq:n_term4}), (\ref{eq:n_term5}), (\ref{eq:n_term6}), (\ref{eq:n_term7}), and (\ref{eq:n_term8}), we obtain the density moment of (\ref{eq:ion_gk_eq}) to first order in $\sqrt{\tau}\vthi h_i/L_B$:
\bea
\label{eq:density_moment_ion_gk}
\frac{\sd}{\sd t}\left(\frac{\delta n_e}{n_e} - \frac{\delta B_{\parallel}}{B}\right) + \nabla_{\parallel}u_{\parallel i} - \frac{\sd}{\sd t}\frac{1}{2}\rho^2_i\nabla^2_{\perp}\frac{Ze\phi}{T_i}\nonumber \\ + \frac{1}{4}\rho_iv_{\text{th}i}\rho^2_i \nabla_{\perp}\cdot\left\lbrace \nabla_{\perp}\frac{Ze\phi}{T_i}, \frac{\delta T_i}{T_i}\right\rbrace \nonumber\\ + \frac{1}{2}\rho_i v_{\text{th}i} \left\lbrace \frac{\delta B_{\parallel}}{B}, \frac{\delta T_i}{T_i}\right\rbrace \nonumber\\ + \frac{\rho_i v_{\text{th}i}}{2L_T}\frac{\partial}{\partial y}\left(\frac{\rho^2_i}{2}\nabla^2_{\perp}\frac{Ze\phi}{T_i} + \frac{\delta B_{\parallel}}{B}\right)\nonumber\\ - \frac{1}{2}\rho_i v_{\text{th}i}\left(\frac{1}{L_B} + \frac{1}{R}\right)\frac{\partial}{\partial y}\left(\frac{Ze\phi}{T_i} + \frac{\delta T_i}{T_i}\right) \nonumber\\ = - \frac{1}{2}\chi\rho^2_i\nabla^4_{\perp}\left(a\frac{Ze\phi}{T_i} - b\frac{\delta T_i}{T_i}\right),
\eea
where $\chi$, $a$, and $b$ are as defined in Section~\ref{sec:system_of_equations} [see (\ref{eq:definition_of_chi})]. These collisional contributions were calculated from the Landau collision operator~\cite{ ivanov_2020,Newton_2010}. Note that we have reused the symbol $\chi$, originally introduced for the generalised potential~(\ref{eq:chi_definition}), to denote the diffusion coefficient. 

Taking the difference between (\ref{eq:density_moment_ion_gk}) and the density moment (\ref{n_e}) of the electron gyrokinetic equation and using (\ref{eq:parallel_ampere2}), we obtain the evolution equation for the vorticity of the $\mathbf{E}\times\mathbf{B}$ flow:
\bea
\label{eq:vorticity}
 - \frac{\sd}{\sd t}\frac{1}{2}\rho^2_i\nabla^2_{\perp}\frac{Ze\phi}{T_i}- \nabla_{\parallel}\frac{c}{4\pi e n_e}\nabla^2_{\perp}A_{\parallel} \nonumber\\+ \frac{1}{4}\rho_iv_{\text{th}i}\rho^2_i \nabla_{\perp}\cdot\left\lbrace \nabla_{\perp}\frac{Ze\phi}{T_i}, \frac{\delta T_i}{T_i}\right\rbrace \nonumber\\+ \frac{\rho_i v_{\text{th}i}}{2L_T}\frac{\partial}{\partial y}\left(\frac{\rho^2_i}{2}\nabla^2_{\perp}\frac{Ze\phi}{T_i} + \frac{\delta B_{\parallel}}{B}\right) \nonumber\\ - \frac{1}{2}\rho_i v_{\text{th}i}\left(\frac{1}{L_B} + \frac{1}{R}\right)\frac{\partial}{\partial y}\left(\frac{1}{\tau}\frac{\delta n_e}{n_e} + \frac{\delta T_i}{T_i}\right)\nonumber\\ = - \frac{1}{2}\chi\rho^2_i\nabla^4_{\perp}\left(a\frac{Ze\phi}{T_i} - b\frac{\delta T_i}{T_i}\right).
\eea

We next take the $v_{\parallel}$ moment of (\ref{eq:ion_gk_eq}),
\beq
\frac{1}{n_i}\int \sd^3\mathbf{v}\,\frac{v_{\parallel}}{v_{\text{th}i}}\left\langle (\mathrm{\ref{eq:ion_gk_eq}})\right\rangle_{\mathbf{r}}.
\eeq
All terms that are even in $\vpa$ will vanish upon velocity integration, so we only need to consider the contributions odd in $\vpa$:
\bea
\frac{\partial h^{(0)}_i}{\partial t} + v_{\parallel}\hat{\mathbf{z}}\cdot\frac{\partial h^{(-1)}_i}{\partial \mathbf{R}_i}\nonumber\\ + \frac{c}{B}\left\lbrace \left\langle \phi\right\rangle_{\mathbf{R}_i}, h^{(0)}_i \right\rbrace + \frac{c}{B}\left\lbrace \left\langle  \frac{-v_{\parallel}A_{\parallel}}{c}\right\rangle_{\mathbf{R}_i}, h^{(-1)}_i\right\rbrace \nonumber\\ + \left(\frac{v^{2}}{v^2_{\text{th}i}}-\frac{3}{2}\right)\frac{c}{B}\hat{\mathbf{z}}\times\frac{\partial}{\partial \mathbf{R}_i}\left\langle \frac{-v_{\parallel}A_{\parallel}}{c}\right\rangle_{\mathbf{R}_i}\cdot\frac{\partial \ln T_i}{\partial \mathbf{R}_i}F_i \nonumber \\= \frac{Ze}{T_i}\frac{\partial}{\partial t}\left\langle \frac{-v_{\parallel}A_{\parallel}}{c}\right\rangle_{\mathbf{R}_i}F_i  + \left\langle C\left[h^{(0)}_i\right]\right\rangle_{\mathbf{R}_i}\nonumber \\ +\; (\mathrm{terms\;even\;in\;}\vpa).
\eea
All the terms here are of the same order except for the collisional term. As a result, except in the latter, we can ignore all FLR corrections and evaluate everything at $\mathbf{r}$. The $v_{\parallel}$ integral then is
\bea
\label{eq:ion_momentum}
\frac{\sd}{\sd t}\frac{u_{\parallel i}}{v_{\text{th}i}} + \frac{v_{\text{th}i}}{2}\nabla_{\parallel}\left(\frac{1}{\tau}\frac{\delta n_e}{n_e} + \frac{\delta T_i}{T_i}\right)& \nonumber\\- \frac{v_{\text{th}i}}{2B}\frac{1}{L_T}\frac{\partial A_{\parallel}}{\partial y} = \frac{9}{10} \chi \nabla^2_{\perp}\frac{u_{\parallel i}}{v_{\text{th}i}},
\eea
where we have used (\ref{eq:hi-1}) and (\ref{eq:hi_0}). Using the fact that ion-ion collisions conserve momentum of the ions, the collisional term was obtained by expanding the gyro-averaged collisional operator up to the lowest nontrivial order.

Finally, we obtain the evolution equation for $\delta T_i$ by taking the energy moment of (\ref{eq:ion_gk_eq}),
\beq
\frac{1}{n_i}\int \sd^3 \mathbf{v}\,\frac{v^2}{v^2_{\text{th}i}}\left\langle (\mathrm{\ref{eq:ion_gk_eq}})
\right\rangle_{\mathbf{r}}.
\eeq
We only need the terms involving $h^{(-1)}_i$ because all other terms are smaller by at least a factor of $\sqrt{\tau}$:
\bea
\label{eq:T_moment_gk}
\frac{\partial h^{(-1)}_i}{\partial t} + \frac{c}{B}\left\lbrace \left\langle \phi\right\rangle_{\mathbf{R}_i}, h^{(-1)}_i\right\rbrace \nonumber\\+ \left(\frac{v^{2}}{v^2_{\text{th}i}}-\frac{3}{2}\right)\frac{c}{B}\hat{\mathbf{z}}\times\frac{\partial \left\langle \phi\right\rangle_{\mathbf{R}_i}}{\partial \mathbf{R}_i} \cdot\frac{\partial \ln T_i}{\partial \mathbf{R}_i}F_i \nonumber\\ = \frac{Ze}{T_i}\frac{\partial \left\langle \phi\right\rangle_{\mathbf{R}_i}}{\partial t}F_i + \left\langle C\left[h^{(-1)}_i \right]\right\rangle_{\mathbf{R}_i}.
\eea
All terms in (\ref{eq:T_moment_gk}) except the collisional term are again of the same size, so that the FLR corrections can be neglected. The result of taking the temperature moment is then
\bea
\label{eq:T_moment}
\frac{\sd}{\sd t}\frac{\delta T_i}{T_i} + \frac{\rho_i v_{\text{th}i}}{2L_T}\frac{\partial}{\partial y}\frac{Ze\phi}{T_i} = \chi \nabla^2_{\perp}\frac{\delta T_i}{T_i},
\eea
where we have used the energy-conservation property of the ion-ion collisions and expanded the gyro-averaged collision operator to lowest nontrivial order in $\sqrt{\tau}$, before taking the energy moment of the term that is next order in $\sqrt{\tau}$.

\subsubsection*{Summary of the fluid equations at $\beta_e \sim 1$.}--
The fluid model that we have derived for the electromagnetic ITG turbulence in a $Z$-pinch consists of the following set of equations:
the parallel electron force balance~(\ref{eq:A_e}):
\begin{equation}
\label{eq:A_full}
\frac{e}{T_e c} \frac{\partial A_{\parallel}}{\partial t} = \nabla_{\parallel} \left(\frac{\delta n_e}{n_e} - \frac{e\phi}{T_e}\right);
\end{equation}
the electron continuity equation (\ref{n_e}):
\bea
\label{eq:n_full}
\frac{\sd}{\sd t}\left(\frac{\delta n_e}{n_e} - \frac{\delta B_{\parallel}}{B}\right) + \nabla_{\parallel} \left(u_{\parallel i} + \frac{c}{4\pi e n_e}\nabla^2_{\perp}\apar\right) \nonumber \\ + \frac{cT_e}{B e} \left\lbrace \frac{\delta n_e}{n_e}, \frac{\delta B_{\parallel}}{B}\right\rbrace \nonumber \\+ \frac{\rho_ev_{\text{th}e}}{2}\left(\frac{1}{L_B} + \frac{1}{R}\right)\frac{\partial}{\partial y}\left(\frac{\delta n_e}{n_e} - \frac{e\phi}{T_e}\right) = 0,
\eea
where we have used (\ref{eq:parallel_ampere2}) to express $u_{\parallel e}$ in terms of $u_{\parallel i}$ and $\apar$; the $\mathbf{E}\times\mathbf{B}$ vorticity equation~(\ref{eq:vorticity}):
\bea
\label{eq:phi_full}
 - \frac{\sd}{\sd t}\frac{1}{2}\rho^2_i\nabla^2_{\perp}\frac{Ze\phi}{T_i}- \nabla_{\parallel}\frac{c}{4\pi e n_e}\nabla^2_{\perp}A_{\parallel}\nonumber\\  + \frac{1}{4}\rho_iv_{\text{th}i}\rho^2_i\nabla_{\perp}\cdot\left\lbrace \nabla_{\perp}\frac{Ze\phi}{T_i},\frac{\delta T_i}{T_i}\right\rbrace \nonumber\\ + \frac{\rho_i v_{\text{th}i}}{2L_T}\frac{\partial}{\partial y}\left(\frac{\rho^2_i}{2}\nabla^2_{\perp}\frac{Ze\phi}{T_i} + \frac{\delta B_{\parallel}}{B}\right) \nonumber\\ - \frac{1}{2}\rho_i v_{\text{th}i}\left(\frac{1}{L_B} + \frac{1}{R}\right)\frac{\partial}{\partial y}\left(\frac{1}{\tau}\frac{\delta n_e}{n_e} + \frac{\delta T_i}{T_i}\right)\nonumber\\ = - \frac{1}{2}\chi\rho^2_i\nabla^4_{\perp}\left(a\frac{Ze\phi}{T_i} - b\frac{\delta T_i}{T_i}\right);
\eea
the parallel ion-momentum equation (\ref{eq:ion_momentum}):
\bea
\label{eq:u_full}
\frac{\sd u_{\parallel i}}{\sd t} &+ \frac{v^2_{\text{th}i}}{2}\nabla_{\parallel}\left(\frac{1}{\tau}\frac{\delta n_e}{n_e} + \frac{\delta T_i}{T_i}\right) - \frac{v^2_{\text{th}i}}{2B}\frac{1}{L_T}\frac{\partial A_{\parallel}}{\partial y} \nonumber\\&= \frac{9}{10}\chi \nabla^2_{\perp}u_{\parallel i};
\eea
the ion-temperature equation (\ref{eq:T_moment}):
\bea
\label{eq:T_full}
\frac{\sd}{\sd t}\frac{\delta T_i}{T_i} + \frac{\rho_i v_{\text{th}i}}{2L_T}\frac{\partial}{\partial y}\frac{Ze\phi}{T_i} = \chi \nabla^2_{\perp}\frac{\delta T_i}{T_i};
\eea
and the pressure balance (\ref{eq:pressure_balance}):
\begin{equation}
\label{eq:B_full}
\frac{\delta B_{\parallel}}{B} = - \frac{\beta_e}{2}\left(\frac{\delta n_e}{n_e} + \tau\frac{\delta T_i}{T_i}\right).
\end{equation}
In the absence of equilibrium gradients, i.e., with $\delta T_i/T_i = 0$, these equations turn into the Hall reduced-MHD system considered in \cite{Schekochihin_2019}.

\subsection{Non-dimensionalised equations and low-$\beta_e$ limit}
\label{low beta limit}
To prepare our equations for numerical solution, we now non-dimensionalise all variables, parameters, and fields as follows:
\begin{equation*}
\hat{t} = \frac{2c_s}{L_B}t,\;\hat{x} = \frac{x}{\rho_s}, \;\hat{y} = \frac{y}{\rho_s},\; \hat{z} = \sqrt{2\beta_e}\frac{z}{L_B},
\end{equation*}
\begin{equation*}
\kappa_T = \frac{\tau}{2}\frac{L_B}{L_T},\;\kappa_c = \frac{L_B}{R},\;\hat{\chi} = \frac{\chi}{2\rho_s^2}\dfrac{L_B}{c_s},
\end{equation*}
\begin{equation*}
\varphi = \frac{e\phi}{T_e}\frac{L_B}{2\rho_s},\; T = \tau\frac{\delta T_i}{T_i}\frac{L_B}{2\rho_s},\;n = \frac{\delta n_e}{n_e}\frac{L_B}{2\rho_s},
\end{equation*}
\bea
\label{eq:normalization_unitybeta}
A = \frac{A_{\parallel}}{\rho_s B}\frac{L_B}{\sqrt{2\beta_e}\rho_s},\;B_{\parallel} = \frac{\delta B_{\parallel}}{B}\frac{L_B}{2\rho_s},\nonumber \\ u = \frac{u_{\parallel i}}{c_s}\frac{L_B}{\sqrt{2}\rho_s}.
\eea
These normalisations have been designed in such a way as to make all normalised quantities order unity. How to do this follows straightforwardly from the ordering assumed in the course of the derivation of (\ref{eq:A_full})-(\ref{eq:B_full}), the only additional non-trivial choices involving the factor of $\sqrt{\beta_e}$. These are motivated by our intention to make a further simplification of our model: filter out sound-wave physics. This is achieved by assuming the sound waves to be slow compared to Alfv\'en waves:
\begin{equation}
\frac{c_s}{v_A} \sim \sqrt{\beta_e} \rightarrow 0.    
\end{equation}
We shall see momentarily that this will eliminate from consideration the perturbation of the magnetic-field strength, $\delta \bpar$ --- we feel safe in doing so as $\delta \bpar$ does not appear to be a necessary ingredient for the high-beta runaway transitions in ITG turbulence~\cite{nonzonal,hysteresis} and indeed also proves not to be in our study. What is necessary is to capture the interplay between the ITG and Alfv\'en-wave physics, which requies $\omega_{*T} \sim \kpar v_A$, where, since also $\omega_{*T} \sim c_s/L_B$ [see (\ref{eq:omega_star})],
\begin{equation}
\kpar L_B \sim \sqrt{\beta_e}.    
\end{equation}
This motivates the $\sqrt{\beta_e}$ factor in the normalisation of the field-parallel coordinate $z$ in (\ref{eq:normalization_unitybeta}). This also leads to the appearance of $\sqrt{\beta_e}$ in the rescaling of the magnetic-fluctuation amplitudes: retaining Alfv\'enic response requires $\delta B_{\perp}/B \sim \uperp/v_A$ and nonlinearity $\kperp \uperp \sim \kpar v_A$, so
\begin{equation}
\frac{\apar}{\rho_s B} \sim \frac{1}{\kperp\rho_s}\frac{\delta B_{\perp}}{B} \sim \frac{\uperp}{v_A} \sim \frac{\kpar}{\kperp} \sim \frac{\rho_s\sqrt{\beta_e}}{L_B}.   
\end{equation}
Hence the normalisation of $\apar$ in (\ref{eq:normalization_unitybeta}). Note that $\sqrt{\beta_e}$ does not appear in the normalisation of $\phi$ or any other amplitudes because, using $\uperp \sim \kpar v_A/\kperp \sim c_s /\kperp L_B$,
\begin{equation}
\frac{e\phi}{T_e} \sim \frac{eB}{cT_e}\frac{\uperp}{\kperp} \sim \frac{1}{\rho_s c_s}\frac{c_s}{\kperp^2L_B} \sim \frac{\rho_s}{L_B}.
\end{equation}

Thus, with the normalisations (\ref{eq:normalization_unitybeta}), rigged to explicit the $\beta_e$ dependence, our equations (\ref{eq:A_full})-(\ref{eq:B_full}) become
\bea
\label{eq:A}
\frac{\partial A}{\partial t} = \nabla_{\parallel} \left(n - \varphi\right),
\eea
\bea
\label{eq:n}
\frac{\sd}{\sd t}\left(n - B_{\parallel}\right) + \nabla_{\parallel}\left(\frac{\sqrt{\beta_e}}{2} u + \nabla^2_{\perp} A\right) \nonumber\\+ \left\lbrace n,B_{\parallel}\right\rbrace + \frac{1 + \kappa_c}{2}\frac{\partial}{\partial y}\left(n - \varphi\right) = 0,
\eea
\bea
\label{eq:phi}
-\frac{\sd}{\sd t} \nabla^2_{\perp}\varphi - \nabla_{\parallel}\nabla^2_{\perp}A + \nabla_{\perp}\cdot\left\lbrace \nabla_{\perp}\varphi,T\right\rbrace \nonumber \\ + \kappa_T\partial_y\left(\nabla^2_{\perp}\varphi + B_{\parallel}\right) - \frac{1 + \kappa_c}{2}\frac{\partial}{\partial y}\left(n+T\right) \nonumber\\ = - \chi \nabla^4_{\perp}\left(a\varphi - bT\right),
\eea
\bea
\label{eq:u}
\frac{\sd u}{\sd t} + \sqrt{\beta_e}\nabla_{\parallel}\left(n + T\right)  - \sqrt{\beta_e}\kappa_T\partial_y A \nonumber \\ = \frac{9}{10}\chi\nabla^2_{\perp} u,
\eea
\beq
\label{eq:T}
\frac{ \sd T}{\sd t} + \kappa_T\frac{\partial \varphi}{\partial y} = \chi \nabla^2_{\perp}T,
\eeq
\beq
\label{eq:B}
B_{\parallel} = -\frac{\beta_e}{2}\left(n + T\right),
\eeq
where the hats on the normalized quantities have been dropped for simplicity of notation and $\sd/\sd t = \partial_t + \left\lbrace \varphi , \cdots\right\rbrace$, $\nabla_{\parallel} = \partial_z - \left\lbrace A , \cdots\right\rbrace$.

By examining the explicit $\beta_e$ dependences in (\ref{eq:A})-(\ref{eq:B}), we see that several simplifications occur for $\beta_e \rightarrow 0$. First, as anticipated, $\bpar$ fluctuations can be neglected in (\ref{eq:n}) and (\ref{eq:phi}) because of (\ref{eq:B}). Secondly, the parallel flow velocity $u$ decouples from the rest of the fields, so can be ignored. Physically, this is a statement that the sound waves are too slow to influence the dynamics. Finally, the equilibrium constraint~(\ref{eq:equilibrium constraint}) at $\beta_e \rightarrow 0$ forces $L_B = R$ (the magnetic tension force balances with the magnetic pressure force). Thus, $\kappa_c = 1$. The resulting four-field equations are (\ref{eq:A_red})-(\ref{eq:T_red}).

\section{The effect of finite dissipation on linear stability}
\label{more linear physics}
In this Appendix, we provide a more detailed derivation than the sketch in Section~\ref{sec:lin_dd} of the DD instabilities and other unstable collisional modes present in our system, as well as of their effect on the stability boundaries.

Let us consider modes with $k_x = 0$ (which are always faster growing than those with $k_x\neq 0$) and assume $\kappa_T$, $\chi$, $k_y$, $k_z \sim 1$. The resulting dispersion relation, obtained from (\ref{eq:n_lin})-(\ref{eq:T_lin}), is  
\bea
\label{eq:disper_colli}
\omega^4 + \left[(\kappa_T - 1)k_y + i(1+a)\chi k^2_y\right]\omega^3 - \lbrace 1 - \kappa_T (1 - k^2_y) \nonumber \\ + k^2_z(1 + k^2_y) + i\chi [1 + a + (b - 1)\kappa_T]k^3_y + a\chi^2k^4_y \rbrace\omega^2 \nonumber \\ - \lbrace [\kappa_T + k^2_z(1+\kappa_T k^2_y)]k_y + i\chi [1 + (1-b)\kappa_T k^2_y \nonumber \\ + k^2_z(1 + (1+a)k^2_y)]k^2_y - a\chi^2k^5_y \rbrace\omega - \kappa_T k^2_z k^2_y \nonumber \\ - i\chi k^2_z k^3_y[1+(1-b)\kappa_T k^2_y] + a \chi^2 k^2_z k^6_y = 0.
\eea
This is a finite-$k_y$, finite-$\chi$ generalisation of the dispersion relation (\ref{eq:kz1_disp}). 

\subsection{Interchange and double-diffusive instabilities}
\label{sec:Interchange and double-diffusive instabilities}
First, setting $k_y = 0$, we recover the IC dispersion relation already derived in Section~\ref{sec:lin_mhd}, but this time including the correction due to finite $k_z$, which describes the stabilising effect of magnetic tension:
\begin{equation}
\omega^2 = 1 - \kappa_T + k^2_z.    
\end{equation}
At large enough $k_z$, the IC modes turn into Alfv\'en waves (Section~\ref{sec:lin_alfven}). At a given $k_z$, the IC stability condition is 
\begin{equation}
\label{eq:ic_stability_condition}
 \kappa_T < 1 + k^2_z.   
\end{equation}

Let us now consider this condition satisfied and, motivated by the simple derivation in Section~\ref{sec:lin_dd}, seek solution of (\ref{eq:disper_colli}) in the form $\omega = \omega_r + i\gamma$, where $\gamma \sim k^2_y \ll \omega_r \sim 1$. To lowest order in $k_y$, the real part of (\ref{eq:disper_colli}) gives us 
\begin{equation}
\label{eq:dd_frequency_low_ky}
\omega_r = \pm \sqrt{1 - \kappa_T + k^2_z},    
\end{equation}
which is indeed a real frequency in view of (\ref{eq:ic_stability_condition}). From the imaginary part of (\ref{eq:disper_colli}), again to lowest order in $k_y$, we get
\begin{equation}
\gamma = \frac{\kappa_T - a\omega^2_r}{2\omega^2_r}\chi k^2_y,    
\end{equation}
which is the same growth rate as in (\ref{eq:dd_omega_a=0}), but with a stabilising correction due to $a\neq 0$ (finite viscosity). Thus, these DD modes are unstable (and unswamped by the ideal IC mode) when
\begin{equation}
\label{eq:dd_stability_boundary}
1 > \frac{\kappa_T}{1 + k^2_z} > \frac{a}{1+ a} \approx 0.184.    
\end{equation}

\subsection{ITG instability and its collisional extension}
\label{sec:ITG instability and its collisional extension}
We have seen that collisional terms can destabilise IC-stable plasma by eroding the effect of the restoring force. We are about to see that they have an analogous effect on the ITG-stable plasmas. 

The dispersion relation (\ref{eq:disper_colli}) with $k_y \ll 1$ and $\chi = 0$ but finite $k_z$ turns into the familiar case (\ref{eq:kz1_disp}), for which the stability condition is (\ref{eq:stab_itg_es_kz_kappaT}). Assuming that this is satisfied and seeking solutions of (\ref{eq:disper_colli}) in the form $\omega = \omega_r + i\gamma$, where $\gamma \sim k^2_y \ll \omega_r \sim k_y \ll 1$ (as this is what the real frequency must turn out to be for a stable ITG mode), we find, to lowest order in $k_y$, 
\begin{equation}
\label{eq:omega_r_finitechi_itg}
\omega_r = - k_y \frac{k^2_z + \kappa_T\pm \sqrt{\Lambda_0}}{2(1 + k^2_z - \kappa_T)},
\end{equation}
where
\begin{eqnarray}
\Lambda_0 &= (k^2_z + \kappa_T)^2 - 4(1+k^2_z - \kappa_T)\kappa_T k^2_z \nonumber \\ &= [(4\kappa_T -1)k^2_z + \kappa_T](\kappa_T - k^2_z).
\end{eqnarray}
The frequency is indeed real because, by (\ref{eq:stab_itg_es_kz_kappaT}), $\Lambda_0 \geq 0$. In the same approximation, the growth (or damping) rate is 
\begin{equation}
\label{eq:gamma_r_finitechi_itg}
\gamma = - \frac{k^2_z + (1 + k^2_z)\omega_r/k_y}{k^2_z + \kappa_T + 2(1 + k^2_z - \kappa_T)\omega_r/k_y} \chi k^2_y.
\end{equation}
Putting (\ref{eq:omega_r_finitechi_itg}) into (\ref{eq:gamma_r_finitechi_itg}), we find the condition under which the `$\pm$' modes are unstable:
\begin{equation}
\Lambda_{\pm} \equiv \pm \left[k^2_z - \frac{(1+k^2_z)\left(k^2_z + \kappa_T\pm\sqrt{\Lambda_0}\right)}{2(1 + k^2_z - \kappa_T)} \right] > 0.
\end{equation}
Note that here, like in \ref{sec:Interchange and double-diffusive instabilities}, we first consider the part of the $(k_z,\,\kappa_T)$ space where these collisional modes are not superceded by ideal IC or ITG instabilities, i.e., where (\ref{eq:ic_stability_condition}) and (\ref{eq:stab_itg_es_kz_kappaT}) ($\Lambda_0 \geq 0$) are satisfied. The combination of these two conditions gives us two disconnected regions in the $(k_z,\,\kappa_T)$ space:
\begin{equation}
\label{eq:kT_region1}
0 < \kappa_T < \frac{k^2_z}{4k^2_z + 1},   
\end{equation}
\begin{equation}
\label{eq:kT_region2}
k^2_z < \kappa_T < 1 + k^2_z.
\end{equation}
Let us observe that, within these regions,
\begin{equation}
\label{eq:product_of_lambdas}
\Lambda_{+}\Lambda_{-} = - \frac{\kappa_T k^6_z}{1 + k^2_z - \kappa_T} < 0,     
\end{equation}
so in each of these, only one of the modes is unstable [$\Lambda_{+}$ and $\Lambda_{-}$ are continuous and, in view of (\ref{eq:product_of_lambdas}), cannot vanish within (\ref{eq:kT_region1}) or (\ref{eq:kT_region2})]. It is not hard to check that $\Lambda_{+} > 0$, $\Lambda_{-} < 0$ in (\ref{eq:kT_region1}) and $\Lambda_{+} < 0$, $\Lambda_{-} > 0$ in (\ref{eq:kT_region2}). Note that the only region where the `$\pm$' modes are unstable simultaneously ($\Lambda_{\pm}>0$) is where the IC modes are also unstable ($\kappa_T > 1 + k^2_z$).

As we have argued before, these modes are to the (stabilised) ITG mode what the DD instabilities are to the (stabilised) IC mode. We shall refer to them as $\chi$ITG$+$ and $\chi$ITG$-$ modes, where `$\pm$' refer to the signs appearing in front of $\sqrt{\Lambda_0}$. The stabilisation in (\ref{eq:kT_region1}), unleashing the $\chi$ITG$+$ mode, is electrostatic ($\kappa_T < 1/4$ when $k_z \rightarrow \infty$), whereas the stabilisation in (\ref{eq:kT_region2}) is the finite-$\beta_e$ effect discussed in Section~\ref{sec:lin_finite_beta}. 

\begin{figure}[]%
    \centering
    \includegraphics[width=7cm]{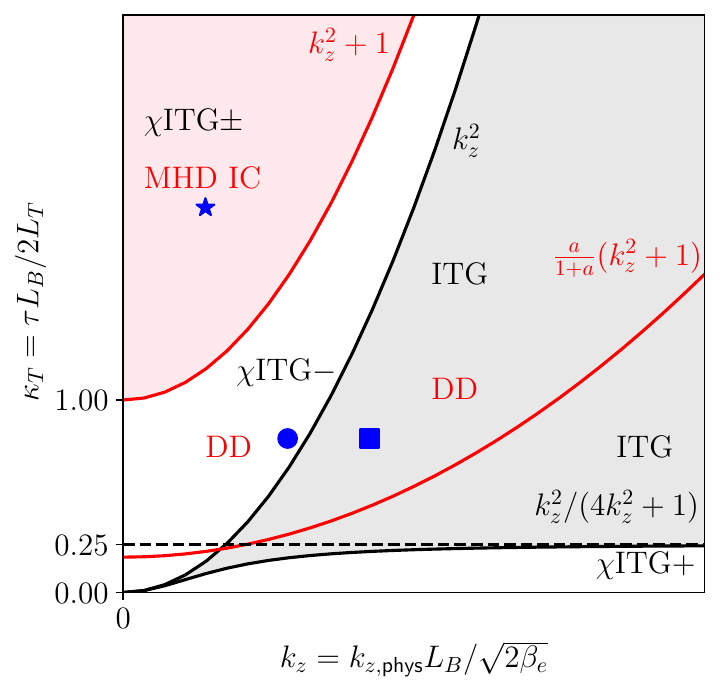}         
    \caption{A sketch of the regions of various instabilities in the $(k_z,\,\kappa_T)$ space. The red lines show the boundaries of the DD instability (\ref{eq:dd_stability_boundary}) discussed in \ref{sec:Interchange and double-diffusive instabilities}. The black lines show the boundaries (\ref{eq:kT_region1}) and (\ref{eq:kT_region2}) of the $\chi$ITG$\pm$ modes discussed in \ref{sec:ITG instability and its collisional extension}. Outside these boundaries, the ideal IC (Section~\ref{sec:lin_mhd}) and ITG (Sections~\ref{sec:lin_ITG} and \ref{sec:lin_finite_beta}) modes are the dominant instabilities. The blue star, dot, and square are the cases discussed further in \ref{more linear numerical results}.}
    \label{fig:stability_boundary}
\end{figure}

The landscape of instabilities in the $(k_z,\,\kappa_T)$ plane is sketched in Fig.~\ref{fig:stability_boundary}.

\subsection{Numerical investigation}
\label{more linear numerical results}
The analytical work presented in \ref{sec:Interchange and double-diffusive instabilities} and \ref{sec:ITG instability and its collisional extension} is limited by the assumption $k_y \ll 1$, which was adopted to simplify the analysis and clarify the underlying physics. To investigate the behaviour of instabilities in the regime $k_y \sim 1$, it is necessary to use numerical tools. In this appendix, we present a numerical study of the linear instabilities discussed in \ref{sec:Interchange and double-diffusive instabilities} and \ref{sec:ITG instability and its collisional extension}, but now based on the full dispersion relation (\ref{eq:disper_colli}). In particular, we will demonstrate how different types of instabilities can coexist at specific locations in the $(k_z,\kappa_T)$ parameter space shown in Fig.~\ref{fig:stability_boundary}.

\begin{figure}[]%
    \centering
    \subfloat[\centering  ]{{\includegraphics[width=8cm]{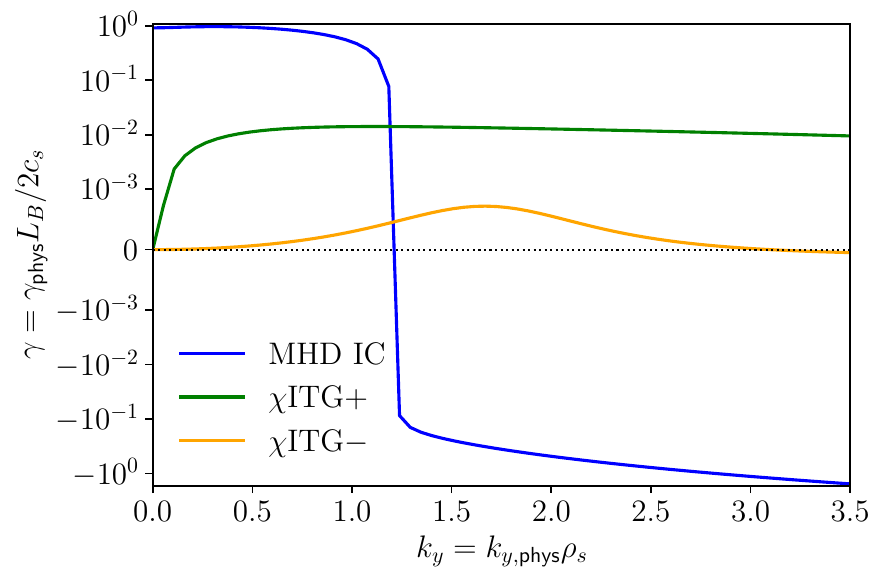} }      \label{fig:g_vs_ky_sub_2.0}}%
    \vfill
    \subfloat[\centering ]{{\includegraphics[width=8cm]{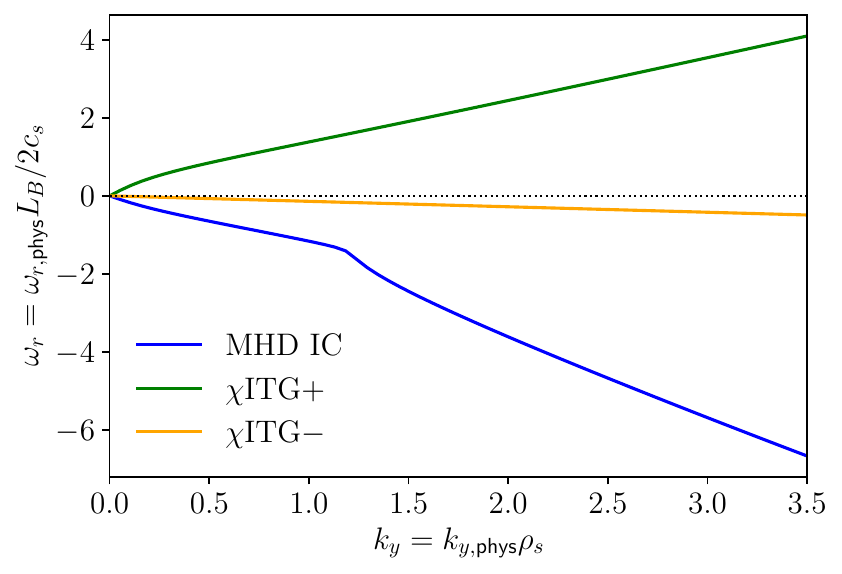} }  \label{f_vs_ky_sub_2.0}}%
    \caption{Growth rate (a) and frequency (b) spectra at $k_z = 0.4$ and $\kappa_T = 2.0$ (indicated by a blue star in Fig.~\ref{fig:stability_boundary}). The $\chi$ITG$\pm$ modes have much smaller growth rates than the IC modes at this value of $k_z$.}%
     \label{fig:plotgky_sub_2.0}
\end{figure}

In Fig.~\ref{fig:plotgky_sub_2.0}, we present the growth rate and real frequency as functions of $k_y$ for all $k_x = 0$ instabilities at $k_z = 0.4$ and $\kappa_T = 2.0$ (indicated by the blue star in Fig.~\ref{fig:stability_boundary}). Three distinct growing modes are found, consistent with the discussion in \ref{sec:ITG instability and its collisional extension}. The IC mode is dominant, while the $\chi$ITG$\pm$ modes are subdominant, exhibiting significantly lower growth rates.

\begin{figure}[]%
    \centering
    \subfloat[\centering  ]{{\includegraphics[width=8cm]{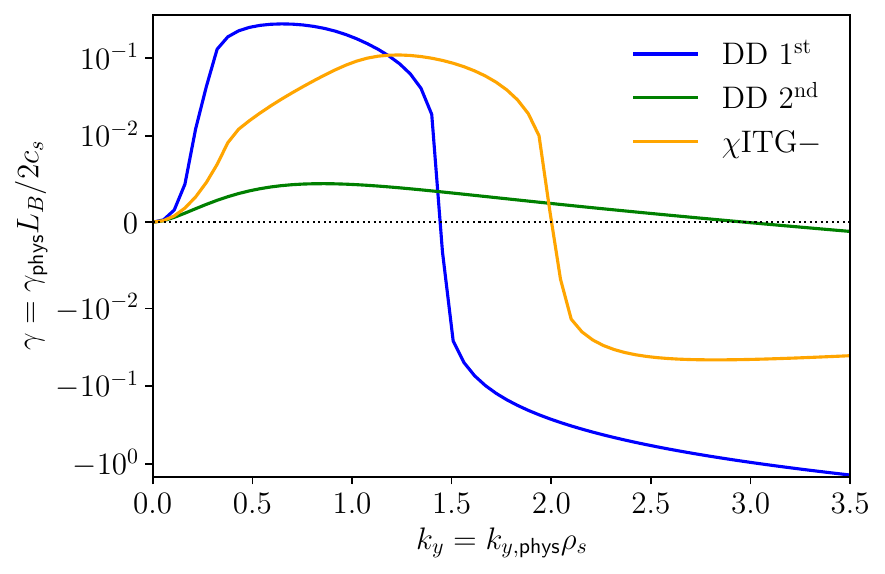} }      \label{fig:g_vs_ky_sub}}%
    \vfill
    \subfloat[\centering ]{{\includegraphics[width=8cm]{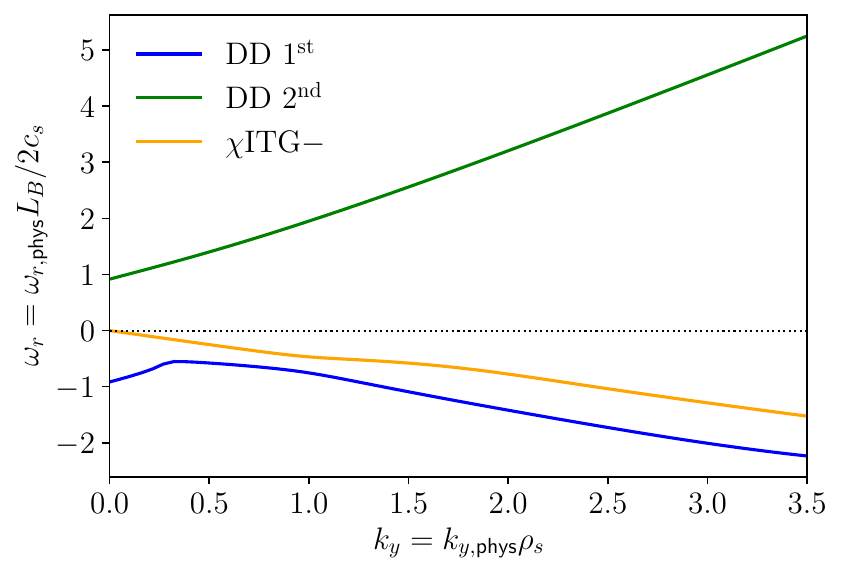} }  \label{f_vs_ky_sub}}%
    \caption{Growth rate (a) and frequency (b) spectra at $k_z = 0.8$ and $\kappa_T = 0.8$ (indicated by a blue dot in Fig.~\ref{fig:stability_boundary}).}%
     \label{fig:plotgky_sub_chi=0.2}
\end{figure}

\begin{figure}[]%
    \centering
    \subfloat[\centering  ]{{\includegraphics[width=8cm]{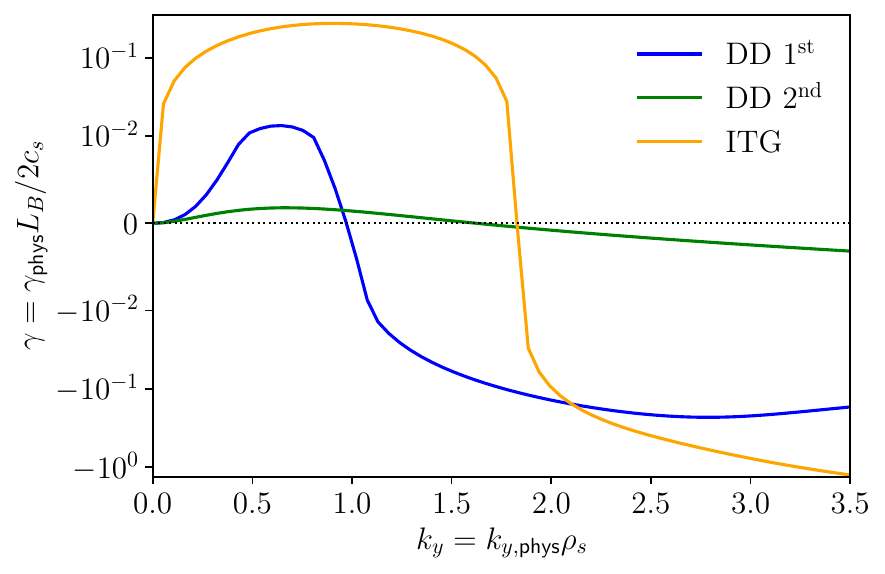} }}     
    \vfill
    \subfloat[\centering ]{{\includegraphics[width=8cm]{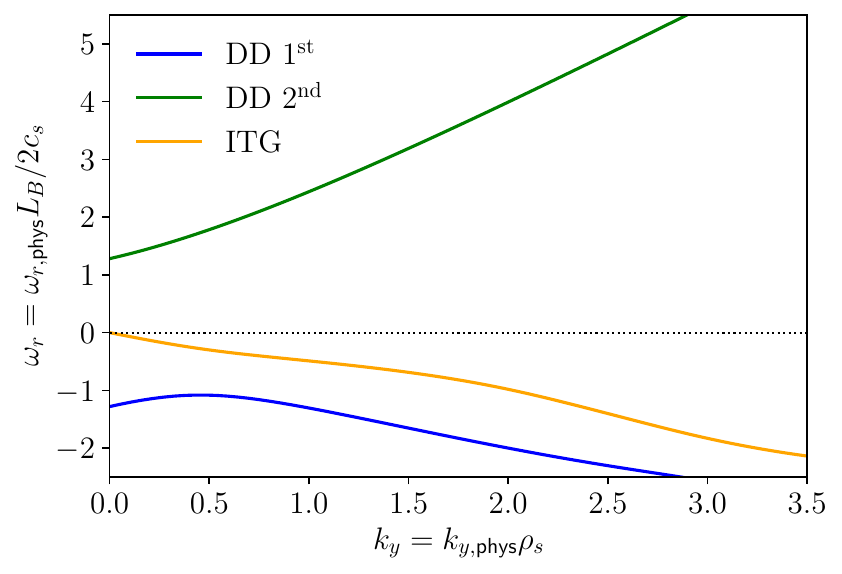}} } 
    \caption{Growth rate (a) and frequency (b) spectra at $k_z = 1.2$ and $\kappa_T = 0.8$ (indicated by a blue square in Fig.~\ref{fig:stability_boundary}).}%
     \label{fig:plotgky_sub_chi=0.2_kz=1.2}
\end{figure}

Similar plots for the growth rate and real frequency as functions of $k_y$ for all $k_x = 0$ instabilities at $k_z = 0.8$ and $\kappa_T = 0.8$ (indicated by the blue dot in Fig.~\ref{fig:stability_boundary}) are shown in Fig.~\ref{fig:plotgky_sub_chi=0.2}. Similarly to the $\kappa_T = 2.0$ case, three distinct instabilities are present here. Two of these have $|\omega_r| \sim 1$, in agreement with the analytical expression (\ref{eq:dd_frequency_low_ky}) for the frequency of the DD modes, with the mode of negative frequency being dominant at finite $k_y$ (at $k_y \ll 1$, the two have the same growth rate). The third instability is the $\chi$ITG$-$ mode, as its frequency spectrum is consistent with the analytical expression in (\ref{eq:omega_r_finitechi_itg}), which predicts $\omega_r \approx -k_y/2$ for this mode at low $k_y$. If $k_z$ is increased to $1.2$ (indicated by the blue square in Fig.~\ref{fig:stability_boundary}), the $\chi$ITG$-$ mode is replaced by the ITG mode, whose growth rate scales linearly with $k_y$ at $k_y \ll 1$ and becomes the dominant instability, as shown in Fig.~\ref{fig:plotgky_sub_chi=0.2_kz=1.2}.

To conclude, our analytical investigation at \mbox{$k_y \ll 1$} appears to have captured the qualitative (in)stability landscape of our problem well. At finite $k_y$, the growth rates peak and then peter out.

\clearpage

\section{Time traces}
\label{time traces}
In this appendix, we present representative time traces of the heat flux and electromagnetic field fluctuations corresponding to the simulations presented in Section~\ref{sec:nonlin_sim}. These time traces are included to illustrate the procedure used to obtain time-averaged quantities and to assess the reliability of those averages.

\begin{figure}[H]%
    \centering
    \subfloat[\centering  $\sum_{k_x,k_z} |\varphi_{\mathbf{k}}|^2$ ]{{\includegraphics[width=7cm]{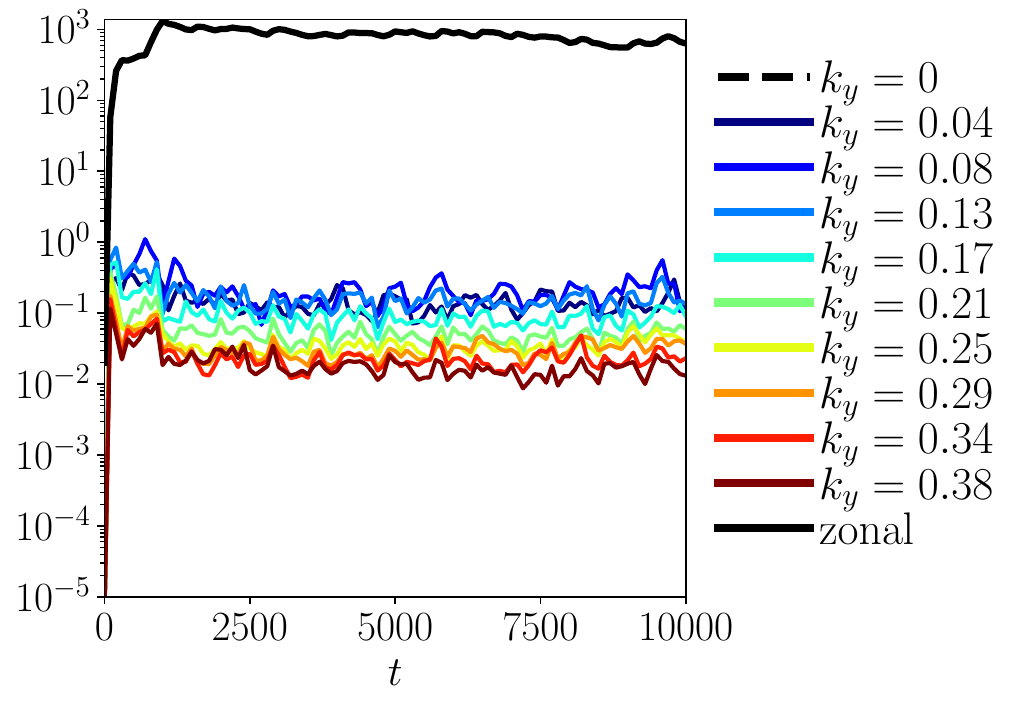}}      \label{fig:phi2_by_ky_vs_t_0.8_4.0}}%
    \vfill
    \subfloat[\centering $\sum_{k_x,k_z} |A_{\mathbf{k}}|^2$]{{\includegraphics[width=7cm]{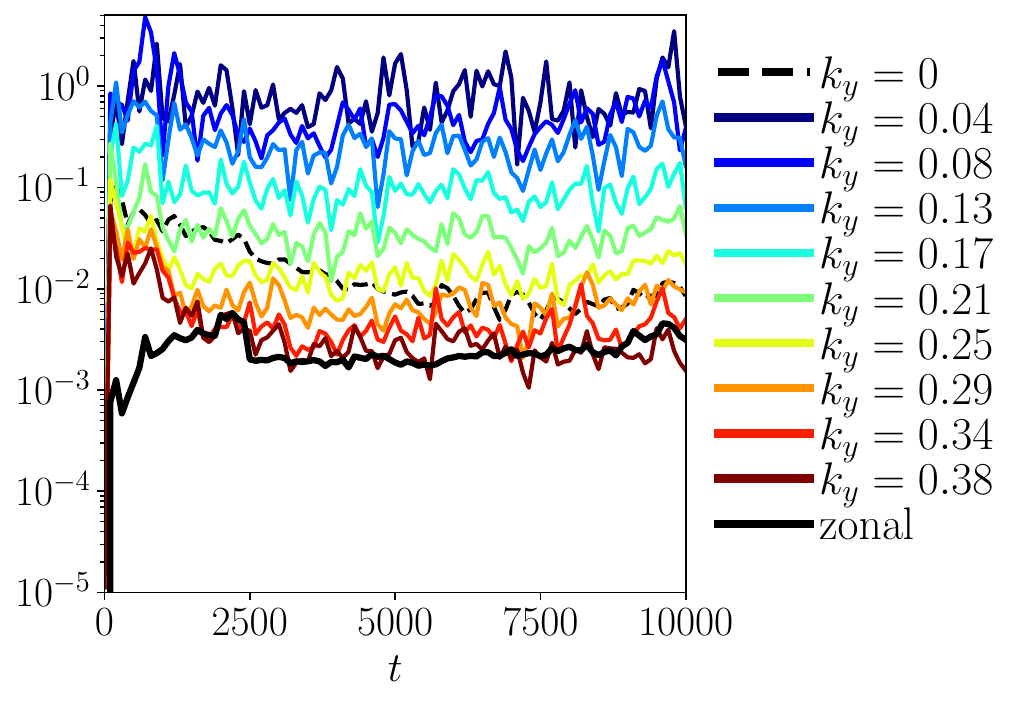}}  \label{fig:A2_by_ky_vs_t_0.8_4.0}}%
    \vfill
    \subfloat[\centering $Q(t)$]{{\includegraphics[width=7cm]{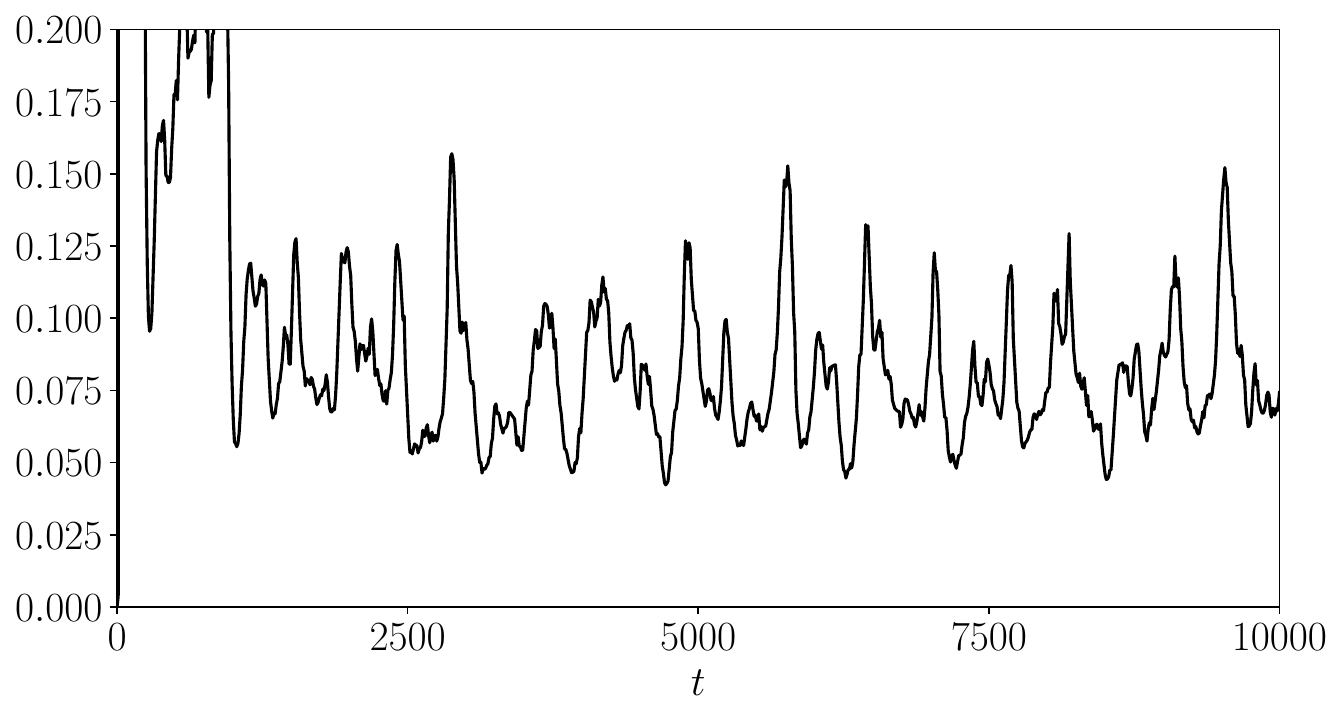}}      \label{fig:Q_flux_0.8_4.0}}%
    \caption{Time traces (up to $t = 10000$) of the perturbations of (a) $\varphi$, (b) $A$ for $\kappa_T = 0.8$, $L^2_z = 7.8$, and $\chi = 0.2$. Only the first $10$ $k_y$ modes are plotted. (c): Heat flux $Q$ versus $t$.}
    \label{fig:plot_timetrace_0.8_4.0}
\end{figure}

\begin{figure}[H]%
    \centering
    \subfloat[\centering  $\sum_{k_x,k_z} |\varphi_{\mathbf{k}}|^2$ ]{{\includegraphics[width=7cm]{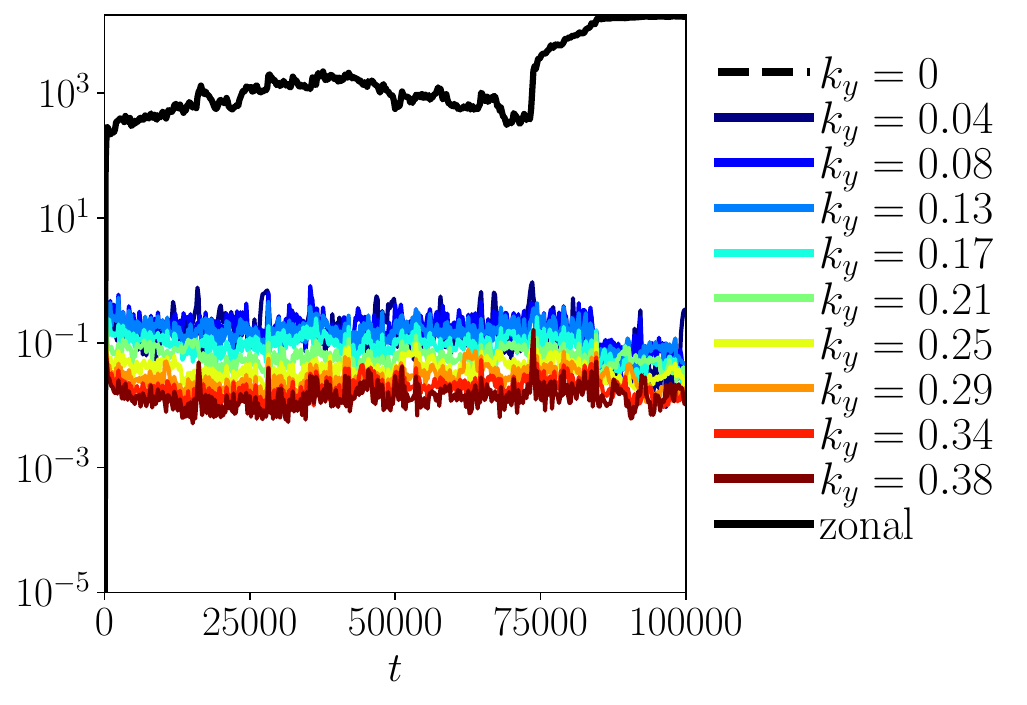}}      \label{fig:phi2_by_ky_vs_t_0.8_4.0_l}}%
    \vfill
    \subfloat[\centering $\sum_{k_x,k_z} |A_{\mathbf{k}}|^2$]{{\includegraphics[width=7cm]{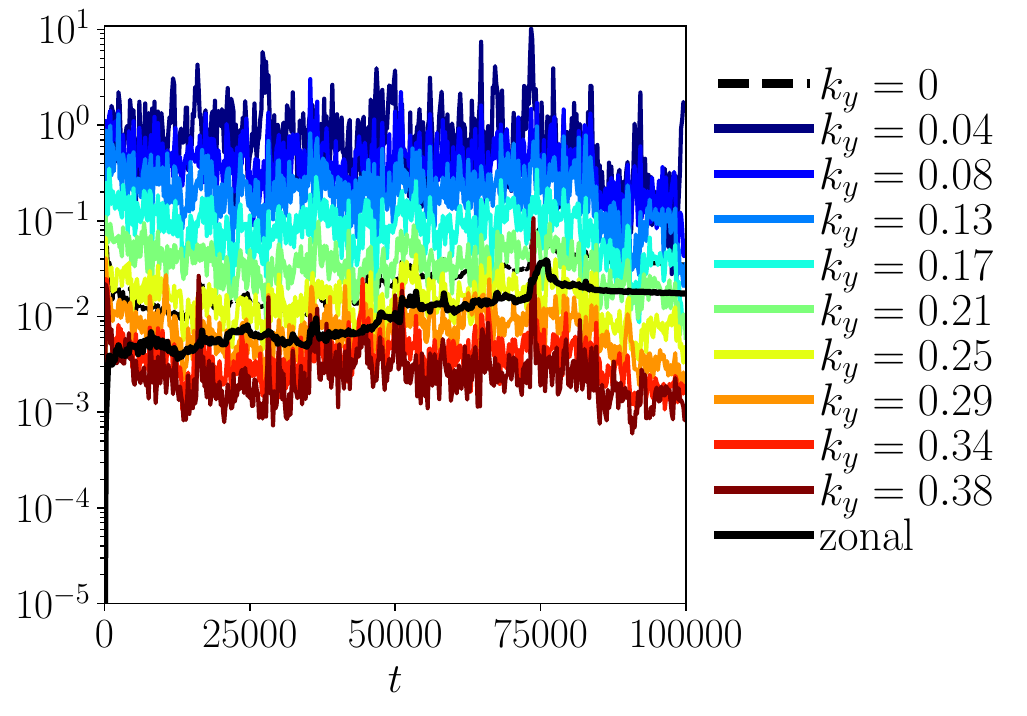}}  \label{fig:A2_by_ky_vs_t_0.8_4.0_l}}%
    \vfill
    \subfloat[\centering $Q(t)$]{{\includegraphics[width=7cm]{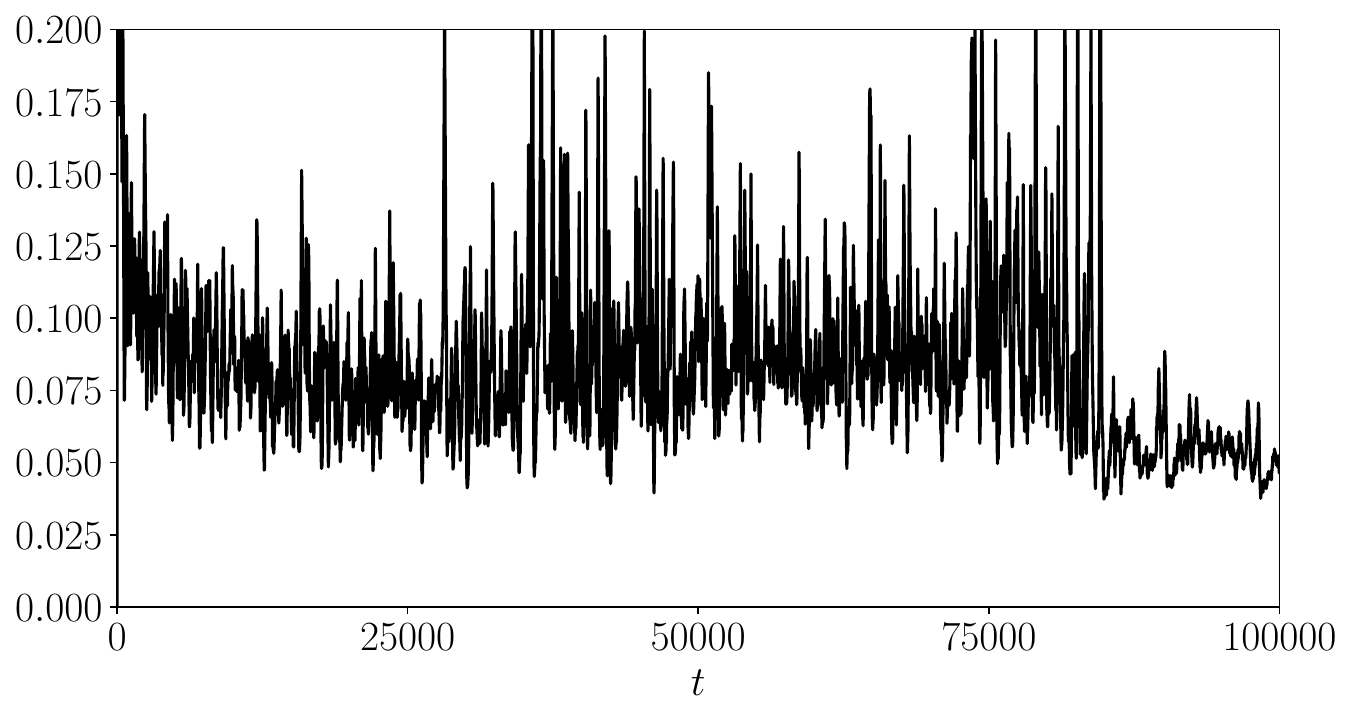}}      \label{fig:Q_flux_0.8_4.0_l}}%
    \caption{Same as Fig.~\ref{fig:plot_timetrace_0.8_4.0} but with time traces extended to $t = 100000$.}
    \label{fig:plot_timetrace_0.8_4.0_l}
\end{figure}

A representative `early-time' trace used for calculating time-averaged fluxes is shown in Fig.~\ref{fig:plot_timetrace_0.8_4.0}. The heat flux and nearly all field components appear to reach a statistically steady state. The zonal part of $\varphi$ dominates over other modes, which is a sign of zonal-flow-dominated turbulence. However, while the zonal component of $A$ remains relatively small, it continues to evolve slowly throughout the simulation. Extending the simulation by approximately an order of magnitude in time (Fig.~\ref{fig:plot_timetrace_0.8_4.0_l}) shows that the zonal component of $A$ continues its growth over a much longer time scale than that required to achieve the initial saturation of the turbulence.

\begin{figure}[H]%
    \centering
    \subfloat[\centering  $\sum_{k_x,k_z} |\varphi_{\mathbf{k}}|^2$ ]{{\includegraphics[width=7cm]{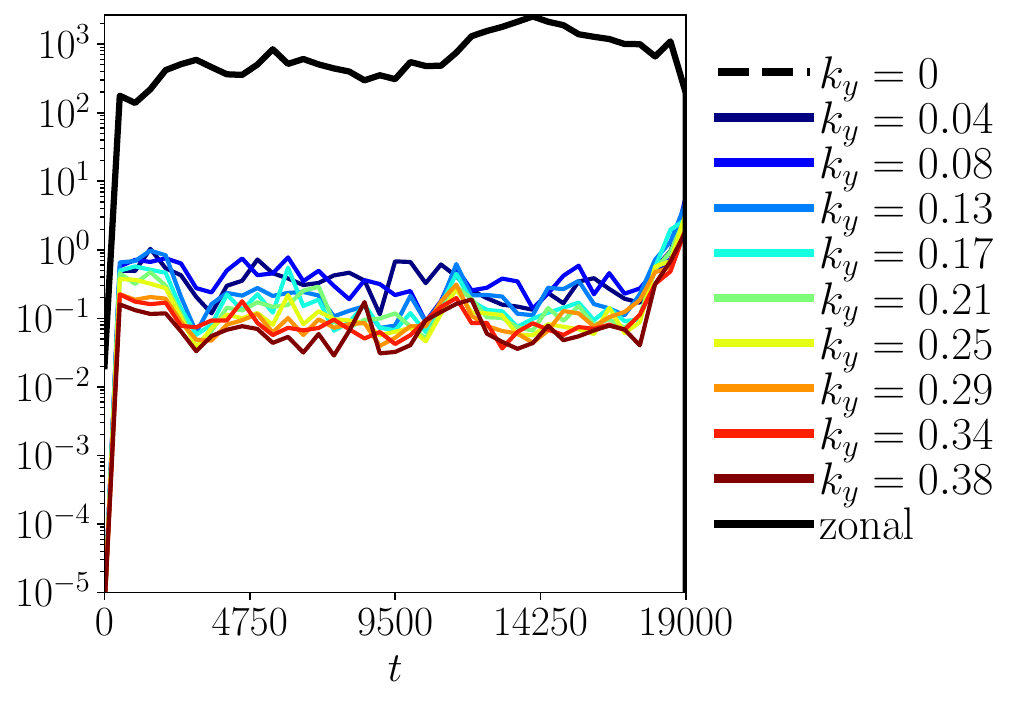}}      \label{fig:phi2_by_ky_vs_t_0.8_6.0}}%
    \vfill
    \subfloat[\centering $\sum_{k_x,k_z} |A_{\mathbf{k}}|^2$]{{\includegraphics[width=7cm]{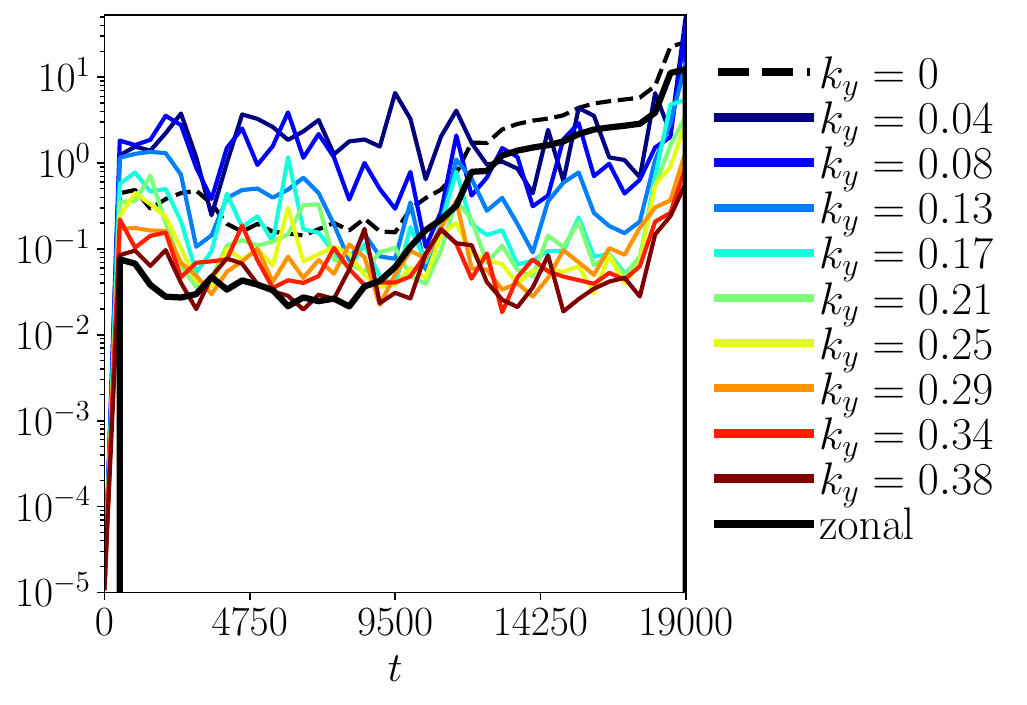}}  \label{fig:A2_by_ky_vs_t_0.8_6.0}}%
    \vfill
    \subfloat[\centering $Q(t)$]{{\includegraphics[width=7cm]{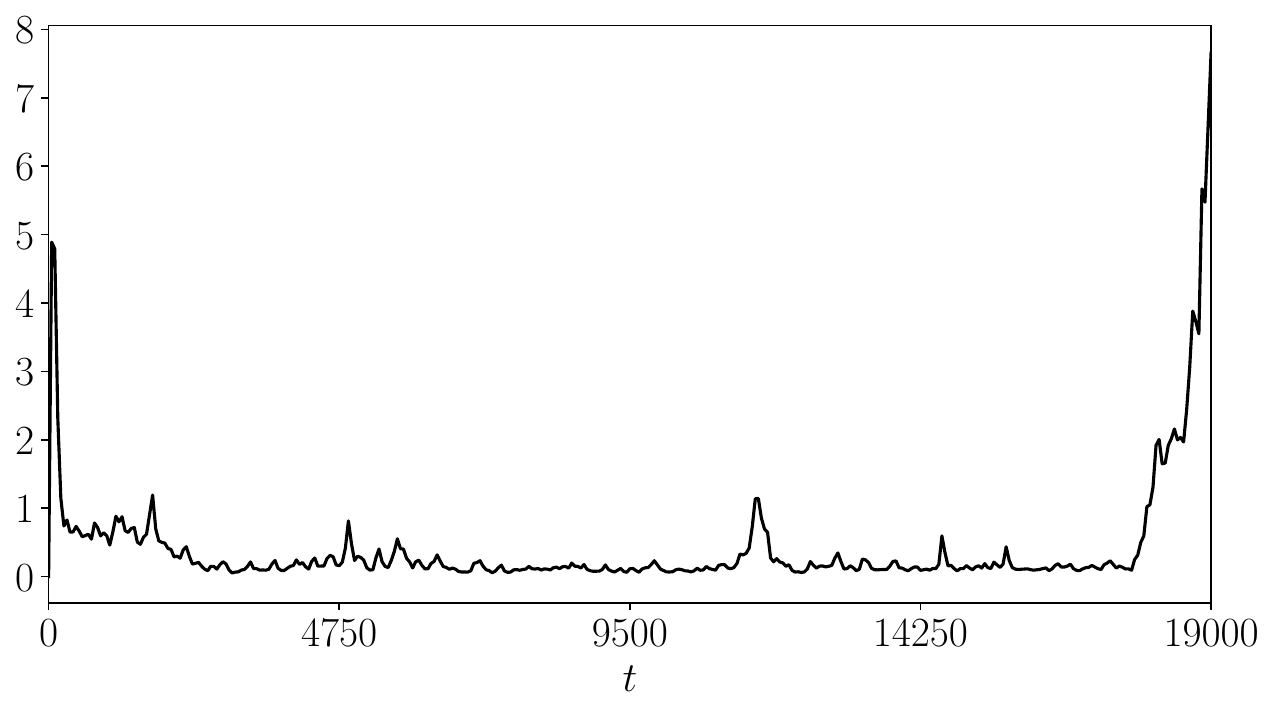}}      \label{fig:Q_flux_0.8_6.0}}%
    \caption{Same as Fig.~\ref{fig:plot_timetrace_0.8_4.0} but for a run with $L^2_z = 17.6$ and time traces up to $t = 19000$.}
    \label{fig:plot_timetrace_0.8_6.0}
\end{figure}

Once the zonal component of $A$ reaches a sufficiently large amplitude, it can disrupt the turbulent steady state. An example of this behaviour is shown in Fig.~\ref{fig:plot_timetrace_0.8_6.0}, which is obtained from a simulation with higher value of $L^2_z$. There, the slow growth of zonal $A$ eventually triggers a runaway process in which the heat flux grows rapidly at the end. It remains unclear whether this disruption would ultimately saturate at a high but finite transport level, as the simulation had to be terminated at approximately \(t = 19000\) when the timestep became prohibitively small.

\begin{figure}[H]%
    \centering
    \subfloat[\centering  $\sum_{k_x,k_z} |\varphi_{\mathbf{k}}|^2$ ]{{\includegraphics[width=7cm]{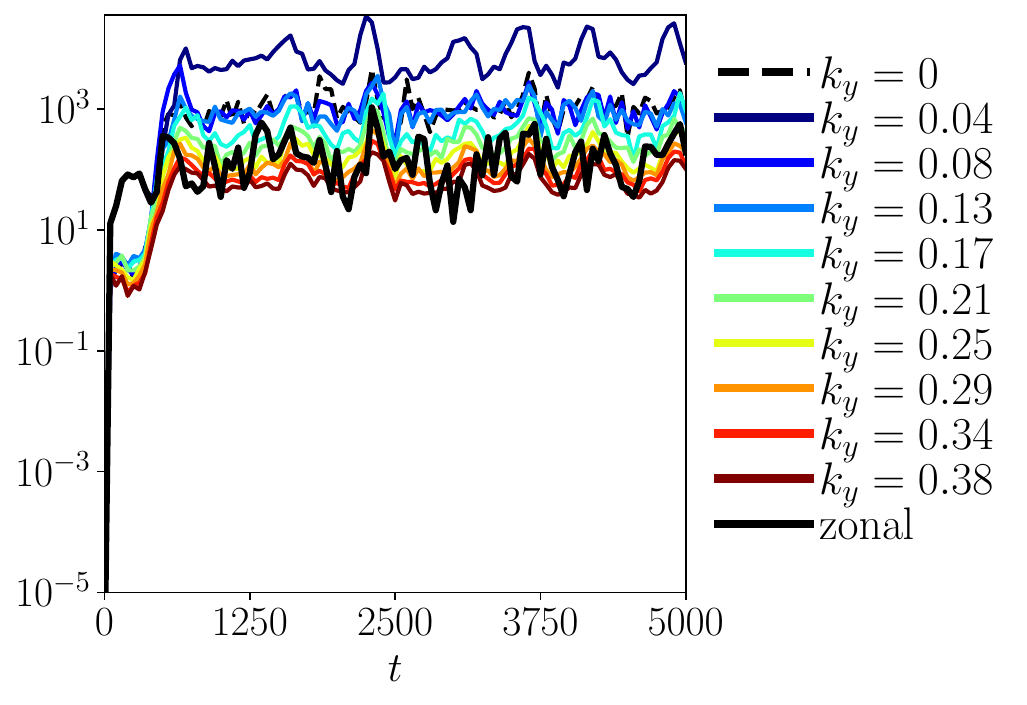}}      \label{fig:phi2_by_ky_vs_t_0.8_6.25}}%
    \vfill
    \subfloat[\centering $\sum_{k_x,k_z} |A_{\mathbf{k}}|^2$]{{\includegraphics[width=7cm]{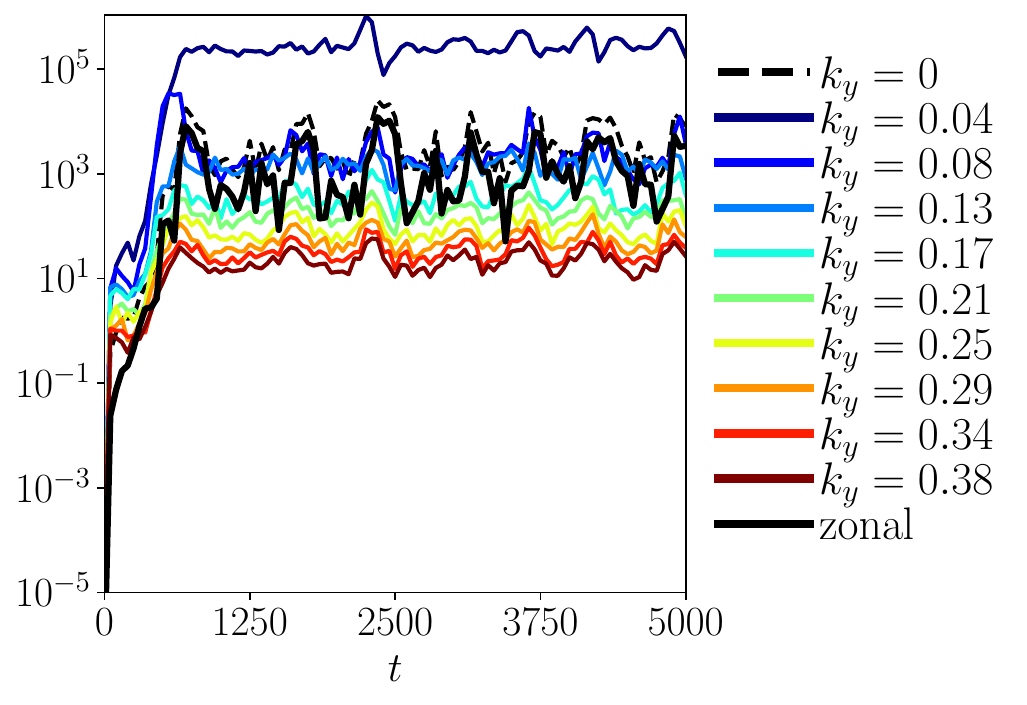}}  \label{fig:A2_by_ky_vs_t_0.8_6.25}}%
    \vfill
    \subfloat[\centering $Q(t)$]{{\includegraphics[width=7cm]{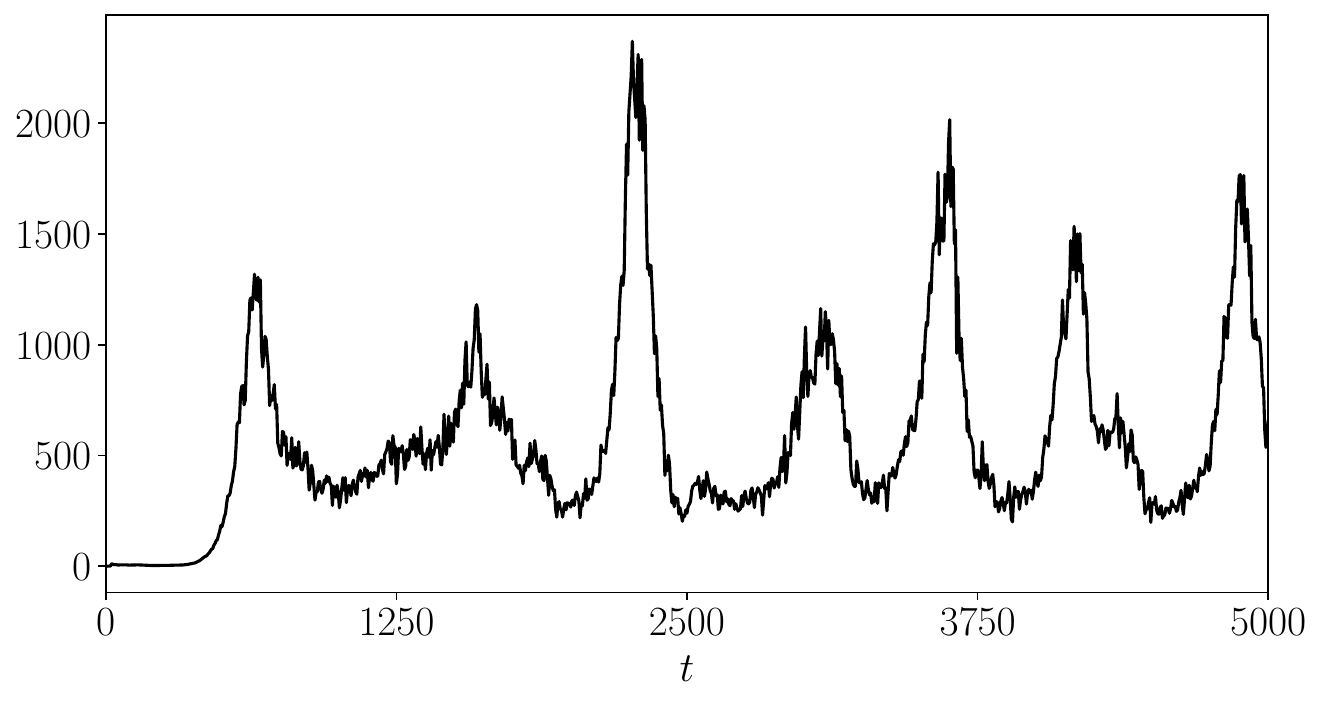}}      \label{fig:Q_flux_0.8_6.25}}%
    \caption{Same as Fig.~\ref{fig:plot_timetrace_0.8_4.0} but for a run with $L^2_z = 19.4$ and time traces up to $t = 5000$.}
    \label{fig:plot_timetrace_0.8_6.25}
\end{figure}

This long-term growth of zonal $A$ can be eliminated by introducing a resistivity term into \eqref{eq:A_red}, which is formally absent from our model because of the ordering assumptions employed. Nevertheless, this long-time behaviour does not present a practical difficulty for identifying the runaway transition discussed in Section~\ref{sec:nonlin_sim}, which is the focus of this study, for two reasons. First, the zonal component of $A$ can be artificially removed while still retaining the runaway transition to the high-flux states. Secondly, simulations exhibiting a runaway transition do so on much shorter time scales, long before zonal $A$ has had sufficient time to grow appreciably, as illustrated by the time traces in Fig.~\ref{fig:plot_timetrace_0.8_6.25}, which is obtained from a simulation with a value of $L^2_z$ beyond the runaway-transition threshold. In this regime, one observes that the zonal part of $\varphi$ has become much smaller than the $k_y\neq 0$ modes, suggesting that the zonal flow is too weak to suppress the turbulence. The system instead enters a non-zonal state with heat fluxes orders of magnitude larger than those obtained for smaller \(L^2_z\). Since the dynamics remain dominated by box-scale streamers, these states should not be interpreted as fully converged turbulence.

\end{document}